\documentclass[aps,prd,twocolumn,amsmath,superscriptaddress,amssymb,showpacs,floatfix,nofootinbib, superscriptaddress, longbibliography]{revtex4-1}
\usepackage{graphicx}
\usepackage{dcolumn}
\usepackage{xcolor}
\usepackage{bm}
\usepackage{natbib}
\usepackage{calligra}
\usepackage{egothic}
\usepackage[T1]{fontenc}
\newfont{\rsfsten}{rsfs10 scaled 1200}
\newfont{\rsfsseven}{rsfs10 scaled 1200}
\newfont{\rsfsfive}{rsfs10 scaled 1200}
\usepackage{color}
\usepackage{epsfig}
\usepackage{units}
\usepackage{hyperref}
\usepackage[utf8]{inputenc}

\newcommand{\apjl}{{Astrophys.~J.~Lett.}}

\newcommand{\aj}{{Astron.~J.}}
\newcommand{\mnras}{{Mon.~Not.~R.~Astron.~Soc.}}

\newcommand{\planss}{{Planetary~and~Space Science}}

\begin{document}

\hspace{13cm}
\space{0.6cm}

\title{Evaluating the Merger Rate of Binary Black Holes from Direct Captures and Third-Body Soft Interactions Using the Milky Way Globular Clusters}

\author{Konstantinos Kritos}
\email{ge16004@central.ntua.gr}
\affiliation{Physics Division, National Technical University of Athens, Zografou, Athens, 15780, Greece}
\author{Ilias Cholis}
\email{cholis@oakland.edu, ORCID: orcid.org/0000-0002-3805-6478}
\affiliation{Department of Physics, Oakland University, Rochester, Michigan, 48309, USA}

\date{\today}

\begin{abstract}

   The multitude of binary black hole  
   coalescence detections in gravitational waves
   has renewed our interest on 
   environments that can be the cradle of these mergers. 
    In this work we study 
    merger rates of binary black holes in globular clusters that 
    are among the most dense stellar environments and a natural 
    place for the creation of black hole 
    binaries. 
    To model these systems with all their variations we rely on the observational properties of the 
    known Milky Way globular clusters. We consider direct capture 
    events between black holes, as well as
    soft interactions of black hole binaries with stars as third bodies  
    that accelerate the evolution of these binaries. 
    We find that binary black holes from direct captures merge at an averaged  rate of 
    $0.3-5 \times 10^{-11}$ yr$^{-1}$ per cluster. Third body soft interactions are a much 
    more prominent channel giving an averaged rate of  
    $2-4 \times 10^{-10}$ yr$^{-1}$ per cluster.
    Those rates in globular clusters 
    can lead to a cumulative merger rate of about 100 mergers per year up to redshift of 1, 
    i.e. a significant fraction of the detectable in the near future binary black hole coalescence events. 
    Further observations of cluster properties both in terms of their 
    masses, profile properties, velocity dispersion of stars 
    and their cosmological distribution,  
    will allow us to better constrain
    the contribution of these environments to the detectable coalescence events rate.

\end{abstract}

\maketitle

\section{Introduction}
\label{intro}

The Laser Interferometer Gravitational wave Observatory (LIGO)  \cite{TheLIGOScientific:2014jea} 
with its first three observing runs concluded allows us a first measurement on the rates of 
binary black hole (BBH) merges in the local Universe \cite{LIGOScientific:2018jsj}; and ask questions 
on the mechanisms/environments responsible for them \cite{Bethe:1998bn, PortegiesZwart:1999nm, 
Belczynski:2001uc, OLeary:2008myb, 2010MNRAS.402..371B, Antonini:2012ad, 2012MNRAS.425..460M, 
Dominik:2013tma, 2013MNRAS.429.2298M, Ziosi:2014sra, Antonini:2016gqe, Bird:2016dcv, 
Sasaki:2016jop, Stone:2016wzz, Carr:2016drx, Askar:2016jwt, Stevenson:2017tfq, Spera:2018wnw, 
Mapelli:2018wys, Gerosa:2019zmo, Baibhav:2020xdf}. Moreover, as LIGO observes more coalescence events, the 
growing statistics will allow us to address their population synthesis \cite{Kovetz:2016kpi, 
Eldridge_2017, Mandel:2018mve, Bouffanais:2019nrw, Baibhav:2019gxm}. Globular clusters (GCs) are 
among the systems with the highest densities in stars, and environments where stellar mass objects 
can undergo multiple dynamical encounters that may lead to merger events \cite{PortegiesZwart:1999nm, 
OLeary:2005vqo, 2010MNRAS.407.1946D, Rodriguez:2015oxa, Rodriguez:2016kxx, Askar:2016jwt, 
Fragione:2018vty}. Extensive N-body simulations regarding the evolution and interactions of black holes 
(BH) inside dense star cluster systems have been performed in the past, \cite{Rodriguez:2016kxx, 
hong2018, haster2016, park2017, pavlik2018}. We utilize this rich dynamics to evaluate merger rates 
of BBH systems in dense stellar clusters throughout cosmic history. Such clusters may 
provide an explanation for the origin of an important component of the total merger rate of BH binaries, 
and also the places where consecutive mergers occur \cite{OLeary:2005vqo,  Fishbach:2017dwv, 
Kovetz:2018vly, Antonini:2018auk}, that can be probed by current observatories \cite{Fishbach:2017zga, 
Fishbach:2019bbm, Gerosa:2020bjb, Kimball:2020opk,Rodriguez:2020viw}.

In this article we study mergers of stellar mass BBH systems formed in GCs. 
We model the mass-distribution  of objects in GCs by the sum of two Dirac delta functions.
One represents a generic light stellar object 
of mass $m_{\textrm{star}}=1 \, M_{\odot}$ and the other a typical BH of mass $m_{BH}=10 \, M_{\odot}$. 
We will refer to this choice as ``two body model''.
Their relative weight is determined by the initial stellar mass function. 
Some of these BHs will participate in binaries. 
Our choice of mass-distribution allows us 
to focus on 
binaries of equal-mass BHs distributed throughout the 
GCs. 
Unequal mass ratios are rare in GCs, as strong interactions of the binary 
with massive third bodies as other BHs lead to the binary 
exchanging its lighter member with the interacting more massive object. 
Furthermore, we assume that the stellar population follows that of the GC's mass density profile with 
the appropriate normalization. Most of the BHs are concentrated instead at the denser core due to mass segregation that 
happens in the first $O(10^{2})$ Myr of the cluster's evolution.

We ignore effects as mass loss during the cluster's evolution as the dynamics of BBHs take place 
mostly at the inner parts of the cluster's profile.
However, we note that as BBHs interact strongly with individual stars they get on orbits that temporarily 
move them further away from the GC core and probe a wider volume of the cluster.
We find that 3rd-body soft interactions, 
do not typically eject the BBHs before they  are already quite tight after which point they will rapidly merge.
While interactions of BHs with stars lead to the depletion of 
stars from the very center of the GCs, $1 \, M_{\odot}$ stars are not 
completely depleted out to the core radius of the GCs. 

An isolated binary in the low density field loses energy into gravitational waves (GW) as predicted by 
the general theory of relativity. This radiation reaction leads to coalescence on a timescale given by 
\cite{peters1964},
\begin{equation}
\label{peters0}
    T_{gw}(e_0=0)\approx1.58\times10^{14}\, \left({a_0\over1 \, \textrm{AU}}\right)^4 \left({10 \, M_{\odot}\over m}\right)^3 \, \textrm{yr} ,
\end{equation}
for a circular binary of equal mass $m$ BHs with initial semi-major axis and eccentricity given 
by the pair $(a_0, e_0)$. For widely separated binaries gravitational radiation is an inefficient process to
lead to coalescence. However, in dense environments dynamical interactions accelerate the binary's evolution. 
This can be achieved in various ways and many different channels have been proposed in the literature 
regarding the coalescence of two BHs, in dense environments, see e.g. \cite{Rodriguez:2016kxx, mandel2008}. 
Relevant in this context, mechanisms include direct capture (DC) merger events \cite{quinlanShapiro1989, 
Mouri:2002mc} and the hardening process of a binary via encounters 
with third bodies, mainly soft third-body 
interactions \cite{1983AJ.....88.1269H, Sesana:2006xw, rodriguez2018, samsing2018}. For the case of  high merger rates this may also lead to runaway growth of 
intermediate-mass BHs in GCs, \cite{Kovetz:2018vly, Antonini:2018auk}.
Other type of effects include stable \cite{miller2002} or meta-stable \cite{arcaSedda2018} triple resonance 
systems, which may form as a consequence of multiple-body interactions, \cite{antognini2016}.

In this work we focus 
on the two dominant contributing channels to our total merger rate from GCs. We start with the direct capture 
 events 
 and the interactions 
involving a BBH system an a third object
(3rd-body channel).
The values describing the GCs' mass, central densities 
and velocity dispersion of stars span orders of magnitude; which 
has a strong impact on their respective  merger rates.
After averaging over the known Milky Way GCs we find that the direct capture merger rate is $0.3-5 \times 10^{-11}$ yr$^{-1}$ per cluster.
Moreover, by evolving the BBHs in the environments of their respective  Milky Way GCs, we evaluate
the merger rate due to 3rd-body soft interactions to be 
 $2-4 \times 10^{-10}$ yr$^{-1}$ per cluster. 
 The rates for the 3rd-body channel are directly proportional 
 to the fraction of BHs that 
 will remain in the 
 cluster and form BBHs hard enough to survive their first encounters with 
 stars. The ratio of the number of these BBHs to the number of the total BHs created from
 stellar evolution is taken to be $0.3 \%$.  

This paper is constructed as follows. In section~\ref{sec:method} we discuss our assumptions on 
the globular cluster properties and the abundance of BHs in them. We also show our methodology for 
calculating the BBH merger rates from direct capture events and from third-body 
hardening processes. 
In section ~\ref{sec:results}, we first present results 
for our example globular cluster 47 Tuc. We then expand our results to a sample of 13 Milky Way GCs 
that encompass the variations between those environments; 
and then include the contribution from all Milky Way globular clusters that we have information for.
Using the Milky Way clusters as a representative sample for all clusters 
of the local Universe we then 
evaluate their BBH
cosmological merger rate. 
Finally, in section~\ref{sec:conclusions} we give our conclusions.

\section{Assumptions and Methodology}
\label{sec:method}

\subsection{The mass profile of globular clusters}

We take Milky Way GCs that have relatively well observed mass distributions, 
with that information publicly available at \cite{GLOBCLUST}. As a reference we
assume that the mass distribution of stars $\rho(r)$ in GCs follows a King profile, \cite{king1962},
\begin{equation}
\label{rhoKing}
    \rho_{\textrm{King}}(r)=\rho_0 {\left[ \left( 1+\left( {r\over r_c} \right)^2\right)^{-{1\over2}} - \left(1+100^c\right)^{-{1\over2}} \over  1-(1+100^c)^{-{1\over2}}  \right]^2},
\end{equation}
with $\rho_{0}$ the central density 
$\rho_{\textrm{King}}(r=0)$. 
The core radius $r_c$ and the concentration parameter $c$ are related to the tidal 
radius $r_t$ of the GC via $c=log_{10}\left({r_t/r_c}\right)$. We rely on Ref.~\cite{GLOBCLUST} 
for the particular values of individual GCs. 
The mass of each of these clusters 
is calculated by integrating their density out to their tidal radius,
\begin{equation}
\label{GCmass}
M_{GC} = \int_{0}^{r_t}dr \, 4\pi r^2 \, \rho_{\textrm{King}}(r).
\end{equation}

For the BHs that are in the GC, mass segregation takes place 
leading to BHs concentrating near the center of each GC. 
We assume for simplicity that all BHs are uniformly distributed within the core radius.
The exact profile of the BHs' density is a detail 
as our final rates will prove to depend mostly on the total number of BHs and BBHs per cluster. 

The velocity dispersion of the stars in the cluster can be calculated by applying the Virial theorem \cite{collins1978},
\begin{equation}
    \sigma(r)=\sqrt{2 G M(r)\over r}, 
\end{equation}
where $G$ is Newton's constant and $M(r)$ is the total mass contained within a sphere of radius $r$ centered at the center of the GC. The dynamics of the GC are characterized by the relaxation timescale ($O(10^2)$ Myr for BHs). 

As an alternative profile for the stars in GCs we take a Plummer model \cite{plummer1911},
\begin{equation}
    \rho_{\textrm{Plummer}}(r)={3\ M_{GC}\over4 \pi  r_{pl}^3} \left[1+\left({r\over r_{pl}}\right)^2\right]^{-{5\over2}},
\end{equation}
where $r_{pl}={r_c\over\sqrt{\sqrt{2}-1}}\approx1.554r_c$ (Eq. 38 from \cite{dejonghe1987}) is a characteristic parameter called the Plummer radius. We ensure that $M_{GC}$ for each cluster is the same regardless of their mass profile 
\footnote{Only up to $3\%$ of the mass of a GC is in BHs and thus $M_{GC}$ is accurate for the mass in the stars.}. 

The velocity dispersion is given by, \cite{dejonghe1987},
\begin{equation}
    \sigma_{\textrm{Plummer}}^2(r)={G M_{GC}\over6 \sqrt{r^2+r_{pl}^2}}.
\end{equation}

In Table~\ref{GCparameters} we show the profile properties for 13 Milky Way GCs. These constitute a representative sample of how much different clusters can contribute to our final results on the merger rate. 
\begin{table}[h]
    \centering
    \begin{tabular}{c c c c c c c}
    \hline
        GC & $r_c\over \textrm{1pc}$ & $c$ & $r_{pl}\over \textrm{1pc}$ & $log_{10}\left({\rho_0\over1M_{\odot}/\textrm{pc}^3}\right)$ & $M_{GC}\over10^{5}M_{\odot}$ & $N_{BH}^{\textrm{ret-max}}$ \\
    \hline
        47 Tuc       & 0.47 & 2.07 & 0.73 & 4.88 & 38.2 & 1145 \\
        $\omega$ Cen & 3.60 & 1.31 & 5.57 & 3.15 & 49.3 & 1477 \\
        M15          & 0.42 & 2.29 & 0.66 & 5.05 & 68.7 & 2060 \\
        M22          & 1.24 & 1.38 & 1.92 & 3.63 & 7.30 & 219 \\
        NGC 6362     & 2.50 & 1.09 & 3.88 & 2.29 & 1.29 & 38 \\
        NGC 5946     & 0.25 & 2.50 & 0.38 & 4.68 & 9.44 & 283 \\
        M 30         & 0.14 & 2.50 & 0.22 & 5.01 & 3.80 & 113 \\
        Terzan 5     & 0.32 & 1.62 & 0.50 & 5.14 & 7.51 & 225 \\
        Pal 2        & 1.35 & 1.53 & 2.09 & 4.06 & 36.8 & 1103 \\
        NGC 6139     & 0.44 & 1.86 & 0.69 & 4.67 & 11.7 & 351 \\
        NGC 2808     & 0.70 & 1.56 & 1.08 & 4.66 & 22.1 & 661 \\
        NGC 5286     & 0.95 & 1.41 & 1.48 & 4.10 & 10.6 & 319 \\
        NGC 6316     & 0.51 & 1.65 & 0.80 & 4.23 & 4.08 & 122 \\
        \hline
    \end{tabular}
    \caption{The parameters of 13 Milky Way GCs relevant in our calculations. The information for the second, third and fifth columns is from Ref.~\cite{harris1996}. 
    For the last column we have used $N_{BH}^{\textrm{ret-max}}=\left({f_{\textrm{ret}}\over0.1}\right) N_{BH}^{\textrm{max}}$ and a BH mass fraction $f_{BH}$ of $0.03$ (see Eqs~\ref{kroupa} and ~\ref{maxBHs}).}
    \label{GCparameters}
\end{table}

\begin{figure}
    \centering
    \includegraphics[width=8.7cm,height=5cm]{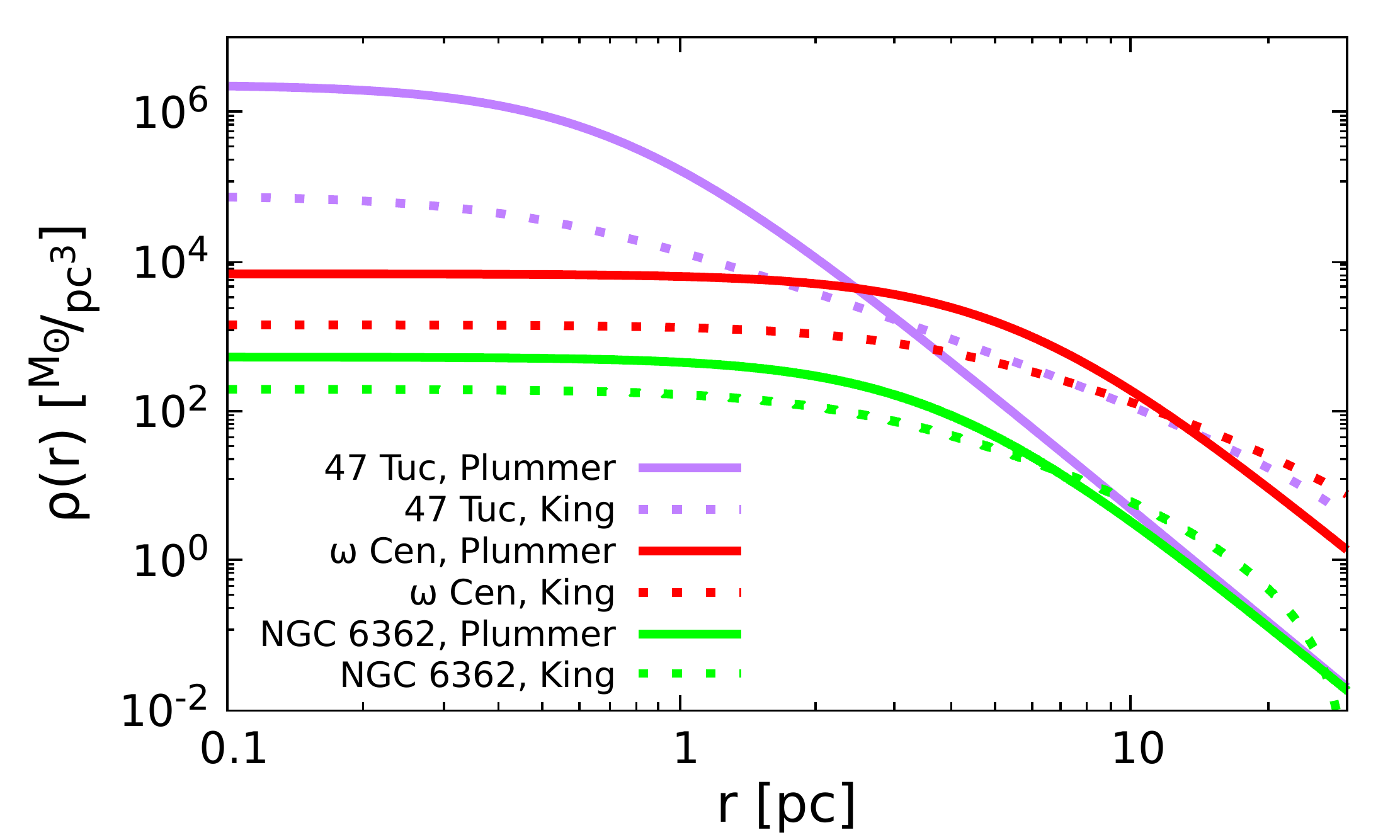} \\
    \vskip 0.15in
    \includegraphics[width=8.7cm,height=5cm]{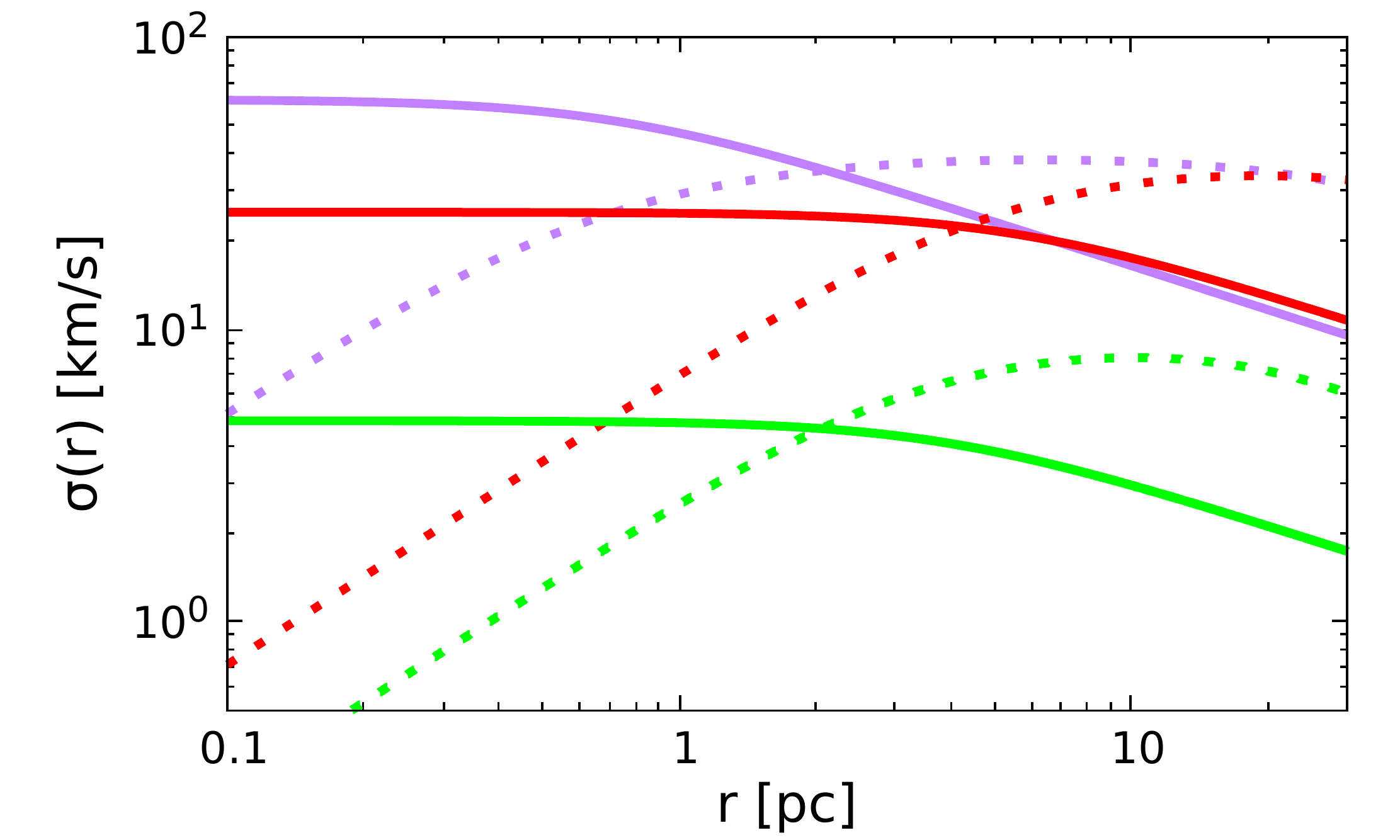}
    \caption{The profiles of mass density and velocity dispersion for 47 Tuc, $\omega$ Cen and NGC 6362. Solid lines represent the Plummer and dashed ones the King model.}
    \label{gcsProfile}
\end{figure}

In Fig.~\ref{gcsProfile} we give the profiles of $\rho(r)$ and $\sigma(r)$ for three Milky Way GCs.

\subsection{Black holes inside globular clusters}

Assuming that stars with mass $\mathcal{M}$ larger than 
$25 \, M_{\odot}$ necessarily give a BH within a few $10^6$ yr, 
and that about 1/3 of the star's original mass is retained by the resulting BH, 
we can estimate the mass fraction of the GC that ends up in BHs,
\begin{equation}
\label{kroupa}
    f_{BH}\simeq \frac{1}{3} {1\over M_{GC}}\,\int_{25 M_{\odot}}^{120 M_{\odot}}d\mathcal{M}\ \mathcal{M}\,\xi(\mathcal{M})\approx0.03,
\end{equation}
where $\xi(\mathcal{M})$ is the Kroupa initial mass function, for which we take the central values of \cite{kroupa2002}.
Taking all BHs to have a mass of $10 \, M_{\odot}$, we can 
estimate their maximum number in a GC to be,  
\begin{equation}
\label{maxBHs}    
    N_{BH}^{\textrm{max}}=f_{BH} {M_{GC}\over10 \, M_{\odot}}.
\end{equation}
As BHs have natal kicks only a fraction, $f_{ret}$, of them is retained in the cluster. The maximum retained number of BHs in the cluster is,
\begin{equation}
\label{eq:NBH_retmax}
    N_{BH}^{\textrm{ret-max}}=f_{\textrm{ret}}\times N_{BH}^{\textrm{max}}.
\end{equation}
Numerical N-body surveys and analytic considerations point out to a value of about $f_{\textrm{ret}}\simeq 10\%$ up to the tidal radius, consistent with a total mass of $M_{GC}\approx10^5 \, M_{\odot}$, virial radius of $r_v=1$ pc and $\sigma_{BH}=50$ km/s, \cite{pavlik2018} (see also \cite{Kruckow:2018slo, Giacobbo:2018etu}).

Our choice of parameters above allows us to assume 
energy equipartition between the BH population and their surrounding stars. 
The Spitzer's criterion in our case is satisfied, as $f_{BH}\times f_{\textrm{ret}}\times(m_{BH}/1 \, M_{\odot})^{1.5}\simeq0.1$ and smaller than 0.16, valid for the two-mass model of the GCs that we consider here  \cite{1969ApJ...158L.139S}. 
Thus, the subsystem of BHs that forms due to mass 
segregation in our GCs is not dynamically decoupled from that of stars.

Some of these BHs will be in binaries while most of them will be isolated.
Consider a Keplerian BBH defined by its orbital semi-major axis 
and eccentricity parameters $(a, e)$ and with binding energy 
$E_{\textrm{bin}}\equiv{G\, m^2\over 2 a}$. 
Then, this binary is said to be hard if its binding energy well 
exceeds the kinetic energy (KE) of its neighboring objects \cite{heggie1975}. When the above condition is not met we will refer to the binaries as soft. 
BH binaries that originate form binary stars i.e. survived both natal kicks and did not lead to a soft binary that would break with third body interactions are defined here as proto-BBH (PBBH)\footnote{Also known in the literature as ``primordial binaries'', see e.g. \cite{ivanova2010}}. We include into the PBBHs, binaries that were created by exchange interactions of BH-star binaries with an isolated BH. 
Some of these isolated BHs may form binaries with ordinary stars via three body induced interactions 
at a high rate, \cite{ivanova2005}, or from a BH sub-population near the core, resulting in the development of a binary population of dynamically assembled BHs, \cite{park2017}. 

The maximum total number of BHs in the GC given in Eq.~\ref{maxBHs} can be decomposed into those that begin in isolation, and those that initially participate into binaries, 
\begin{equation}
\label{eq:ALL_BH_comp}
N_{BH}^{\textrm{ret-max}}=N_{BH,\textrm{isol}}+N_{BH,\textrm{bin}}. 
\end{equation}
In turn the number of PBBHs is, 
\begin{eqnarray}
\label{eq:PBBH}
N_{\textrm{PBBH}}&=&\frac{1}{2}\times f_{\textrm{hard}} \times N_{BH,\textrm{bin}} \nonumber \\
&=&\frac{1}{2}\times f_{\textrm{hard}} \times \left({N_{BH,\textrm{bin}}\over N_{BH}^{\textrm{ret-max}}}\right) N_{BH}^{\textrm{ret-max}} \nonumber \\ 
&=&\frac{1}{2}\times f_{\textrm{hard}} \times f_{\textrm{bin}}\times N_{BH}^{\textrm{ret-max}}.
\end{eqnarray}
$f_{\textrm{bin}}$ refers to the fraction of BHs in binaries. Since only hard binaries will make it ``unscathed'' from early 
interactions with neighboring bodies, only a fraction $f_{\textrm{hard}}$ of binaries will be our PBBHs.  
Those may eventually merge after further hardening 3rd-body interactions. 

Combining with Eq.~\ref{eq:NBH_retmax} we get,
\begin{eqnarray}
\label{eq:PBBH_2}
N_{\textrm{PBBH}} &=& \frac{1}{2}\times f_{\textrm{hard}}\times f_{\textrm{bin}}\times f_{\textrm{ret}} \times N_{BH}^{\textrm{max}} \nonumber \\ 
         &=& f_{\textrm{eff}}\times N_{BH}^{\textrm{max}}.
\end{eqnarray}
where we have combined all previously mentioned factors 
into a single effective factor $f_{\textrm{eff}} \equiv 1/2 \times f_{\textrm{hard}}\times f_{\textrm{bin}}\times f_{\textrm{ret}}$.
From this point on when discussing the 
3rd-body channel we will only care about the PBBHs' evolution and refer to these
binaries as just BBHs.

Monte Carlo simulations from \cite{OLeary:2005vqo, 2010MNRAS.402..371B} 
that study BBHs in clusters and probe the $f_{\textrm{eff}}/f_{\textrm{ret}}$ ratio, suggest a range 
of $0.01 \lesssim f_{\textrm{eff}}/f_{\textrm{ret}} \lesssim 0.1$. 
Since we take $f_{\textrm{ret}} \simeq 0.1$ we will consider the range for the effective factor to be
$1\times 10^{-3} \lesssim f_{\textrm{eff}} \lesssim 1\times 10^{-2}$.  

The remaining isolated BHs may contribute to the DC channel or participate in hard binaries\footnote{There may be soft binaries that survive and contribute to the 3rd-body channel. The disruption of hard binaries due to some very energetic interaction giving off two isolated BHs is a rare event, \cite{heggie1975, hills1980} %\ic{@KK add Heggie's-Hill citation here.}
.}. Their number is $\left(1-f_{\textrm{hard}}\times f_{\textrm{bin}}\right)\times f_{\textrm{ret}}\times N_{BH}^{\textrm{max}} \simeq f_{\textrm{ret}}\times N_{BH}^{\textrm{max}}$.

\subsection{The merger rate}
\label{theMergerRate_part}

Assuming that the GC properties depend 
only on the radial distance from the center, 
and ignoring mass distributions, we can write a universal scheme for the differential merger 
rate in the GC regarding a single merger channel as,
\begin{equation}
\label{UniScem}
    {d\Gamma_{\textrm{ch.}}\over d^3r}=n_{\textrm{ch.}}(r) \, \left\langle{1\over T_{m}(r)}\right\rangle.
\end{equation}
Here, $n_{\textrm{ch.}}$ is the number density of merger events in that channel 
and $\langle{1\over T_{m}(r)}\rangle$ is the average rate for such an event at radius $r$.
In the following we will apply this formula in the cases of the DC channel and the 3rd-body channel, as these are the dominant contributing channels to our total merger rate from GCs. For an N-object configuration with N$\ge3$ the probability for interaction gets significantly suppressed with increasing N.

\subsubsection{Direct Capture events}

In \cite{quinlanShapiro1989, mouri2002} the cross section for the 
DC channel is calculated assuming that a pair of objects $A$ and $B$, with reduced mass $\mu_{AB}$, 
can form a bound system as long as they interact at such a 
small pericenter so that the energy lost in GWs exceeds the total 
energy of the reduced system; i.e. $\delta E_{GW}\ge{1\over2}\, \mu_{AB}\, \sigma_{AB}^2$, where $\sigma_{AB} = \sqrt{1 \, M_{\odot}/\mu_{AB}}\ \sigma_{\textrm{star}}$ 
since we have assumed energy equipartition. $\delta E_{GW}$ was calculated in \cite{turner1977}. $\sigma_{\textrm{star}}$ is the velocity dispersion of the stars as shown in Fig.~\ref{gcsProfile}. The cross section for this interaction is,
\begin{multline}
\label{SigmaDC_express}
    \Sigma_{DC}\simeq 16.8\times\left({\sigma_{AB}\over c}\right)^{-{18/7}}\\\times\left({G^2 \, m_A^{2/7}\, m_B^{2/7}\,(m_A+m_B)^{10/7}\over c^4}\right),
\end{multline}
in the gravitational focusing approximation and under the hypothesis that $A-B$ interactions are nearly parabolic, \cite{quinlanShapiro1989}. In this work we care for both $A$ and $B$ to be $10 \, M_{\odot}$ BHs.
The total DC merger rate over a GC, is just,
\begin{equation}
\label{DCtotMergerRate}    
    \Gamma_{DC}=\int_{r_{\textrm{min}}}^{r_{\textrm{max}}}dr \, 4\pi r^2 \, {1\over2} \, n_{BH}^2 \, \Sigma_{DC} \, v_{BH,BH}.
\end{equation}
We take for the relative velocity of two BHs
 $v_{BH,BH} = \sigma_{BH,BH} = \sqrt{2}\sigma_{BH}$;
 $\sigma_{BH}$ is the velocity dispersion of the BHs.

In our calculations we use a lower limit of radius $r_{\textrm{min}}$,
\begin{equation}
    r_{\textrm{min}} = \left({4\pi\over3} n_{BH}^{\textrm{ret-max}}(r=0)\right)^{-1/3},
\end{equation}
i.e down to the radius where only one 10 $M_{\odot}$ BH is included. 
We also take  $r_{\textrm{max}}=r_c$, as we assume all BHs within the core radius. 

Every DC event leads to a merger as the timescale for isolated radiation reaction coalescence 
is very small compared to the interaction timescale with an object that might disturb the BH binary. 
The merger timescale for a newly DC-formed BBH is bounded above by, \cite{cholis2016, oleary2009},
\begin{equation}
    T_{\textrm{m}}^{DC}\lesssim376 \times\left({m_{BH}\over10 \, M_{\odot}}\right)\times\left({10 \,  \textrm{km/s}\over\sigma_{\textrm{star}}}\right)^3 \textrm{yr}.
\end{equation}
The interaction timescale in our context is given by,
\begin{eqnarray}
\label{interTime}
    T_{\textrm{int}}&=&15.6 \times\left({\sigma_{\textrm{star}}\over10 \, \textrm{km/s}}\right)\times \left({0.4\times10^5 \, M_{\odot}/\textrm{pc}^3\over\rho_{\textrm{star}}}\right)\nonumber \\
    &\times& \left({20 \, M_{\odot}\over m_{\textrm{tot}}}\right)\times\left({6 \, \textrm{AU}\over a_h}\right) \textrm{Myr},
\end{eqnarray}
which is much larger than $T_{\textrm{m}}^{DC}$. 

\subsubsection{3rd-body hardening process on BBH}

A hard circular BBH with SMA of 0.1 AU or larger will merge on a timescale 
that is larger than the Hubble time (Eq.\ref{peters0}). However, interactions
of the binary with stars can lead to the hardening of the BBH, 
with the stars gaining kinetic energy out of the binary and thus increasing 
its binding energy. A few 3rd-body interactions inside of dense stellar clusters may 
be enough to accelerate the merger \cite{heggie1975, hills1980}.

We take a hard semi-major axis to be defined by,
\footnote{The definition of a hard binary of Eq.~\ref{hardnessSMA} corresponds to 
setting its semimajor axis a factor of
 $\sim0.025\times\left({\langle m_{\textrm{star}}\rangle\over1 \, M_{\odot}}\right)\ \left({10 \, M_{\odot}\over m_{BH}}\right)$ to match Ref.~\cite{quinlan1996}. This is smaller than the value of semi-major axis evaluated by setting $\frac{G m_{BH}^{2}}{2 a} \simeq \frac{1}{2} m_{\textrm{star}} \sigma_{\textrm{star}}^2$.}
\cite{quinlan1996},
\begin{equation}
\label{hardnessSMA}
    a_h={G \, m_{BH}\over4\, \sigma^2} \simeq 5.58 \times\left({m_{BH}\over10 \, M_{\odot}}\right) \, \left({20 \,  \textrm{km/s}\over \sigma_{\textrm{star}}}\right)^2 \, \textrm{AU}.
\end{equation}

Considering an energetic interaction with a  point of closest approach of the
order of the binary's semi-major axis, i.e. $\sim a$, the average fractional energy 
variation per encounter is given by,
\begin{equation}
\label{avgEnergyVar}
    {\langle\Delta E_b\rangle\over E_b}\simeq 0.12\times\left({H\over15}\right) \, \left({m_{\textrm{star}}\over1 \, M_{\odot}}\right) \, \left({10 \, M_{\odot}\over m_{BH}}\right).
\end{equation}
$H$ is the hardening rate (not to be confused with the Hubble rate), \cite{quinlan1996}. 
$H$ is best determined by numerical 3rd-body experiments, as in \cite{sesana2006}, where is was approximated by,
\begin{equation}
\label{eq:H}
    H=14.55\times\left(1+0.287 \, {a\over a_h}\right)^{-0.95},
\end{equation}
for a unit mass ratio BBH and independent of $e$\footnote{
There is a weak dependence of $H$ on the eccentricity. 
The coefficient at Eq.~\ref{eq:H} varies from 14.5
at $e=0$ to $17$ at $e=0.9$.}. 
Furthermore, we use an averaged time 
rate of change of the binary's internal energy is given 
by\footnote{The over-dot denotes time derivative.},
\begin{equation}
\label{avgEbDeriv}
    \left\langle\dot{E_b}\right\rangle=\langle\Delta E_b\rangle \, n_{\textrm{star}}\pi b^2\sigma_{\textrm{rel}},
\end{equation}
where the impact parameter can be shown to be, \cite{sigurdsson1993},
$b^2=r_p^2\ \left(1+{2G m_{\textrm{tot}}\over r_p \sigma_{\textrm{rel}}}\right)$, with a relative velocity
of $\sigma_{\textrm{rel}}\simeq \sigma_{\textrm{star}}$.

The effective merger timescale can be estimated from the evolution of the 
semi-major axis alongside with that of eccentricity. 
The total semi-major axis evolution is given by,
\begin{equation}
\label{SMAevol}
    \dot{a}=-{G \, H \, \rho_{\textrm{star}}\over\sigma_{\textrm{star}}} \, a^2
    -{128\over5} \, {G^3 \, m_{BH}^3\over c^5 \, a^3} \, F(e).
\end{equation}
The first term describes the averaged effect of hardening interactions while the 
second term is the Peters secular 
evolution due to GW emission \cite{peters1964}. $F(e)$ is given by,
\begin{equation}
\label{eq:Fe}
F(e) = (1-e^2)^{-7/2} \cdot\left(1+{73\over24} e^2+{37\over96} e^4\right) \; \textrm{for} \; e\in[0,1).
\end{equation}

Similarly the eccentricity evolution equation is, \cite{biava2019, sesana2006},
\begin{equation}
\label{ECCevol}
    \dot{e}=+{G \, H \, K \, \rho_{\textrm{star}}\over\sigma_{\textrm{star}}} \, a-{608\over15} \, {G^3 \, m_{BH}^3\over c^5 \, a^4} D(e).
\end{equation}
$K$ is called the ``eccentricity growth rate'', \cite{quinlan1996} 
which is also determined by numerical 3rd-body
experiments. 
We use the fitting function provided in \cite{sesana2006} (their equation 18).
The second term in Eq.~\ref{ECCevol} represents the GW Peters secular 
evolution of the eccentricity with $D(e)=(1-e^2)^{-{5/2}}\cdot\left(e+{121\over304}  e^3\right)$, \cite{peters1964}. 
We solve the differential system of equations \ref{SMAevol} and \ref{ECCevol} for a hard binary 
and for a few pairs of initial conditions $(a_0, e_0)$. 
We use as reference 47 Tucanae or just 47 Tuc (NGC 104) at its core radius and later on expand our analysis on other clusters. 

The merger for the 3rd-body channel is calculated from,
\begin{equation}
\label{3bodyTotMergerRate}
    \Gamma_{\textrm{3rd-body}}=f_{\textrm{eff}}\times f_{BH}\\\times\int_{r_{\textrm{min}}}^{r_{\textrm{max}}} dr\ 4\pi r^2 \, {\rho(r)\over m_{BH}} \, \frac{1}{T_{GC}} \left\langle f_{e}(r)\right\rangle,
\end{equation}
where we have used $n_{\textrm{3rd-body}}=N_{\textrm{PBBH}} \, \rho(r) \, M_{GC}^{-1}$. $\rho(r)$ is the mass density of the cluster and $\simeq \rho_{\textrm{star}}(r)$.  
$T_{GC}$ is the age of the cluster and $f_{e}$ 
refers to the fraction of BBHs that will merge within 
a time of $T_{GC}$ and obtained by evolving Eqs~\ref{SMAevol} 
and~\ref{ECCevol} for a hard BBH. 
In our case this reduces to $\Gamma_{\textrm{3rd-body}}
= f_{\textrm{eff}}\times N_{BH}^{\textrm{max}}\times\langle f_e(r_c)\rangle/T_{GC}$.

\section{Results}
\label{sec:results}

We use equations \ref{DCtotMergerRate} and \ref{3bodyTotMergerRate} to calculate the DC and 3rd-body merger rates for the Milky Way GCs.

\subsection{The Merger Timescale of Binary Black Holes in 47 Tuc; our example cluster}
\label{sec:Merger_Time_47Tuc}

We start with the BBH evolution on a single cluster
to show our treatment of these objects near the GC core. We use 47 Tuc as reference. 

\subsubsection{The Averaged Direct Capture Merger Timescale}

We know that the interaction timescale of two 
objects inside a dense environment depends on their 
number densities, 
their relative velocity, 
as well as their impact parameter. 
We assume that the typical 
relative velocity of two BHs is just $\sqrt{2}\sigma_{BH}$, with $\sigma_{BH}$ their velocity dispersion. 
Relying on Eq.~\ref{SigmaDC_express} 
we can estimate the time required for a $10 \, M_{\odot}$ BH to interact with a $10 \, M_{\odot}$ and lead to a DC event. 
For future reference, we call this time $T_{DC}$.
It is approximately equal to,
\begin{eqnarray}
\label{DC_timescale_}    
    T_{DC}\simeq8.2\times10^{13}\cdot\left({100 \, \textrm{pc}^{-3}\over n_{BH}^{\textrm{ret-max}}(r)}\right)\left({\sigma_{\textrm{star}}(r)\over \textrm{10 \, km/s}}\right)^{11/7}
    \nonumber
    \\\times\left({10 \, M_{\odot}\over m_{BH}}\right)^2 \textrm{yr}.\ \ \
\end{eqnarray}
We take the velocity dispersion of the BHs to be that of the stars multiplied by the factor of $1/\sqrt{10}$ assuming energy equipartition\footnote{The velocity dispersion of BHs in the inner sub-cluster is expected to be at around 14.5 km/s for 47 Tuc, as is predicted by the Virial theorem at core radius and neglecting the mass contribution from the stars.}. 
Mass segregation will also result in 
lower velocity dispersion for the BHs. Our choice gives an upper bound on the $T_{DC}$, which in turn will give a lower/conservative limit for the merger rate for this channel. 
Also, in our case, $n_{BH}^{\textrm{ret-max}}\simeq \left({4\over3}\pi r_c^3\right)^{-1}N_{BH}^{\textrm{ret-max}}$.

In the case of 47 Tuc, we calculate this time near the center of the GC at the core radius, where most of the BHs reside.
We find that such a BH encounters another BH in about $4.7\times10^{13}$ yr according to the Plummer model, and in about $0.9\times10^{13}$ yr in the King model.
The Plummer timescale is only slightly bigger than the King one, due to its higher velocity dispersion, (see Fig.~\ref{gcsProfile}). 
Both of these timescales are larger than the Hubble time in the case of 47 Tuc, which indicates DC are rare events inside of 47 Tuc alike GCs.
However these numbers are to be understood as averages and the probability that a BH encounters another BH follows the Poisson distribution.

\subsubsection{The BBH Merger Timescale due to 3rd-body interactions}

For the BBH interactions with stars, we solve Eqs.~\ref{SMAevol} and~\ref{ECCevol} numerically. 
Our results are to be understood in a statistical sense and as we show later the evolution 
of BBHs depends on the specific environment defined by the mass profile of the clusters.

Let us consider a BBH in the environment of 47 Tuc, at core radius. 
For illustrative purposes at first we ignore the variation of the binary's radius 
from the center of the GC and study its evolution for a few pairs of initial conditions $(a_0, e_0)$. 
We solve the system of equations (\ref{SMAevol}) and (\ref{ECCevol}) implementing a 
fourth-order Runge-Kutta type numerical algorithm and iterate until the BBH attains a 
semi-major axis of $a_{\textrm{end}}=0.01$ AU. 
For the time-step we choose $dt = 30$ Myr which is the typical timescale for 3rd-body 
interactions at the core of 47 Tuc (see appendix~\ref{Appendix_numerical} for further details). 
 
\begin{figure}
    \centering
    \includegraphics[width=8.5cm,height=5.2cm]{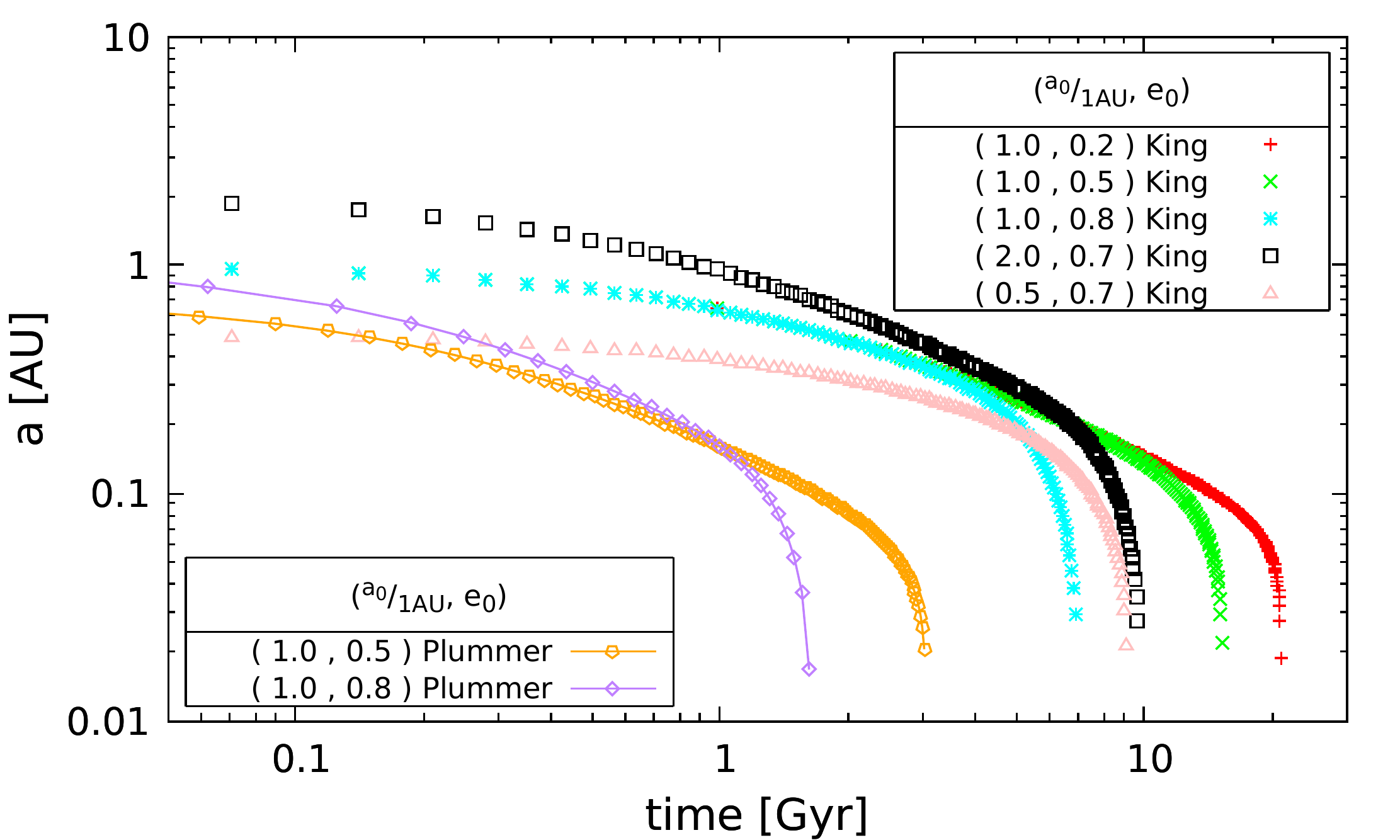}   
\vskip 0.05in
\hskip 0.2cm
    \includegraphics[width=8.5cm,height=5.2cm]{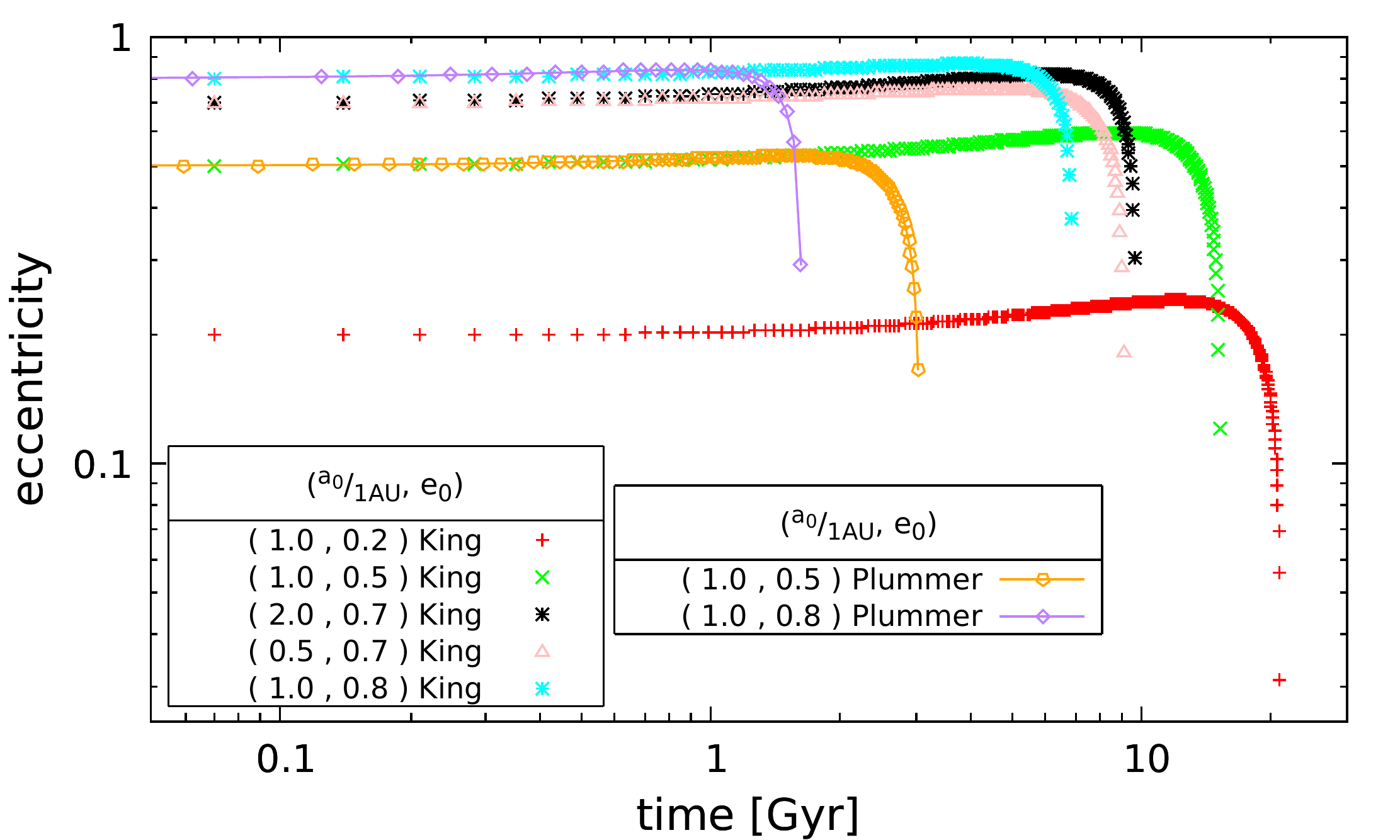}
    \caption{The evolution of a hard BBH in 47 Tuc at its core radius. The red (plus) and green (cross) overlap with the cyan (star) SMA data because they have the same evolution until 5 Gyr. The black (square) and pink (triangle) data have approximately the same merger times even though their initial semi-major axis differs by a factor of four. We also include two curves, the purple (diamonds) and orange (hexagonal) data, that correspond to a Plummer profile. 
    }
    \label{5simulations47Tuc}
\end{figure}

In Fig.~\ref{5simulations47Tuc} we show the evolution of BBHs undergoing 3rd-body 
interactions with starts starting from different $(a_0, e_0)$ conditions. 
We show five simulations under the assumption of a King density 
profile and two simulations under a Plummer profile.  
 
In the binary's early evolution, the positive sign of the first term in Eq.~\ref{ECCevol} (relating to the 3rd-body hardening interactions)
dominates\footnote{When $e_0$ is smaller than $0.4$ for $a_0=a_h$, initially the binary experiences a slight eccentricity decrease which is due 
a negative sign in the eccentricity growth rate, supported by numerical surveys \cite{sesana2006}. 
This is a small effect and does not significantly affect the overall growing 
statistical character of eccentricity with 3rd-body encounters.}. 
As a result the eccentricity of the BBH increases at first. 
That increased eccentricity in turn enhances  the GW emission, which accelerates the binary's coalescence. 
Throughout the binary's evolution, both the GW emission and the 3rd-body interactions cause a reduction of the 
semi-major axis of the binary (Eq.~\ref{SMAevol}). As $a$ 
becomes smaller the binary's interactions with stars become 
more separated in time, suppressing the positive $\dot{e}$ first term 
in Eq.~\ref{ECCevol}; while the GW emission becomes more prominent, amplifying 
the negative $\dot{e}$ second term in Eq.~\ref{ECCevol}. At some point the binary enters the GW domination regime, leading to the coalescence and circularization of the binary.

For a BBH in a Plummer environment the merger timescale is smaller 
compared to a King type core. The ratio $\rho/\sigma$ in the Plummer profile is larger by about one order of 
magnitude enhancing the first term in Eq.~\ref{SMAevol} which drives the binary to shrink more efficiently. 
This also translates to a larger 3rd-body merger rate for the Plummer profile compared to the King one. 

Given that GCs have ages up to $\simeq 10$ Gyr, 
any of our simulated BBHs that requite more than that 
time to coalesce can not contribute to the BBH merger 
rates.
For a given GC environment, that upper limit on $T_{m}$ 
constrains the combination of initial BBH properties 
($a_{0}, e_{0}$) required for a binary to merge (within 10 Gyr).  

\begin{figure}
    \centering
    \includegraphics[width=8.5cm,height=5cm]{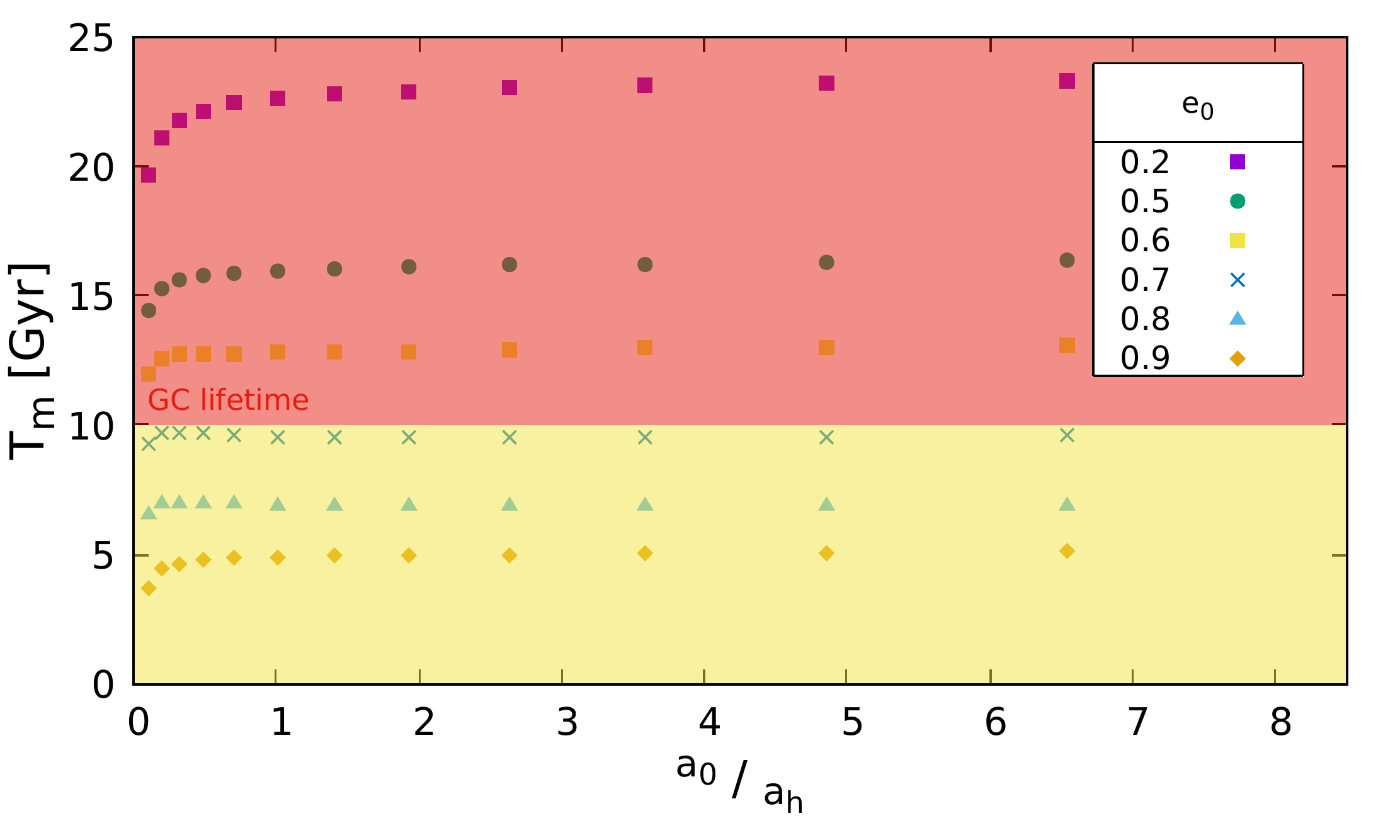}
    \caption{The 3rd-body merger timescale versus the BBHs initial semimajor axis $a_{0}$. We show cases of $a_0 \in [0.1  a_h, 10 a_h]$ and plot lines for $e_{0} \in [0.2, 0.9]$. Any point that falls inside the yellow region refers to a BBH with initial conditions that in the environment of 47 Tuc will merge at a value of $T_{m}$ depicted by its y-axis value. We observe a weak dependence on $a_0\in[a_h, 8 a_h]$. For 47 Tuc $a_h=5.7$ AU at core radius. Each point in this figure is a single simulation.}
    \label{TmergePlot47Tuc}
\end{figure}

In Fig.~\ref{TmergePlot47Tuc}, we show 
for combinations of ($a_{0}, e_{0}$) the 
3rd-body interactions merger timescale versus the dimensionless 
semi-major axis $a_{0}/a_{h}$. Every symbol represents the 
$T_{m}$ for a unique set of initial conditions. Any point inside the 
yellow region satisfies $T_{m} < 10$ Gyr, and its $y-$axis value 
gives the predicted $T_{m}$. In Fig.~\ref{TmergePlot47Tuc}, to evaluate $T_{m}$ for 47 Tuc, we take as a reference the $\rho_{\textrm{star}}(r)$ and $\sigma_{\textrm{star}}(r)$ at its core radius $r_{c}$. For 47 Tuc 
$a_h=5.7$ AU. 
In different GC environments 
one needs to change not only the value of $a_{h}$ but also those of the $\rho_{\textrm{star}}(r)$ and $\sigma_{\textrm{star}}(r)$ profiles. 
In the following we will generalize our result for all GCs. 
We also remind the reader that we evolve Eqs.~\ref{SMAevol} 
and \ref{ECCevol} until the binary reaches $a_{\textrm{end}}=0.01$ AU.
We make sure the evolution lies within the Newtonian regime at all
times until the end of the simulation. 

While BBHs interacting with close-by passing stars 
is inherently a stochastic process, we treat it here in a 
deterministic manner. The fact that from this point on, we exclude
all BBHs with initial conditions ($a_{0}, e_{0}$) that for their given environment do not lead to merger in more than 10 Gyr, leads in  
a lower/conservative value for their merger rate once averaging over all 
GCs.

In our calculations of the merger timescale we have used up to here the radial distance of $r=r_{c}$ as a reference. This choice provides a good description of the 3rd-body interactions inside the GC
for a sequence of reasons. For both the King and Plummer profiles, most 
BHs are included within the core radius. Moreover, for the Plummer profile and for $r \leq r_{c}$ 
as we showed in Fig.~\ref{gcsProfile} the total matter density $\rho$, and the 
velocity dispersion $\sigma$ of stellar objects are approximately constant. Thus the 3rd-body interaction terms in Eqs.~\ref{SMAevol} 
and~\ref{ECCevol} that scale as $\rho/\sigma$ are also constant for $r \lesssim r_{c}$. 
Showing the $T_{m}$ at $r=r_{c}$ for the Plummer profile is representative of the entire core. 
For the King profile the $\rho/\sigma$ drops significantly beyond the $r > r_{c}$, thus the 3rd-body interactions 
get suppressed at larger radii. Plotting $T_{m}$ at $r=r_{c}$ provides a 
consecutive but still representative estimate. 
Finally, we clarify again that in evaluating the total 3rd-body merger rate 
from Eq.~\ref{3bodyTotMergerRate} under a general mass profile 
for the binaries, $T_{m}$  should be evaluated for different radii $r$. 

While mass segregation also takes place in GCs, forcing stars away from the very center of the clusters the 1~$M_{\odot}$ stars 
remain within the observed core radii. Given simulation results where BBHs get excited to higher orbits (e.g. \cite{samsing2019}), 
we expect that the $r \sim r_{c}$ region is still the place where BBHs interact most often with stars.

Our results for the 3rd-body interactions 
are limited to soft interactions of BBHs with stars. 
Taking 47 Tuc, the number of hard binaries that survive in the core is expected to be about
${f_{\textrm{eff}}\over f_{ret}}\times N_{\textrm{BH}}^{\textrm{ret-max}}\simeq30$ BBHs. 
This corresponds to a BBH number core density of about $70$ pc$^{-3}$. 
Also, the BH number core density for 47 Tuc is $\simeq3000$ pc$^{-3}$. 
However, the density of stars near its core is $\simeq10^{5}$ pc$^{-3}$ or $\simeq10^{6}$ pc$^{-3}$,
depending on whether we choose the King or the Plummer profile respectively (see Fig.~\ref{gcsProfile}). 
Therefore, BBH-star soft interactions at $r\simeq r_{c}$ are the most common type with BBH-BH interactions following them (for a recent study of BBH-BH interactions see \cite{Samsing:2020qqd}).
Binary-binary interactions are rare as the number of BBHs is relatively small in GCs.

\subsection{Binary Black Holes in any GC}

\subsubsection{The Averaged Direct Capture Merger Timescale}

In table~\ref{DC4threeGCs} 
we calculate the direct capture timscale for three Milky Way GCs 
with similar structural properties calculated for BHs at core 
radius and for the Plummer and King profiles.

\begin{table}[]
    \centering
    \begin{tabular}{c c c}
    \hline
        GC & \multicolumn{2}{c}{$T_{GC}\times10^{13}$ [yr]} \\
         & Plummer & King \\
    \hline
        47 Tuc   & 4.7 & 0.9 \\
        M 30     & 0.5 & 0.05 \\
        NGC 5946 & 1.5 & 0.1 \\
    \hline
    \end{tabular}
    \caption{
    Timescales for a $10 \, M_{\odot}$ BH to capture and merge with another one at core radius for a few cases. We have assumed the Plummer and King profiles. In all of these cases the numbers exceed the threshold GC lifetime of about $10^{10}$ yr.}
    \label{DC4threeGCs}
\end{table}

Having examined the DC timescale on three dense GCs, we conclude that for those environments this timescale exceeds the GC lifetime by at least one order of magnitude.
We find that no known Milky Way GC has a $T_{DC}$ that is smaller than 10 Gyr.
This indicates that DC events are very rare in Milky Way GCs.
This will show up even in the final merger rates that we calculate in the next section. We expect the DC channel to dominate only in exotic environments with very high densities and with a very low velocity dispersion.

\subsubsection{The 3rd-body Merger Timescale}

In section~\ref{sec:Merger_Time_47Tuc} we showed in Figure~\ref{TmergePlot47Tuc} 
that a BBH undergoing 3rd-body interactions with stars in 
47 Tuc will merge within 10 Gyr only if its initial eccentricity is $e_{0} \gtrsim 0.7$; with weak dependence on its initial 
semi-major axis $a_{0}$. That result came for the specific environment of 47 Tuc. 
In this section we will generalize our results for any GC. 
Knowing what initial eccentricity conditions are required in order 
for a BBH to merge is a crucial element in our calculations. We can simulate the BBH's initial eccentricity distribution in a GC environment, and in turn derive 
what fraction of those BBHs will merge through the 3rd-body channel. 

In the following we will show a sequence of approximations that lead 
to a semi-analytical answer on the 
required initial eccentricity conditions 
for a BBH to merge at any given cluster. 
That answer we will compare to our answer from the numerical evolution 
of Eqs.~\ref{SMAevol} and~\ref{ECCevol}.

Given Eq.~\ref{SMAevol} we can calculate the 
time after which the Peters GW emission 
dominates over the interaction term. 
We call this moment in the binary's evolution the equality point. 
This corresponds to the state $(a_{eq}, e_{eq})$. 
We evaluate the semi-major axis at that point $a_{eq}$ by equating the two rates, $\left(\dot{a}\right)_{\textrm{3rd-body}}=\left(\dot{a}\right)_{\textrm{Peters}}$.
The $e_{eq}$ could be approximated with $\approx e_{max}$,
as the point of equality occurs 
only slightly after the point of maximum eccentricity $e_{max}$.  
However, the condition $\dot{e}=0$ leads to a transcendental equation 
for $e_{max}$ which does not have a closed form solution. 
Thus, instead we set $e_{eq}\simeq e_{0}$, i.e. at the moment when the binary's eccentricity becomes again $e_0$. 
We checked for the ensemble of clusters at Table~\ref{GCparameters} and 
for initial eccentricities 
$e_{0}$ between 0.4 and 0.9 that 
the ratio $e_{eq}/e_{0} = e_{\textrm{max}}/e_{0}$ falls always within the region of 1.03 up to 1.30. 

The time required for the 
binary to reach equality $T_{eq}(r)$ if located at radius $r$, 
is approximated by integrating Eq.~\ref{SMAevol} 
from $a_0$ to $a_{eq}$, and ignoring the Peters term which dominates after equality.  
The result is,
\begin{equation}
\label{eqTime}
    T_{eq}(r)=\left({a_0\over a_{eq}}-1\right) \frac{\sigma_{\textrm{star}}(r)}{G H \,\rho_{\textrm{star}}(r) \, a_0}.
\end{equation}
 Since in all cases where multiple 3rd-body interactions take place before GW emission starts having an impact to the binary's evolution we can take in Eq.~\ref{eqTime}
 $(a_{0}/a_{eq} - 1)(1/a_{0}) \simeq 1/a_{eq}$
 \footnote{$a_{eq}$ is just evaluated by equating the two terms of the right-hand side of Eq.~\ref{SMAevol}.}.
 We then get,
 \begin{eqnarray}
 \label{semiAnal}
     T_{eq}(r) \simeq T_{m}^{s-a}(r)&=&5.75 \left({\sigma_{\textrm{star}}(r)\over10  \, \textrm{km/s}}\right)^{4/5}
     \left({15\over H}\right)^{4/5} \\
      &\times&
    \left({10^5 \, M_{\odot}/\textrm{pc}^3\over\rho_{\textrm{star}}(r)}\right)^{4/5} 
     \left({10 \, M_{\odot}\over m_{BH}}\right)^{3/5} \nonumber \\
     &\times&
     F^{-{1/5}}(e_0) \; \; \textrm{Gyr}. \nonumber
 \end{eqnarray}
 $T_{m}^{s-a}(r)$ is our semi-analytical evaluation of $T_{m}$. 
 To the limit that $a_{0} \gg a_{eq}$ satisfied when 
 multiple 3rd-body interactions take place before the binary merges, 
 $T_{m}$ is independent of $a_{0}$. This is in qualitative agreement with our simulations for 47 Tuc.  
 We also tested the accuracy of Eq.~\ref{semiAnal} 
 with the numerical results. This is shown in Table~\ref{compareTmerges} 
 for 47 Tuc, M 30, and NGC 5946. 
 We picked those three clusters as they 
 envelope the $\rho_{\textrm{star}}(r)$ and $\sigma_{\textrm{star}}(r)$ 
 profile properties of the ensemble of clusters that can contribute to the merger. 
 We find these values quite satisfactory for our purposes. 
\begin{table}[h]
    \centering
    \begin{tabular}{c c c c}
    \hline
        GC & $e_0$ & $T_{m}^{\textrm{num}}$ [Gyr] & $T_{m}^{\textrm{s-a}}$ [Gyr] \\
        \hline 
        47 Tuc & 0.2 & 22.6 & 20.5 \\
        47 Tuc & 0.8 & 7.0 & 8.4 \\
        \hline
        M 30 & 0.2 & 7.1 & 6.9 \\
        M 30 & 0.8 & 1.2 & 2.8 \\
        \hline
        NGC 5946 & 0.2 & 15.4 & 14.6 \\
        NGC 5946 & 0.9 & 2.5 & 3.7 \\
         \hline
    \end{tabular}
    \caption{A few numerical and the corresponding semi-analytical values of $T_{m}$ as obtained from Eq.~\ref{semiAnal}. We checked 47 Tuc, M 30 and NGC 5946 with two values of $e_0$. $a_0$ is set to the $a_{h}$, which is, $a_h =5.7$ AU for 47 Tuc, $a_h=47$ AU for M 30, and $a_h=33$ AU for NGC 5946. Variables are assumed in the context of the King profile at core radius.}
    \label{compareTmerges}
\end{table} 

We note that for high eccentricities the semi-analytical answer is larger than the numerical result. 
At higher eccentricities the GW emission 
from the Peters term is dominant and Eq.~\ref{eqTime} becomes less accurate. 
Moreover, we use $F(e_{0})$ instead of $F(e_{eq})$ in Eq.~\ref{semiAnal} which 
at larger eccentricities leads to overestimating $T_{m}^{s-a}$.
This semi-analytical approach works best at smaller eccentricities, but with satisfactory results for up to $e\approx0.9$ (see Table~\ref{compareTmerges}). 

The semi-analytical expression of Eq.~\ref{semiAnal} connects 
the merger time of the BBH to its initial eccentricity and 
to the mass density $\rho_{\textrm{star}}(r)$ and velocity dispersion $\sigma_{\textrm{star}}(r)$ of 
the GCs' stars at a radial distance of $r$.  
That is, we have a dependence scheme of the form 
$T_{m}\propto \left({\sigma/\rho}\right)^{4/5} \, F^{-{1/5}}(e_0)$. 
We can therefore, set constraints on the minimum initial 
eccentricity the BBH should have in order for it to merge 
within the lifetime of the GC, $T_{GC}$ 
and connect to GC quantities that can be inferred from observations. We remind that 
we can evaluate $\rho_{\textrm{star}}$ and $\sigma_{\textrm{star}}$ at
the GC's core radius, $r_c$. I.e. $\sigma_c\equiv\sigma_{\textrm{star}}(r_c)$ and $\rho_c\equiv\rho_{\textrm{star}}(r_c)$.  We get (see again Eq.~\ref{eq:Fe} for the definition of $F(e)$),
 \begin{eqnarray}
 \label{ECCconstraint}
     F(e_0) &>& \left(\frac{5.75 \; \textrm{Gyr}}{T_{GC}}\right)^{5} \left({\sigma_c\over10 \,  \textrm{km/s}}\right)^{4} \left({10^5 \, M_{\odot}/\textrm{pc}^3\over\rho_c}\right)^{4} \nonumber \\
     &\times& \left(\frac{15}{H}\right)^{4}
     \left(\frac{10 \,  M_{\odot}}{m_{BH}}\right)^{3}.
 \end{eqnarray}
 
 Eq.~\ref{ECCconstraint} can also be thought of as a constraint 
 on the ratio $\sigma_{c}/\rho_{c}$ given an initial eccentricity $e_0$. 
 Therefore, we can probe for an appropriate value of $e_0$ which environments allow 
 a BBH to merge in $T_{GC}$ given the pair $(\rho_c,\sigma_c)$,
 \begin{eqnarray}
 \label{loglogLine}
     \log_{10}\left({\sigma_c\over10 \,  \textrm{km/s}}\right)&=&\log_{10}\left({\rho_c\over10^5 \, M_{\odot}/\textrm{pc}^3}\right) +  \log_{10}\left({\frac{H}{15}}\right) \nonumber \\
     &+& 0.75 \,\log_{10}\left({ \frac{m_{BH}}{10 \, M_{\odot}} }\right) \nonumber \\ 
     &+ &1.25 \log_{10}\left({\frac{T_{GC}}{5.75 \,  \textrm{Gyr}}}\right) \nonumber \\ &+& 0.25\,\log_{10}\left({F(e_0)}\right).  %+1.301. 
 \end{eqnarray}
 As we can see this condition is fairly insensitive to either the exact age of the
 GCs and to the exact mass of the BHs.
 
\begin{figure*}[t]
    \centering
    \includegraphics[width=16.4cm,height=10cm]{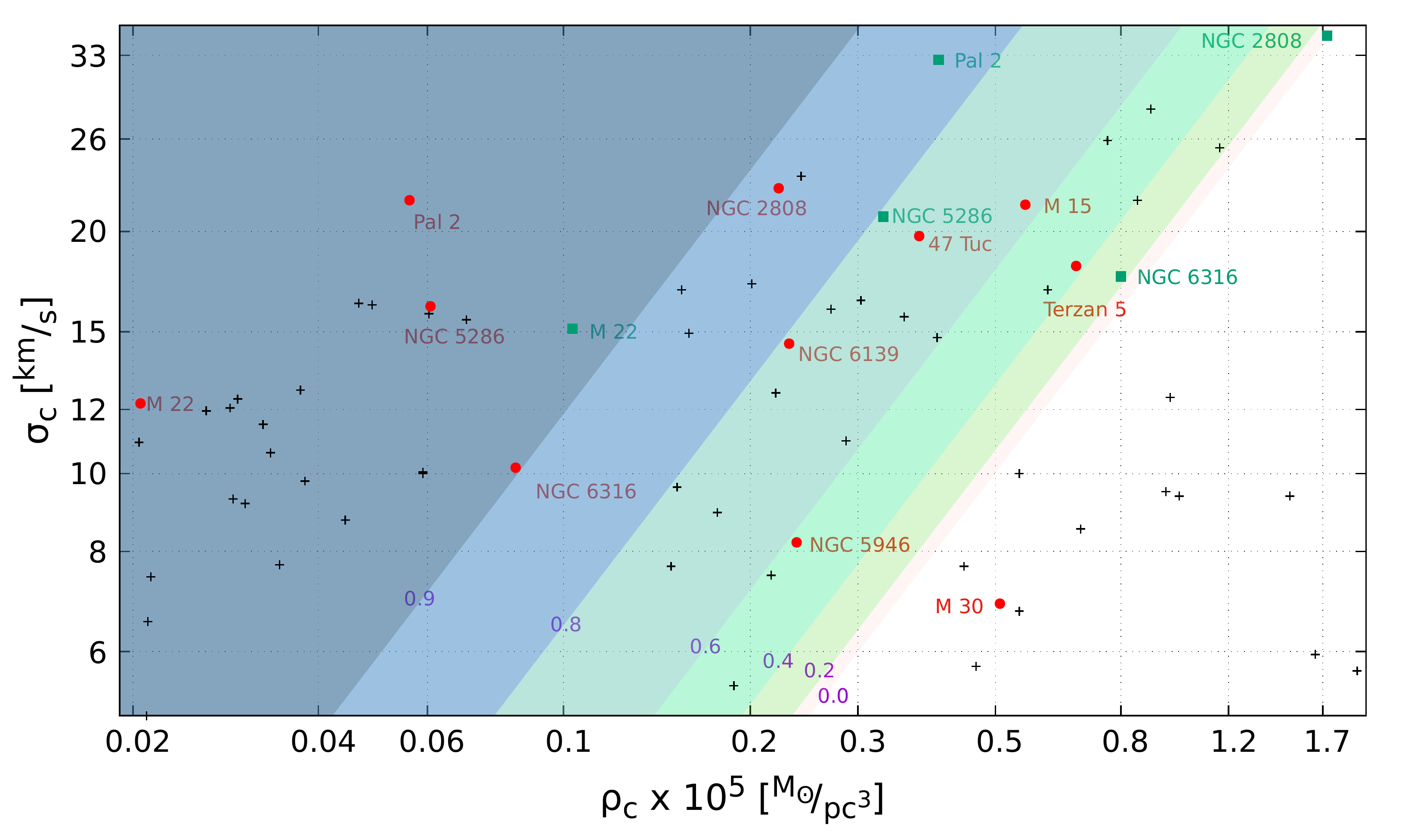}
    \caption{Milky Way GCs positioned in the $\rho_{c}$ vs $\sigma_{c}$ parameter space for a small part of the observed space (see also Fig.~\ref{allMWGCs}). We plot the GCs and partition the selected area into seven regions, with color boundary lines defined by Eq.~\ref{loglogLine}. Labels on the boundary lines represent values of $e_0$. For a given GC, its BBHs with $e_{0}$ larger than the $e_{0}$-line to its left will merge within 10 Gyr. As an example Terzan 5 sits very close to the line of $e_{0} = 0.4$. All its BBHs with $e_{0}$ larger than 0.4 will merge. Red dots are GCs from Table~\ref{GCparameters}, and green boxes represent GCs under the Plummer profile, while other points are other known GCs.
    }
    \label{eccCHART}
\end{figure*}

 In Fig.~\ref{eccCHART} we map out in the observable 
 $(\rho_c,\sigma_c)$-parameter space some Milky Way GCs, 
 along with a few specific curves on the required 
 initial eccentricity for binaries of 10 $M_{\odot}$ BHs to 
 merge within 10 Gyr. We take also $H = 15$.
These curves correspond to straight lines in a log-log plot. 
For a given GC its BBHs with initial eccentricity $e_{0}$ 
larger than the line to its left will merge within 10 Gyr.

 \begin{figure}
     \centering
     \includegraphics[width=8.5cm,height=5cm]{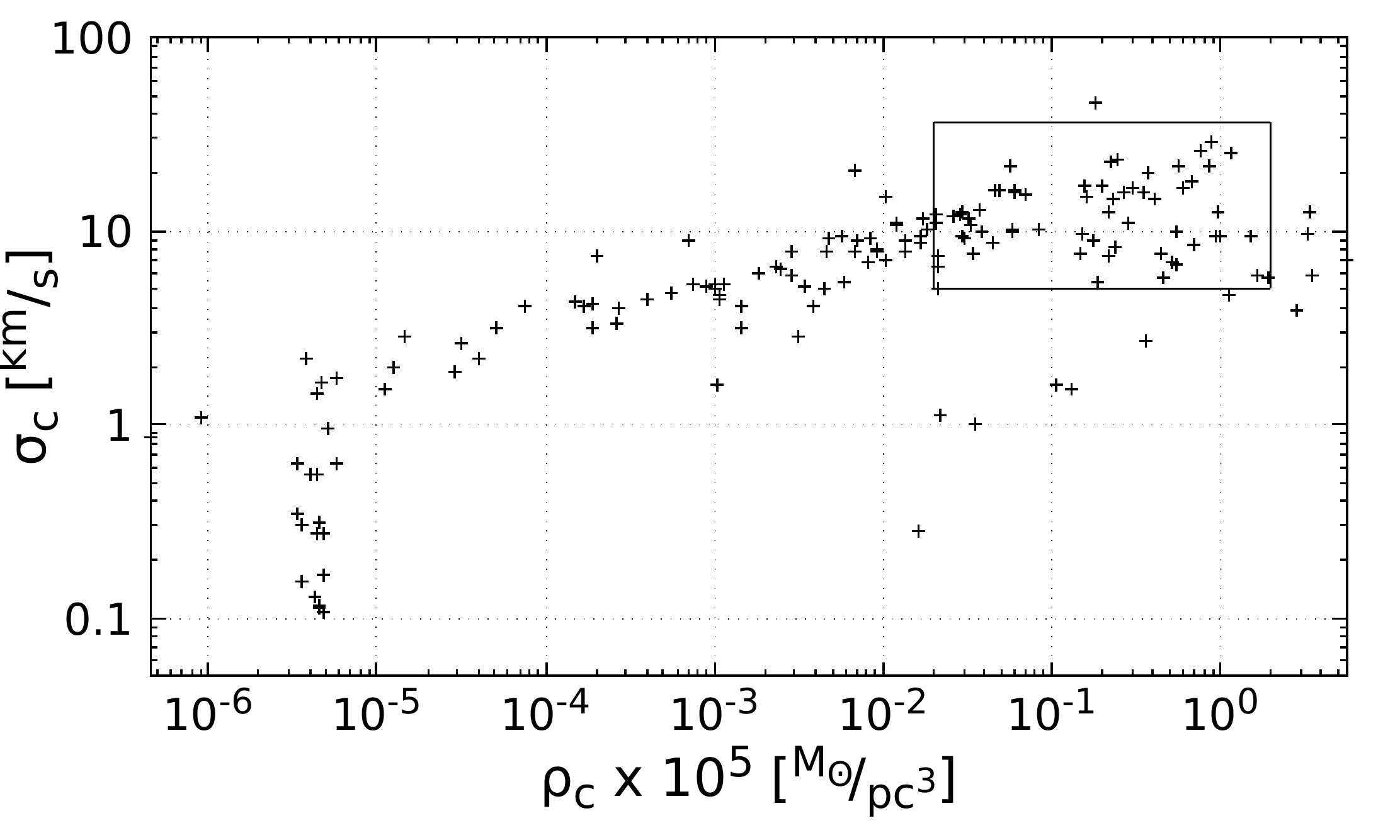}
     \caption{All observed Milky Way GCs in the full $\rho_{c}$ vs $\sigma_{c}$ parameter space. The box on the top left represents the parameter space of the Fig.~\ref{eccCHART}.
     The black "+" points represent GCs under the King profile.
     }
     \label{allMWGCs}
 \end{figure}

 As shown in Fig.~\ref{eccCHART} there are Milky Way GCs, as M 30, that pose no constraints on 
 the initial eccentricity.
 In these GCs BBHs merge within at most 10 Gyr independent of what their $e_0$.
 This is attributed to their high core densities.
 However, GCs like M 30 host a small number of BHs and as we will 
 show do not contribute too much on the final merger rates. 
Instead, in clusters as 47 Tuc, the imposed  constraints are only mild. 
47 Tuc falls near the $e_0=0.7$ line, a result consistent with Figure~\ref{TmergePlot47Tuc}. 

In Fig.~\ref{allMWGCs}, we show all GC for which we have information to place them on a 
more extended $\rho_{c}$ vs $\sigma_{c}$ space. 
With the exception of a few very dense systems, GCs that fall outside the reduced 
space plotted on Fig.~\ref{eccCHART} will not contribute significantly 
to the merger rate even if there are many more of these objects. 
On general grounds, for environments 
with $\rho_c\in[0.1,0.5]\times10^5 M_{\odot}/$pc$^3$ there is a strong dependence 
of the minimal $e_{0}$ value that leads to a merger and the local velocity dispersion. 
High velocity environments will host BBH merger only for relatively high 
$e_{0}$ values. In Fig.~\ref{eccCHART} we take a King profile for the GCs.
For a Plummer profile since $\sigma_c^{\textrm{Plummer}}>\sigma_c^{\textrm{King}}$ (see Fig.~\ref{gcsProfile})
the minimal $e_{0}$ values are typically higher. GCs points move to up and to the right. 
We show that effect in Fig.~\ref{eccCHART} with green boxes for a small number of GCs.
BBHs in GC environments situated on the left of the $e_0=0.9$ 
line should necessarily have an initial eccentricity greater 
than 0.9 if they are to satisfy Eq.~\ref{ECCconstraint}. 
Such binaries are expected to form dynamically 
and contribute to the dynamical component of the total merger rate.

\subsection{Merger Rates of the Milky Way Globular Clusters} 
 
 Here we first evaluate the merger rates for individual clusters. 
 We do that 
 separately for the direct capture and for the 3rd-body interactions channel. 
 We rely on Eqs.~\ref{DCtotMergerRate} and~\ref{3bodyTotMergerRate} respectively. 
 Then we sum the merger rates from all 
 Milky Way globular clusters to evaluate the
 expected rate from a collection such as that of our own galaxy.

\subsubsection{DC merger rate}

The direct capture merger is a stochastic process.
Calculating the DC merger rate from Eq.~\ref{DCtotMergerRate}, 
we show our results in the second and fourth 
columns of table~\ref{totMergerRateS}, under the Plummer 
and King profiles respectively.
We use the set of 13 GCs given in table~\ref{GCparameters}.

Since the DC merger rate scales as $\Gamma_{DC}\propto n_{BH}^2\sigma_{BH}^{-11/7}$, 
the GC with the largest rates will be those that have 
a small velocity dispersion 
and large BH densities. 
Among the cases we have examined, M 30 is one of 
those dense GCs which also have a small 
velocity dispersion under the King profile.
This GC is a good candidate for a high DC merger 
rate, as we also saw in table~\ref{DC4threeGCs} where the DC timescale 
was calculated to be relatively small compared to other GCs.
The King DC rate of M 30 is about $2.1\times10^{-10}$yr$^{-1}$ 
and is among the highest on the list.

We perform an average on the DC merger 
rate over the known 139 Milky Way GCs of the Harris catalogue \cite{harris1996, GLOBCLUST}, 
that we have $\rho$ and $\sigma$ information for. 
We find an averaged DC rate of
$4.9\times10^{-11}$ yr$^{-1}$ per cluster relying on the King profile.
Using the Plummer profile for the stars we find instead a per cluster rate of 
$3.3\times10^{-12}$ yr$^{-1}$. 
Typically and on average the King DC rate is higher than the Plummer one by about an order of magnitude.
This can be seen by comparing for the individual GC the DC rates in Table~\ref{totMergerRateS}.
The King profile typically  predicts a lower velocity dispersion 
compared to the Plummer model around the core radius of each 
GC, (see Fig.~\ref{gcsProfile}).

\subsubsection{3rd-body merger rate}

For the 3rd-body interactions as we showed, the initial eccentricity of a BBH hosted on a GC
plays a crucial role on whether that binary will have enough time to merge.
To calculate the merger rate we assume a thermal distribution of the BBH's eccentricity, 
$P(e) = 2\cdot e$, with a mean of $\langle e\rangle=0.7$. We then evaluate for each Milky Way GC the $f_{e}$ of Eq.~\ref{3bodyTotMergerRate} from,
\begin{equation}
    f_{e}(r)\equiv\int_{e_0(r)}^{1}de_0'\ P(e_0')=1-e_0(r)^2.
\end{equation}
$e_{0}$ is the minimal initial eccentricity required 
for a BBH to merge within 10 Gyr, calculated for each GC by equating the two sides 
of Eq.~\ref{ECCconstraint} with $T_{GC} = 10$ Gyr 
and using the unique GC's $\sigma_{c}$ and
$\rho_{c}$ properties.
A thermal distribution is consistent with a dynamical assembly 
scenario of our BBHs, \cite{heggie1975}. 
Proto-binaries may acquire high initial eccentricities 
by the internal dynamics during the formation of the BBH pair. 
We concentrate on hard BBHs. We take for them $a_{0} = a_{h}$, 
with $a_{h}$ having a unique value for each mass profile (and each GC). 
Binaries hosted in an environment described by a Plummer mass profile, 
begin tighter ($a_{h}$ is smaller) compared to a same mass cluster with a King mass profile.  
The smaller $a_{h}$ is a consequence 
of the larger velocity dispersion of these objects in a Plummer profile, (see Fig.~\ref{gcsProfile}). 
However, our results are fairly insensitive to 
the exact choice of $a_{0}$ (see Fig.~\ref{TmergePlot47Tuc} and  Eq.~\ref{semiAnal}).
Initial semi-major axes smaller than 
$\sim0.3a_h$ have a smaller merger time than we consider in this work. 
Yet, these ultra-hard binaries are rare.

\begin{table}[h]
    \centering
    \begin{tabular}{c c c c c}
    \hline
    GC & \multicolumn{2}{c}{Plummer} & \multicolumn{2}{c}{King} \\
         & DC & 3rd-body & DC & 3rd-body \\
    \hline
        \; 47 Tuc   \;     &  \; 1.2e-11 \; &  \; 3.4e-09  \;&  \; 7.6e-11  \; &  \; 1.5e-09  \; \\
        $\omega$ Cen & 1.8e-13 & 9.8e-11 & 3.5e-13 & 2.2e-11 \\
        M 15         & 3.0e-11 & 6.2e-09 & 3.1e-10 & 3.7e-09 \\
        M 22         & 1.8e-13 & 9.9e-11 & 3.4e-13 & 2.1e-11 \\
        NGC 6362     & 4.3e-15 & 9.0e-13 & 3.5e-15 & 2.7e-13 \\
        NGC 5946     & 8.8e-12 & 8.5e-10 & 1.9e-10 & 5.6e-10 \\
        M 30         & 9.8e-12 & 3.4e-10 & 2.1e-10 & 3.4e-10 \\
        Terzan 5     & 3.7e-12 & 6.8e-10 & 9.4e-12 & 5.5e-10 \\
        Pal 2        & 1.1e-12 & 9.5e-10 & 3.2e-12 & 1.7e-10 \\
        NGC 6139     & 3.2e-12 & 1.1e-09 & 1.2e-11 & 3.9e-10 \\
        NGC 2808     & 2.5e-12 & 2.0e-09 & 7.2e-12 & 4.5e-10 \\
        NGC 5286     & 5.1e-13 & 3.5e-10 & 1.1e-12 & 7.5e-11 \\
        NGC 6316     & 6.1e-13 & 3.5e-10 & 1.2e-12 & 6.8e-11 \\
    \hline
    \end{tabular}
    \caption{The integrated DC and 3rd-body merger rates, in yr$^{-1}$, for our 13 Milky Way GCs. The parameter $f_{\textrm{eff}}$ is set to $0.3\%$.}
    \label{totMergerRateS}
\end{table}

The 3rd-body merger rates for the GCs of Table~\ref{GCparameters} are 
given in Table~\ref{totMergerRateS}, for the two choices of mass profiles.
The 3rd-body merger rate surpasses the 
DC rate by typically one to two orders of magnitude. 
This is despite the fact that 
BBHs are less abundant by a factor of $3\times10^{2}$ than 
isolated BHs that seed the DC events. 
47 Tuc falls in the category of a massive GC and NGC 6362 represents a small GC. 

Following the King profile we get typically a factor of two smaller rates, 
even though in some cases as M 30 the 
rate is the same for both assumed profiles on the distribution of the stars. 
We find that the average per cluster rate of the 3rd-body channel is $2.0 \times 10^{-10}$yr$^{-1}$
under the King profile assumptions for the stars and 
$4.2 \times 10^{-10}$yr$^{-1}$ per cluster under the Plummer.

In the numbers of Table~\ref{totMergerRateS} we have taken $f_{\textrm{eff}} = 0.3 \%$. 
$f_{\textrm{eff}}$ is in the range of $0.1\%\le f_{\textrm{eff}}\le 1\%$ which directly translates to one order of magnitude in range of 
the 3rd-body merger rate.
That uncertainty 
is the main source of uncertainty on evaluating the 3rd-body merger rate from a given GC. 

We make a last note here regarding the ejection of  binaries from the globular clusters. 
During each encounter of a BBH with a perturbing object (a star in our case), 
the perturbing object acquires on average 
a high kick and by momentum conservation the binary recoils. 
We have checked that in order for the BBH to receive a large enough recoil 
to get ejected from the 
GC environment the interaction has to be very close and in turn the binary 
already at that stage a tight one. 
BBHs encountering $1 \, M_{\odot}$ stars 
tend to be retained near the GC core as long as they have a semi-major axis of 
$\gtrsim 0.01$ AU before the interaction. 
We remind the reader that for our 3rd-body calculations we use that value of semi-major 
axis to end our evolution of the binaries' orbital properties. 
Once the binary has a semi-major axis smaller than 0.01 AU 
it only takes a few Myr until it merges via GW emission. 
Regardless of such a binary still being 
in the cluster or having been ejected from it, 
its merger contributes to our 
calculation of the merger rate. 

\subsection{The Cosmological Merger Rate}

Having evaluated the merger rate in the GCs of the Milky Way 
and the averaged per cluster merger rate 
$\langle\Gamma_{GC}\rangle$
we will evaluate the cumulative merger rate at 
redshift $z$ from the local Universe, $\mathcal{R}_c(z)$. 
Formally $\langle\Gamma_{GC}\rangle$ is a function of redshift. However, 
we will take it a constant over 
a period of 10 Gyrs and will not extend our analysis beyond a redshift of 5. 
This can be calculated by \cite{Ye:2019xvf, Rodriguez:2016kxx},
\begin{equation}
    \mathcal{R}_c(z)=\int_0^z dz'\ \langle\Gamma_{GC}(z')\rangle\ n_{GC}\ {dV_c\over dz'}\ (1+z')^{-1}.
\end{equation}
where $dV_c/dz$ is the comoving volume,
$n_{GC}(z)$ is the GC number density and the  $(1+z)^{-1}$ factor accounts for 
the time dilation. 
$\langle\Gamma_{GC}\rangle$, for the combination of the DC and the 3rd-body channels, 
we have already estimated to be 
$2.5 \times 10^{-10}$ yr$^{-1}$ and $4.2 \times 10^{-10}$ yr$^{-1}$ per cluster 
for the King and Plummer profiles respectively. 
At low redshift $n_{GC}$ has a 
range from a conservative minimum of $0.33 \times 10^{9}$ Gpc$^{-3}$ up to an optimistic value of $\simeq 3 \times 10^{9}$ Gpc$^{-3}$, with a more conventional value of $0.77 \times 10^{9}$ Gpc$^{-3}$ \cite{Rodriguez:2016kxx}. 
We assume that
Milky Way GCs constitute a representative ensemble of GCs in the local Universe.
In the following, we 
take 
$H_0=100h\ $ km/s/Mpc, $h=0.7$, $\Omega_{K}=0$, $\Omega_{M}=0.3$ and $\Omega_{\Lambda}=0.7$, based on Planck data (2015) \cite{Ade:2015xua}.
We can rewrite the comoving volume as, \cite{Hogg:1999ad},
\begin{eqnarray}
    {dV_c\over dz'}={4\pi c^3\over H_0^3}\ {1\over E(z')}\ \left(\int_0^{z'}{dz''\over E(z'')}\right)^2,
\end{eqnarray}
where we have used the function, \cite{Xu:2016grp,Hogg:1999ad},
\begin{equation}
    E(z)=\sqrt{\Omega_M\cdot(1+z)^3+\Omega_{\Lambda}}.
\end{equation}

We take the more conventional value of $n_{GC} = 0.77 \times 10^{9}$ Gpc$^{-3}$ constant in redshift or evolving as $n_{GC}(z) = 0.77 \times 10^{9}  \cdot E(z)$ Gpc$^{-3}$. 
Integrating up to a redshift of 1, we find a rate of 20-65 mergers 
per year for the choice of a King profile and 33-110 
mergers for the choice of a Plummer profile. 
For each of the profiles the higher values for the mergers come from assuming $n_{GC} \propto E(z)$. 
Extending to higher redshifts comes with a significant increase associated to the  uncertainties on the exact choice of $n_{GC}(z)$. 
The total number of BBHs mergers up redshift of 5 can 
be as large as $10^4$ 
mergers per year for the King and Plummer profiles respectively. 
However for a nearly constant in redshift $n_{GC}$  
these numbers are much more suppressed and only $O(100)$ per year.

We note that the DC merger rate per GC at source is taken to be  constant with redshift up to $z=5$ while for
the 3rd-body merger rate its redshift evolution is described by 
the term $\langle f_e(r)\rangle/T_{GC}$ of Eq.~\ref{3bodyTotMergerRate} 
through the lifetime of the GC evaluated at redshift $z'$.

\section{Conclusions}
\label{sec:conclusions}

In this paper we evaluate the 
BH-BH merger rates in Milky Way type 
GCs considering BH-BH direct capure and 3rd-body BBH-star soft interaction mechanisms. 
Our calculations are based under the assumption of a two-mass 
model of $m_{\textrm{star}}=1 \, M_{\odot}$ and $m_{BH}=10 \, M_{\odot}$. 
We consider King and 
Plummer profiles for  the distribution of the stars in the GCs, 
and take a segregated state on the BHs concentrating all of them uniformly inside the core of each GC.

The averaged 3rd-body merger rate per cluster 
was found to be 
$2.0 \times 10^{-10}$yr$^{-1}$ and $4.2 \times 10^{-10}$yr$^{-1}$ for the King and Plummer profiles.
These rates are more significant than 
the rates from the DC channel 
by one to two orders of magnitude.
For the DC channel we get instead $4.9 \times 10^{-11}$yr$^{-1}$ and $3.3 \times 10^{-12}$yr$^{-1}$
for the King and Plummer profiles respectively. 

The largest uncertainty with respect to the BBH merger rates is in the fraction $f_{eff}$ of
BBHs that can be formed from the original population of BHs created by stellar evolution, and then undergo multiple 3rd-body interactions with stars. 
Most BHs will be ejected from the clusters
due to their natal kicks. Also most BHs that remain in the cluster will not be 
in binaries that are tight enough to survive the 3rd-body interactions even 
with regular stars. Throughout this 
work we take $f_{\textrm{eff}} = 0.3 \%$. That fraction's range is between $0.1 \%$ to $1 \%$. 
Our results on the merger 
rates from the 3rd-body channel are proportional 
to $(f_{\textrm{eff}}/0.3 \%)$.

Once integrating over cosmological distances we find in total 
between 20 and 110 mergers 
per year up to a redshift of $z=1$.
We note that in calculating the cumulative  
rates, the numbers are dominated by the 
most dense and 
massive clusters. These clusters are also the easiest to observe. 
Clusters with small core densities 
or masses maybe difficult to observe but also 
do not contribute to the our rates in any significant manner.  

In this work we have quantified and important aspect of the 3rd-body channel in GCs. 
The initial eccentricity of the BBH plays a crucial role in the 
evolution of the binary given its surroundings and whether it will 
evolve fast enough to merge within a Hubble time. 
Only very compact GCs allow the presence of low eccentricity mergers. 
As an example, an environment like 
47 Tuc requires BBHs to have an 
eccentricity of no less than $\sim0.7$ 
which is the average value of the thermal eccentricity distribution.
Instead, the BBH's merger time depends weakly on 
the initial semi-major axis. 

Our results are limited to soft interactions of BBHs with stars, as these are the most common type of interactions.
As this project was being completed Ref.~\cite{Samsing:2020qqd} 
appeared studying BBH-BH hard interactions. 
Other possible channels leading to BBH 
merges that we have not included in this work as  
they are not dominant, relate 
to four body effects. 
One such mechanism is the Kozai resonance, \cite{kozai1962, 1962P&SS....9..719L}, 
as known in the literature in the context of triple systems (hierarchical or not) \cite{miller2002, arcaSedda2018, antognini2016}. 
This channel requires a triple system where two of the objects are BHs forming an inner compact binary and a third object, 
a BH or a star, that orbits in an outer orbit. 
The outer object, causes the eccentricity of the inner pair to 
increase near unity on a timescale of a few hundred years, 
on condition that the two orbits have a relative inclination close to 90 
degrees and that no other object 
perturbs the system in the meantime \cite{Naoz_2016}. The two inner BHs can 
merge very quickly if the GW emission 
peaks near the phase of high eccentricity. 
The contribution from this channel to the BBH merger rate is highly suppressed, 
due to the small number of 
BBHs inside of GCs. 
The statistics of such a mechanism 
work best inside of denser 
and larger environments like the cores of galaxies where the number of BHs and BBHs are higher.

\textit{Acknowledgements:} The authors are  grateful to A. Kehagias for his support in facilitating this collaboration. 

\bibliography{literature}

%merlin.mbs apsrev4-1.bst 2010-07-25 4.21a (PWD, AO, DPC) hacked
%Control: key (0)
%Control: author (0) dotless jnrlst
%Control: editor formatted (1) identically to author
%Control: production of article title (0) allowed
%Control: page (1) range
%Control: year (0) verbatim
%Control: production of eprint (0) enabled
\begin{thebibliography}{87}%
\makeatletter
\providecommand \@ifxundefined [1]{%
 \@ifx{#1\undefined}
}%
\providecommand \@ifnum [1]{%
 \ifnum #1\expandafter \@firstoftwo
 \else \expandafter \@secondoftwo
 \fi
}%
\providecommand \@ifx [1]{%
 \ifx #1\expandafter \@firstoftwo
 \else \expandafter \@secondoftwo
 \fi
}%
\providecommand \natexlab [1]{#1}%
\providecommand \enquote  [1]{``#1''}%
\providecommand \bibnamefont  [1]{#1}%
\providecommand \bibfnamefont [1]{#1}%
\providecommand \citenamefont [1]{#1}%
\providecommand \href@noop [0]{\@secondoftwo}%
\providecommand \href [0]{\begingroup \@sanitize@url \@href}%
\providecommand \@href[1]{\@@startlink{#1}\@@href}%
\providecommand \@@href[1]{\endgroup#1\@@endlink}%
\providecommand \@sanitize@url [0]{\catcode `\\12\catcode `\$12\catcode
  `\&12\catcode `\#12\catcode `\^12\catcode `\_12\catcode `\%12\relax}%
\providecommand \@@startlink[1]{}%
\providecommand \@@endlink[0]{}%
\providecommand \url  [0]{\begingroup\@sanitize@url \@url }%
\providecommand \@url [1]{\endgroup\@href {#1}{\urlprefix }}%
\providecommand \urlprefix  [0]{URL }%
\providecommand \Eprint [0]{\href }%
\providecommand \doibase [0]{http://dx.doi.org/}%
\providecommand \selectlanguage [0]{\@gobble}%
\providecommand \bibinfo  [0]{\@secondoftwo}%
\providecommand \bibfield  [0]{\@secondoftwo}%
\providecommand \translation [1]{[#1]}%
\providecommand \BibitemOpen [0]{}%
\providecommand \bibitemStop [0]{}%
\providecommand \bibitemNoStop [0]{.\EOS\space}%
\providecommand \EOS [0]{\spacefactor3000\relax}%
\providecommand \BibitemShut  [1]{\csname bibitem#1\endcsname}%
\let\auto@bib@innerbib\@empty
%</preamble>
\bibitem [{\citenamefont {Aasi}\ \emph {et~al.}(2015)\citenamefont {Aasi} \emph
  {et~al.}}]{TheLIGOScientific:2014jea}%
  \BibitemOpen
  \bibfield  {author} {\bibinfo {author} {\bibfnamefont {J.}~\bibnamefont
  {Aasi}} \emph {et~al.} (\bibinfo {collaboration} {LIGO Scientific}),\
  }\bibfield  {title} {\enquote {\bibinfo {title} {{Advanced LIGO}},}\ }\href
  {\doibase 10.1088/0264-9381/32/7/074001} {\bibfield  {journal} {\bibinfo
  {journal} {Class. Quant. Grav.}\ }\textbf {\bibinfo {volume} {32}},\ \bibinfo
  {pages} {074001} (\bibinfo {year} {2015})},\ \Eprint
  {http://arxiv.org/abs/1411.4547} {arXiv:1411.4547 [gr-qc]} \BibitemShut
  {NoStop}%
\bibitem [{\citenamefont {Abbott}\ \emph {et~al.}(2019)\citenamefont {Abbott}
  \emph {et~al.}}]{LIGOScientific:2018jsj}%
  \BibitemOpen
  \bibfield  {author} {\bibinfo {author} {\bibfnamefont {B.P.}\ \bibnamefont
  {Abbott}} \emph {et~al.} (\bibinfo {collaboration} {LIGO Scientific,
  Virgo}),\ }\bibfield  {title} {\enquote {\bibinfo {title} {{Binary Black Hole
  Population Properties Inferred from the First and Second Observing Runs of
  Advanced LIGO and Advanced Virgo}},}\ }\href {\doibase
  10.3847/2041-8213/ab3800} {\bibfield  {journal} {\bibinfo  {journal}
  {Astrophys. J. Lett.}\ }\textbf {\bibinfo {volume} {882}},\ \bibinfo {pages}
  {L24} (\bibinfo {year} {2019})},\ \Eprint {http://arxiv.org/abs/1811.12940}
  {arXiv:1811.12940 [astro-ph.HE]} \BibitemShut {NoStop}%
\bibitem [{\citenamefont {Bethe}\ and\ \citenamefont
  {Brown}(1998)}]{Bethe:1998bn}%
  \BibitemOpen
  \bibfield  {author} {\bibinfo {author} {\bibfnamefont {Hans~A.}\ \bibnamefont
  {Bethe}}\ and\ \bibinfo {author} {\bibfnamefont {G.E.}\ \bibnamefont
  {Brown}},\ }\bibfield  {title} {\enquote {\bibinfo {title} {{Evolution of
  binary compact objects which merge}},}\ }\href {\doibase 10.1086/306265}
  {\bibfield  {journal} {\bibinfo  {journal} {Astrophys. J.}\ }\textbf
  {\bibinfo {volume} {506}},\ \bibinfo {pages} {780--789} (\bibinfo {year}
  {1998})},\ \Eprint {http://arxiv.org/abs/astro-ph/9802084}
  {arXiv:astro-ph/9802084} \BibitemShut {NoStop}%
\bibitem [{\citenamefont {Portegies~Zwart}\ and\ \citenamefont
  {McMillan}(2000)}]{PortegiesZwart:1999nm}%
  \BibitemOpen
  \bibfield  {author} {\bibinfo {author} {\bibfnamefont {Simon~F.}\
  \bibnamefont {Portegies~Zwart}}\ and\ \bibinfo {author} {\bibfnamefont
  {Stephen}\ \bibnamefont {McMillan}},\ }\bibfield  {title} {\enquote {\bibinfo
  {title} {{Black hole mergers in the universe}},}\ }\href {\doibase
  10.1086/312422} {\bibfield  {journal} {\bibinfo  {journal} {Astrophys. J.
  Lett.}\ }\textbf {\bibinfo {volume} {528}},\ \bibinfo {pages} {L17} (\bibinfo
  {year} {2000})},\ \Eprint {http://arxiv.org/abs/astro-ph/9910061}
  {arXiv:astro-ph/9910061} \BibitemShut {NoStop}%
\bibitem [{\citenamefont {Belczynski}\ \emph {et~al.}(2001)\citenamefont
  {Belczynski}, \citenamefont {Kalogera},\ and\ \citenamefont
  {Bulik}}]{Belczynski:2001uc}%
  \BibitemOpen
  \bibfield  {author} {\bibinfo {author} {\bibfnamefont {Krzysztof}\
  \bibnamefont {Belczynski}}, \bibinfo {author} {\bibfnamefont {Vassiliki}\
  \bibnamefont {Kalogera}}, \ and\ \bibinfo {author} {\bibfnamefont {Tomasz}\
  \bibnamefont {Bulik}},\ }\bibfield  {title} {\enquote {\bibinfo {title} {{A
  Comprehensive study of binary compact objects as gravitational wave sources:
  Evolutionary channels, rates, and physical properties}},}\ }\href {\doibase
  10.1086/340304} {\bibfield  {journal} {\bibinfo  {journal} {Astrophys. J.}\
  }\textbf {\bibinfo {volume} {572}},\ \bibinfo {pages} {407--431} (\bibinfo
  {year} {2001})},\ \Eprint {http://arxiv.org/abs/astro-ph/0111452}
  {arXiv:astro-ph/0111452} \BibitemShut {NoStop}%
\bibitem [{\citenamefont {O'Leary}\ \emph {et~al.}(2009)\citenamefont
  {O'Leary}, \citenamefont {Kocsis},\ and\ \citenamefont
  {Loeb}}]{OLeary:2008myb}%
  \BibitemOpen
  \bibfield  {author} {\bibinfo {author} {\bibfnamefont {Ryan~M.}\ \bibnamefont
  {O'Leary}}, \bibinfo {author} {\bibfnamefont {Bence}\ \bibnamefont {Kocsis}},
  \ and\ \bibinfo {author} {\bibfnamefont {Abraham}\ \bibnamefont {Loeb}},\
  }\bibfield  {title} {\enquote {\bibinfo {title} {{Gravitational waves from
  scattering of stellar-mass black holes in galactic nuclei}},}\ }\href
  {\doibase 10.1111/j.1365-2966.2009.14653.x} {\bibfield  {journal} {\bibinfo
  {journal} {Mon. Not. Roy. Astron. Soc.}\ }\textbf {\bibinfo {volume} {395}},\
  \bibinfo {pages} {2127--2146} (\bibinfo {year} {2009})},\ \Eprint
  {http://arxiv.org/abs/0807.2638} {arXiv:0807.2638 [astro-ph]} \BibitemShut
  {NoStop}%
\bibitem [{\citenamefont {{Banerjee}}\ \emph {et~al.}(2010)\citenamefont
  {{Banerjee}}, \citenamefont {{Baumgardt}},\ and\ \citenamefont
  {{Kroupa}}}]{2010MNRAS.402..371B}%
  \BibitemOpen
  \bibfield  {author} {\bibinfo {author} {\bibfnamefont {Sambaran}\
  \bibnamefont {{Banerjee}}}, \bibinfo {author} {\bibfnamefont {Holger}\
  \bibnamefont {{Baumgardt}}}, \ and\ \bibinfo {author} {\bibfnamefont {Pavel}\
  \bibnamefont {{Kroupa}}},\ }\bibfield  {title} {\enquote {\bibinfo {title}
  {{Stellar-mass black holes in star clusters: implications for gravitational
  wave radiation}},}\ }\href {\doibase 10.1111/j.1365-2966.2009.15880.x}
  {\bibfield  {journal} {\bibinfo  {journal} {\mnras}\ }\textbf {\bibinfo
  {volume} {402}},\ \bibinfo {pages} {371--380} (\bibinfo {year} {2010})},\
  \Eprint {http://arxiv.org/abs/0910.3954} {arXiv:0910.3954 [astro-ph.SR]}
  \BibitemShut {NoStop}%
\bibitem [{\citenamefont {Antonini}\ and\ \citenamefont
  {Perets}(2012)}]{Antonini:2012ad}%
  \BibitemOpen
  \bibfield  {author} {\bibinfo {author} {\bibfnamefont {Fabio}\ \bibnamefont
  {Antonini}}\ and\ \bibinfo {author} {\bibfnamefont {Hagai~B.}\ \bibnamefont
  {Perets}},\ }\bibfield  {title} {\enquote {\bibinfo {title} {{Secular
  evolution of compact binaries near massive black holes: Gravitational wave
  sources and other exotica}},}\ }\href {\doibase 10.1088/0004-637X/757/1/27}
  {\bibfield  {journal} {\bibinfo  {journal} {Astrophys. J.}\ }\textbf
  {\bibinfo {volume} {757}},\ \bibinfo {pages} {27} (\bibinfo {year} {2012})},\
  \Eprint {http://arxiv.org/abs/1203.2938} {arXiv:1203.2938 [astro-ph.GA]}
  \BibitemShut {NoStop}%
\bibitem [{\citenamefont {{McKernan}}\ \emph {et~al.}(2012)\citenamefont
  {{McKernan}}, \citenamefont {{Ford}}, \citenamefont {{Lyra}},\ and\
  \citenamefont {{Perets}}}]{2012MNRAS.425..460M}%
  \BibitemOpen
  \bibfield  {author} {\bibinfo {author} {\bibfnamefont {B.}~\bibnamefont
  {{McKernan}}}, \bibinfo {author} {\bibfnamefont {K.~E.~S.}\ \bibnamefont
  {{Ford}}}, \bibinfo {author} {\bibfnamefont {W.}~\bibnamefont {{Lyra}}}, \
  and\ \bibinfo {author} {\bibfnamefont {H.~B.}\ \bibnamefont {{Perets}}},\
  }\bibfield  {title} {\enquote {\bibinfo {title} {{Intermediate mass black
  holes in AGN discs - I. Production and growth}},}\ }\href {\doibase
  10.1111/j.1365-2966.2012.21486.x} {\bibfield  {journal} {\bibinfo  {journal}
  {\mnras}\ }\textbf {\bibinfo {volume} {425}},\ \bibinfo {pages} {460--469}
  (\bibinfo {year} {2012})},\ \Eprint {http://arxiv.org/abs/1206.2309}
  {arXiv:1206.2309 [astro-ph.GA]} \BibitemShut {NoStop}%
\bibitem [{\citenamefont {Dominik}\ \emph {et~al.}(2013)\citenamefont
  {Dominik}, \citenamefont {Belczynski}, \citenamefont {Fryer}, \citenamefont
  {Holz}, \citenamefont {Berti}, \citenamefont {Bulik}, \citenamefont
  {Mandel},\ and\ \citenamefont {O'Shaughnessy}}]{Dominik:2013tma}%
  \BibitemOpen
  \bibfield  {author} {\bibinfo {author} {\bibfnamefont {Michal}\ \bibnamefont
  {Dominik}}, \bibinfo {author} {\bibfnamefont {Krzysztof}\ \bibnamefont
  {Belczynski}}, \bibinfo {author} {\bibfnamefont {Christopher}\ \bibnamefont
  {Fryer}}, \bibinfo {author} {\bibfnamefont {Daniel~E.}\ \bibnamefont {Holz}},
  \bibinfo {author} {\bibfnamefont {Emanuele}\ \bibnamefont {Berti}}, \bibinfo
  {author} {\bibfnamefont {Tomasz}\ \bibnamefont {Bulik}}, \bibinfo {author}
  {\bibfnamefont {Ilya}\ \bibnamefont {Mandel}}, \ and\ \bibinfo {author}
  {\bibfnamefont {Richard}\ \bibnamefont {O'Shaughnessy}},\ }\bibfield  {title}
  {\enquote {\bibinfo {title} {{Double Compact Objects II: Cosmological Merger
  Rates}},}\ }\href {\doibase 10.1088/0004-637X/779/1/72} {\bibfield  {journal}
  {\bibinfo  {journal} {Astrophys. J.}\ }\textbf {\bibinfo {volume} {779}},\
  \bibinfo {pages} {72} (\bibinfo {year} {2013})},\ \Eprint
  {http://arxiv.org/abs/1308.1546} {arXiv:1308.1546 [astro-ph.HE]} \BibitemShut
  {NoStop}%
\bibitem [{\citenamefont {{Mapelli}}\ \emph {et~al.}(2013)\citenamefont
  {{Mapelli}}, \citenamefont {{Zampieri}}, \citenamefont {{Ripamonti}},\ and\
  \citenamefont {{Bressan}}}]{2013MNRAS.429.2298M}%
  \BibitemOpen
  \bibfield  {author} {\bibinfo {author} {\bibfnamefont {M.}~\bibnamefont
  {{Mapelli}}}, \bibinfo {author} {\bibfnamefont {L.}~\bibnamefont
  {{Zampieri}}}, \bibinfo {author} {\bibfnamefont {E.}~\bibnamefont
  {{Ripamonti}}}, \ and\ \bibinfo {author} {\bibfnamefont {A.}~\bibnamefont
  {{Bressan}}},\ }\bibfield  {title} {\enquote {\bibinfo {title} {{Dynamics of
  stellar black holes in young star clusters with different metallicities - I.
  Implications for X-ray binaries}},}\ }\href {\doibase 10.1093/mnras/sts500}
  {\bibfield  {journal} {\bibinfo  {journal} {\mnras}\ }\textbf {\bibinfo
  {volume} {429}},\ \bibinfo {pages} {2298--2314} (\bibinfo {year} {2013})},\
  \Eprint {http://arxiv.org/abs/1211.6441} {arXiv:1211.6441 [astro-ph.HE]}
  \BibitemShut {NoStop}%
\bibitem [{\citenamefont {Ziosi}\ \emph {et~al.}(2014)\citenamefont {Ziosi},
  \citenamefont {Mapelli}, \citenamefont {Branchesi},\ and\ \citenamefont
  {Tormen}}]{Ziosi:2014sra}%
  \BibitemOpen
  \bibfield  {author} {\bibinfo {author} {\bibfnamefont {Brunetto~Marco}\
  \bibnamefont {Ziosi}}, \bibinfo {author} {\bibfnamefont {Michela}\
  \bibnamefont {Mapelli}}, \bibinfo {author} {\bibfnamefont {Marica}\
  \bibnamefont {Branchesi}}, \ and\ \bibinfo {author} {\bibfnamefont
  {Giuseppe}\ \bibnamefont {Tormen}},\ }\bibfield  {title} {\enquote {\bibinfo
  {title} {{Dynamics of stellar black holes in young star clusters with
  different metallicities -- II. Black hole--black hole binaries}},}\ }\href
  {\doibase 10.1093/mnras/stu824} {\bibfield  {journal} {\bibinfo  {journal}
  {Mon. Not. Roy. Astron. Soc.}\ }\textbf {\bibinfo {volume} {441}},\ \bibinfo
  {pages} {3703--3717} (\bibinfo {year} {2014})},\ \Eprint
  {http://arxiv.org/abs/1404.7147} {arXiv:1404.7147 [astro-ph.GA]} \BibitemShut
  {NoStop}%
\bibitem [{\citenamefont {Antonini}\ and\ \citenamefont
  {Rasio}(2016)}]{Antonini:2016gqe}%
  \BibitemOpen
  \bibfield  {author} {\bibinfo {author} {\bibfnamefont {Fabio}\ \bibnamefont
  {Antonini}}\ and\ \bibinfo {author} {\bibfnamefont {Frederic~A.}\
  \bibnamefont {Rasio}},\ }\bibfield  {title} {\enquote {\bibinfo {title}
  {{Merging black hole binaries in galactic nuclei: implications for
  advanced-LIGO detections}},}\ }\href {\doibase 10.3847/0004-637X/831/2/187}
  {\bibfield  {journal} {\bibinfo  {journal} {Astrophys. J.}\ }\textbf
  {\bibinfo {volume} {831}},\ \bibinfo {pages} {187} (\bibinfo {year}
  {2016})},\ \Eprint {http://arxiv.org/abs/1606.04889} {arXiv:1606.04889
  [astro-ph.HE]} \BibitemShut {NoStop}%
\bibitem [{\citenamefont {Bird}\ \emph {et~al.}(2016)\citenamefont {Bird},
  \citenamefont {Cholis}, \citenamefont {Muñoz}, \citenamefont {Ali-Haïmoud},
  \citenamefont {Kamionkowski}, \citenamefont {Kovetz}, \citenamefont
  {Raccanelli},\ and\ \citenamefont {Riess}}]{Bird:2016dcv}%
  \BibitemOpen
  \bibfield  {author} {\bibinfo {author} {\bibfnamefont {Simeon}\ \bibnamefont
  {Bird}}, \bibinfo {author} {\bibfnamefont {Ilias}\ \bibnamefont {Cholis}},
  \bibinfo {author} {\bibfnamefont {Julian~B.}\ \bibnamefont {Muñoz}},
  \bibinfo {author} {\bibfnamefont {Yacine}\ \bibnamefont {Ali-Haïmoud}},
  \bibinfo {author} {\bibfnamefont {Marc}\ \bibnamefont {Kamionkowski}},
  \bibinfo {author} {\bibfnamefont {Ely~D.}\ \bibnamefont {Kovetz}}, \bibinfo
  {author} {\bibfnamefont {Alvise}\ \bibnamefont {Raccanelli}}, \ and\ \bibinfo
  {author} {\bibfnamefont {Adam~G.}\ \bibnamefont {Riess}},\ }\bibfield
  {title} {\enquote {\bibinfo {title} {{Did LIGO detect dark matter?}}}\ }\href
  {\doibase 10.1103/PhysRevLett.116.201301} {\bibfield  {journal} {\bibinfo
  {journal} {Phys. Rev. Lett.}\ }\textbf {\bibinfo {volume} {116}},\ \bibinfo
  {pages} {201301} (\bibinfo {year} {2016})},\ \Eprint
  {http://arxiv.org/abs/1603.00464} {arXiv:1603.00464 [astro-ph.CO]}
  \BibitemShut {NoStop}%
\bibitem [{\citenamefont {Sasaki}\ \emph {et~al.}(2016)\citenamefont {Sasaki},
  \citenamefont {Suyama}, \citenamefont {Tanaka},\ and\ \citenamefont
  {Yokoyama}}]{Sasaki:2016jop}%
  \BibitemOpen
  \bibfield  {author} {\bibinfo {author} {\bibfnamefont {Misao}\ \bibnamefont
  {Sasaki}}, \bibinfo {author} {\bibfnamefont {Teruaki}\ \bibnamefont
  {Suyama}}, \bibinfo {author} {\bibfnamefont {Takahiro}\ \bibnamefont
  {Tanaka}}, \ and\ \bibinfo {author} {\bibfnamefont {Shuichiro}\ \bibnamefont
  {Yokoyama}},\ }\bibfield  {title} {\enquote {\bibinfo {title} {{Primordial
  Black Hole Scenario for the Gravitational-Wave Event GW150914}},}\ }\href
  {\doibase 10.1103/PhysRevLett.117.061101} {\bibfield  {journal} {\bibinfo
  {journal} {Phys. Rev. Lett.}\ }\textbf {\bibinfo {volume} {117}},\ \bibinfo
  {pages} {061101} (\bibinfo {year} {2016})},\ \bibinfo {note} {[Erratum:
  Phys.Rev.Lett. 121, 059901 (2018)]},\ \Eprint
  {http://arxiv.org/abs/1603.08338} {arXiv:1603.08338 [astro-ph.CO]}
  \BibitemShut {NoStop}%
\bibitem [{\citenamefont {Stone}\ \emph {et~al.}(2017)\citenamefont {Stone},
  \citenamefont {Metzger},\ and\ \citenamefont {Haiman}}]{Stone:2016wzz}%
  \BibitemOpen
  \bibfield  {author} {\bibinfo {author} {\bibfnamefont {Nicholas~C.}\
  \bibnamefont {Stone}}, \bibinfo {author} {\bibfnamefont {Brian~D.}\
  \bibnamefont {Metzger}}, \ and\ \bibinfo {author} {\bibfnamefont {Zoltán}\
  \bibnamefont {Haiman}},\ }\bibfield  {title} {\enquote {\bibinfo {title}
  {{Assisted inspirals of stellar mass black holes embedded in AGN discs:
  solving the `final au problem'}},}\ }\href {\doibase 10.1093/mnras/stw2260}
  {\bibfield  {journal} {\bibinfo  {journal} {Mon. Not. Roy. Astron. Soc.}\
  }\textbf {\bibinfo {volume} {464}},\ \bibinfo {pages} {946--954} (\bibinfo
  {year} {2017})},\ \Eprint {http://arxiv.org/abs/1602.04226} {arXiv:1602.04226
  [astro-ph.GA]} \BibitemShut {NoStop}%
\bibitem [{\citenamefont {Carr}\ \emph {et~al.}(2016)\citenamefont {Carr},
  \citenamefont {Kuhnel},\ and\ \citenamefont {Sandstad}}]{Carr:2016drx}%
  \BibitemOpen
  \bibfield  {author} {\bibinfo {author} {\bibfnamefont {Bernard}\ \bibnamefont
  {Carr}}, \bibinfo {author} {\bibfnamefont {Florian}\ \bibnamefont {Kuhnel}},
  \ and\ \bibinfo {author} {\bibfnamefont {Marit}\ \bibnamefont {Sandstad}},\
  }\bibfield  {title} {\enquote {\bibinfo {title} {{Primordial Black Holes as
  Dark Matter}},}\ }\href {\doibase 10.1103/PhysRevD.94.083504} {\bibfield
  {journal} {\bibinfo  {journal} {Phys. Rev. D}\ }\textbf {\bibinfo {volume}
  {94}},\ \bibinfo {pages} {083504} (\bibinfo {year} {2016})},\ \Eprint
  {http://arxiv.org/abs/1607.06077} {arXiv:1607.06077 [astro-ph.CO]}
  \BibitemShut {NoStop}%
\bibitem [{\citenamefont {Askar}\ \emph {et~al.}(2017)\citenamefont {Askar},
  \citenamefont {Szkudlarek}, \citenamefont {Gondek-Rosi\'nska}, \citenamefont
  {Giersz},\ and\ \citenamefont {Bulik}}]{Askar:2016jwt}%
  \BibitemOpen
  \bibfield  {author} {\bibinfo {author} {\bibfnamefont {Abbas}\ \bibnamefont
  {Askar}}, \bibinfo {author} {\bibfnamefont {Magdalena}\ \bibnamefont
  {Szkudlarek}}, \bibinfo {author} {\bibfnamefont {Dorota}\ \bibnamefont
  {Gondek-Rosi\'nska}}, \bibinfo {author} {\bibfnamefont {Mirek}\ \bibnamefont
  {Giersz}}, \ and\ \bibinfo {author} {\bibfnamefont {Tomasz}\ \bibnamefont
  {Bulik}},\ }\bibfield  {title} {\enquote {\bibinfo {title} {{MOCCA-SURVEY
  Database -- I. Coalescing binary black holes originating from globular
  clusters}},}\ }\href {\doibase 10.1093/mnrasl/slw177} {\bibfield  {journal}
  {\bibinfo  {journal} {Mon. Not. Roy. Astron. Soc.}\ }\textbf {\bibinfo
  {volume} {464}},\ \bibinfo {pages} {L36--L40} (\bibinfo {year} {2017})},\
  \Eprint {http://arxiv.org/abs/1608.02520} {arXiv:1608.02520 [astro-ph.HE]}
  \BibitemShut {NoStop}%
\bibitem [{\citenamefont {Stevenson}\ \emph {et~al.}(2017)\citenamefont
  {Stevenson}, \citenamefont {Vigna-Gómez}, \citenamefont {Mandel},
  \citenamefont {Barrett}, \citenamefont {Neijssel}, \citenamefont {Perkins},\
  and\ \citenamefont {de~Mink}}]{Stevenson:2017tfq}%
  \BibitemOpen
  \bibfield  {author} {\bibinfo {author} {\bibfnamefont {Simon}\ \bibnamefont
  {Stevenson}}, \bibinfo {author} {\bibfnamefont {Alejandro}\ \bibnamefont
  {Vigna-Gómez}}, \bibinfo {author} {\bibfnamefont {Ilya}\ \bibnamefont
  {Mandel}}, \bibinfo {author} {\bibfnamefont {Jim~W.}\ \bibnamefont
  {Barrett}}, \bibinfo {author} {\bibfnamefont {Coenraad~J.}\ \bibnamefont
  {Neijssel}}, \bibinfo {author} {\bibfnamefont {David}\ \bibnamefont
  {Perkins}}, \ and\ \bibinfo {author} {\bibfnamefont {Selma~E.}\ \bibnamefont
  {de~Mink}},\ }\bibfield  {title} {\enquote {\bibinfo {title} {{Formation of
  the first three gravitational-wave observations through isolated binary
  evolution}},}\ }\href {\doibase 10.1038/ncomms14906} {\bibfield  {journal}
  {\bibinfo  {journal} {Nature Commun.}\ }\textbf {\bibinfo {volume} {8}},\
  \bibinfo {pages} {14906} (\bibinfo {year} {2017})},\ \Eprint
  {http://arxiv.org/abs/1704.01352} {arXiv:1704.01352 [astro-ph.HE]}
  \BibitemShut {NoStop}%
\bibitem [{\citenamefont {Spera}\ \emph {et~al.}(2018)\citenamefont {Spera},
  \citenamefont {Mapelli}, \citenamefont {Giacobbo}, \citenamefont {Trani},
  \citenamefont {Bressan},\ and\ \citenamefont {Costa}}]{Spera:2018wnw}%
  \BibitemOpen
  \bibfield  {author} {\bibinfo {author} {\bibfnamefont {Mario}\ \bibnamefont
  {Spera}}, \bibinfo {author} {\bibfnamefont {Michela}\ \bibnamefont
  {Mapelli}}, \bibinfo {author} {\bibfnamefont {Nicola}\ \bibnamefont
  {Giacobbo}}, \bibinfo {author} {\bibfnamefont {Alessandro~Alberto}\
  \bibnamefont {Trani}}, \bibinfo {author} {\bibfnamefont {Alessandro}\
  \bibnamefont {Bressan}}, \ and\ \bibinfo {author} {\bibfnamefont {Guglielmo}\
  \bibnamefont {Costa}},\ }\bibfield  {title} {\enquote {\bibinfo {title}
  {{Merging black hole binaries with the SEVN code}},}\ }\href {\doibase
  10.1093/mnras/stz359} {\  (\bibinfo {year} {2018}),\ 10.1093/mnras/stz359},\
  \Eprint {http://arxiv.org/abs/1809.04605} {arXiv:1809.04605 [astro-ph.HE]}
  \BibitemShut {NoStop}%
\bibitem [{\citenamefont {Mapelli}\ and\ \citenamefont
  {Giacobbo}(2018)}]{Mapelli:2018wys}%
  \BibitemOpen
  \bibfield  {author} {\bibinfo {author} {\bibfnamefont {Michela}\ \bibnamefont
  {Mapelli}}\ and\ \bibinfo {author} {\bibfnamefont {Nicola}\ \bibnamefont
  {Giacobbo}},\ }\bibfield  {title} {\enquote {\bibinfo {title} {{The cosmic
  merger rate of neutron stars and black holes}},}\ }\href {\doibase
  10.1093/mnras/sty1613} {\bibfield  {journal} {\bibinfo  {journal} {Mon. Not.
  Roy. Astron. Soc.}\ }\textbf {\bibinfo {volume} {479}},\ \bibinfo {pages}
  {4391--4398} (\bibinfo {year} {2018})},\ \Eprint
  {http://arxiv.org/abs/1806.04866} {arXiv:1806.04866 [astro-ph.HE]}
  \BibitemShut {NoStop}%
\bibitem [{\citenamefont {Gerosa}\ and\ \citenamefont
  {Berti}(2019)}]{Gerosa:2019zmo}%
  \BibitemOpen
  \bibfield  {author} {\bibinfo {author} {\bibfnamefont {Davide}\ \bibnamefont
  {Gerosa}}\ and\ \bibinfo {author} {\bibfnamefont {Emanuele}\ \bibnamefont
  {Berti}},\ }\bibfield  {title} {\enquote {\bibinfo {title} {{Escape speed of
  stellar clusters from multiple-generation black-hole mergers in the upper
  mass gap}},}\ }\href {\doibase 10.1103/PhysRevD.100.041301} {\bibfield
  {journal} {\bibinfo  {journal} {Phys. Rev. D}\ }\textbf {\bibinfo {volume}
  {100}},\ \bibinfo {pages} {041301} (\bibinfo {year} {2019})},\ \Eprint
  {http://arxiv.org/abs/1906.05295} {arXiv:1906.05295 [astro-ph.HE]}
  \BibitemShut {NoStop}%
\bibitem [{\citenamefont {Baibhav}\ \emph {et~al.}(2020)\citenamefont
  {Baibhav}, \citenamefont {Gerosa}, \citenamefont {Berti}, \citenamefont
  {Wong}, \citenamefont {Helfer},\ and\ \citenamefont
  {Mould}}]{Baibhav:2020xdf}%
  \BibitemOpen
  \bibfield  {author} {\bibinfo {author} {\bibfnamefont {Vishal}\ \bibnamefont
  {Baibhav}}, \bibinfo {author} {\bibfnamefont {Davide}\ \bibnamefont
  {Gerosa}}, \bibinfo {author} {\bibfnamefont {Emanuele}\ \bibnamefont
  {Berti}}, \bibinfo {author} {\bibfnamefont {Kaze~W.K.}\ \bibnamefont {Wong}},
  \bibinfo {author} {\bibfnamefont {Thomas}\ \bibnamefont {Helfer}}, \ and\
  \bibinfo {author} {\bibfnamefont {Matthew}\ \bibnamefont {Mould}},\
  }\bibfield  {title} {\enquote {\bibinfo {title} {{The mass gap, the spin gap,
  and the origin of merging binary black holes}},}\ }\href@noop {} {\
  (\bibinfo {year} {2020})},\ \Eprint {http://arxiv.org/abs/2004.00650}
  {arXiv:2004.00650 [astro-ph.HE]} \BibitemShut {NoStop}%
\bibitem [{\citenamefont {Kovetz}\ \emph {et~al.}(2017)\citenamefont {Kovetz},
  \citenamefont {Cholis}, \citenamefont {Breysse},\ and\ \citenamefont
  {Kamionkowski}}]{Kovetz:2016kpi}%
  \BibitemOpen
  \bibfield  {author} {\bibinfo {author} {\bibfnamefont {Ely~D.}\ \bibnamefont
  {Kovetz}}, \bibinfo {author} {\bibfnamefont {Ilias}\ \bibnamefont {Cholis}},
  \bibinfo {author} {\bibfnamefont {Patrick~C.}\ \bibnamefont {Breysse}}, \
  and\ \bibinfo {author} {\bibfnamefont {Marc}\ \bibnamefont {Kamionkowski}},\
  }\bibfield  {title} {\enquote {\bibinfo {title} {{Black hole mass function
  from gravitational wave measurements}},}\ }\href {\doibase
  10.1103/PhysRevD.95.103010} {\bibfield  {journal} {\bibinfo  {journal} {Phys.
  Rev. D}\ }\textbf {\bibinfo {volume} {95}},\ \bibinfo {pages} {103010}
  (\bibinfo {year} {2017})},\ \Eprint {http://arxiv.org/abs/1611.01157}
  {arXiv:1611.01157 [astro-ph.CO]} \BibitemShut {NoStop}%
\bibitem [{\citenamefont {Eldridge}\ \emph {et~al.}(2017)\citenamefont
  {Eldridge}, \citenamefont {Stanway}, \citenamefont {Xiao}, \citenamefont
  {McClelland}, \citenamefont {Taylor}, \citenamefont {Ng}, \citenamefont
  {Greis},\ and\ \citenamefont {Bray}}]{Eldridge_2017}%
  \BibitemOpen
  \bibfield  {author} {\bibinfo {author} {\bibfnamefont {J.~J.}\ \bibnamefont
  {Eldridge}}, \bibinfo {author} {\bibfnamefont {E.~R.}\ \bibnamefont
  {Stanway}}, \bibinfo {author} {\bibfnamefont {L.}~\bibnamefont {Xiao}},
  \bibinfo {author} {\bibfnamefont {L.~A.~S.}\ \bibnamefont {McClelland}},
  \bibinfo {author} {\bibfnamefont {G.}~\bibnamefont {Taylor}}, \bibinfo
  {author} {\bibfnamefont {M.}~\bibnamefont {Ng}}, \bibinfo {author}
  {\bibfnamefont {S.~M.~L.}\ \bibnamefont {Greis}}, \ and\ \bibinfo {author}
  {\bibfnamefont {J.~C.}\ \bibnamefont {Bray}},\ }\bibfield  {title} {\enquote
  {\bibinfo {title} {Binary population and spectral synthesis version 2.1:
  Construction, observational verification, and new results},}\ }\href
  {\doibase 10.1017/pasa.2017.51} {\bibfield  {journal} {\bibinfo  {journal}
  {Publications of the Astronomical Society of Australia}\ }\textbf {\bibinfo
  {volume} {34}} (\bibinfo {year} {2017}),\ 10.1017/pasa.2017.51}\BibitemShut
  {NoStop}%
\bibitem [{\citenamefont {Mandel}\ \emph {et~al.}(2019)\citenamefont {Mandel},
  \citenamefont {Farr},\ and\ \citenamefont {Gair}}]{Mandel:2018mve}%
  \BibitemOpen
  \bibfield  {author} {\bibinfo {author} {\bibfnamefont {Ilya}\ \bibnamefont
  {Mandel}}, \bibinfo {author} {\bibfnamefont {Will~M.}\ \bibnamefont {Farr}},
  \ and\ \bibinfo {author} {\bibfnamefont {Jonathan~R.}\ \bibnamefont {Gair}},\
  }\bibfield  {title} {\enquote {\bibinfo {title} {{Extracting distribution
  parameters from multiple uncertain observations with selection biases}},}\
  }\href {\doibase 10.1093/mnras/stz896} {\bibfield  {journal} {\bibinfo
  {journal} {Mon. Not. Roy. Astron. Soc.}\ }\textbf {\bibinfo {volume} {486}},\
  \bibinfo {pages} {1086--1093} (\bibinfo {year} {2019})},\ \Eprint
  {http://arxiv.org/abs/1809.02063} {arXiv:1809.02063 [physics.data-an]}
  \BibitemShut {NoStop}%
\bibitem [{\citenamefont {Bouffanais}\ \emph {et~al.}(2019)\citenamefont
  {Bouffanais}, \citenamefont {Mapelli}, \citenamefont {Gerosa}, \citenamefont
  {Di~Carlo}, \citenamefont {Giacobbo}, \citenamefont {Berti},\ and\
  \citenamefont {Baibhav}}]{Bouffanais:2019nrw}%
  \BibitemOpen
  \bibfield  {author} {\bibinfo {author} {\bibfnamefont {Yann}\ \bibnamefont
  {Bouffanais}}, \bibinfo {author} {\bibfnamefont {Michela}\ \bibnamefont
  {Mapelli}}, \bibinfo {author} {\bibfnamefont {Davide}\ \bibnamefont
  {Gerosa}}, \bibinfo {author} {\bibfnamefont {Ugo~N.}\ \bibnamefont
  {Di~Carlo}}, \bibinfo {author} {\bibfnamefont {Nicola}\ \bibnamefont
  {Giacobbo}}, \bibinfo {author} {\bibfnamefont {Emanuele}\ \bibnamefont
  {Berti}}, \ and\ \bibinfo {author} {\bibfnamefont {Vishal}\ \bibnamefont
  {Baibhav}},\ }\bibfield  {title} {\enquote {\bibinfo {title} {{Constraining
  the fraction of binary black holes formed in isolation and young star
  clusters with gravitational-wave data}},}\ }\href {\doibase
  10.3847/1538-4357/ab4a79} {\bibfield  {journal} {\bibinfo  {journal}
  {Astrophys. J.}\ }\textbf {\bibinfo {volume} {886}} (\bibinfo {year}
  {2019}),\ 10.3847/1538-4357/ab4a79},\ \Eprint
  {http://arxiv.org/abs/1905.11054} {arXiv:1905.11054 [astro-ph.HE]}
  \BibitemShut {NoStop}%
\bibitem [{\citenamefont {Baibhav}\ \emph {et~al.}(2019)\citenamefont
  {Baibhav}, \citenamefont {Berti}, \citenamefont {Gerosa}, \citenamefont
  {Mapelli}, \citenamefont {Giacobbo}, \citenamefont {Bouffanais},\ and\
  \citenamefont {Di~Carlo}}]{Baibhav:2019gxm}%
  \BibitemOpen
  \bibfield  {author} {\bibinfo {author} {\bibfnamefont {Vishal}\ \bibnamefont
  {Baibhav}}, \bibinfo {author} {\bibfnamefont {Emanuele}\ \bibnamefont
  {Berti}}, \bibinfo {author} {\bibfnamefont {Davide}\ \bibnamefont {Gerosa}},
  \bibinfo {author} {\bibfnamefont {Michela}\ \bibnamefont {Mapelli}}, \bibinfo
  {author} {\bibfnamefont {Nicola}\ \bibnamefont {Giacobbo}}, \bibinfo {author}
  {\bibfnamefont {Yann}\ \bibnamefont {Bouffanais}}, \ and\ \bibinfo {author}
  {\bibfnamefont {Ugo~N.}\ \bibnamefont {Di~Carlo}},\ }\bibfield  {title}
  {\enquote {\bibinfo {title} {{Gravitational-wave detection rates for compact
  binaries formed in isolation: LIGO/Virgo O3 and beyond}},}\ }\href {\doibase
  10.1103/PhysRevD.100.064060} {\bibfield  {journal} {\bibinfo  {journal}
  {Phys. Rev. D}\ }\textbf {\bibinfo {volume} {100}},\ \bibinfo {pages}
  {064060} (\bibinfo {year} {2019})},\ \Eprint
  {http://arxiv.org/abs/1906.04197} {arXiv:1906.04197 [gr-qc]} \BibitemShut
  {NoStop}%
\bibitem [{\citenamefont {O'Leary}\ \emph {et~al.}(2006)\citenamefont
  {O'Leary}, \citenamefont {Rasio}, \citenamefont {Fregeau}, \citenamefont
  {Ivanova},\ and\ \citenamefont {O'Shaughnessy}}]{OLeary:2005vqo}%
  \BibitemOpen
  \bibfield  {author} {\bibinfo {author} {\bibfnamefont {Ryan~M.}\ \bibnamefont
  {O'Leary}}, \bibinfo {author} {\bibfnamefont {Frederic~A.}\ \bibnamefont
  {Rasio}}, \bibinfo {author} {\bibfnamefont {John~M.}\ \bibnamefont
  {Fregeau}}, \bibinfo {author} {\bibfnamefont {Natalia}\ \bibnamefont
  {Ivanova}}, \ and\ \bibinfo {author} {\bibfnamefont {Richard~W.}\
  \bibnamefont {O'Shaughnessy}},\ }\bibfield  {title} {\enquote {\bibinfo
  {title} {{Binary mergers and growth of black holes in dense star
  clusters}},}\ }\href {\doibase 10.1086/498446} {\bibfield  {journal}
  {\bibinfo  {journal} {Astrophys. J.}\ }\textbf {\bibinfo {volume} {637}},\
  \bibinfo {pages} {937--951} (\bibinfo {year} {2006})},\ \Eprint
  {http://arxiv.org/abs/astro-ph/0508224} {arXiv:astro-ph/0508224} \BibitemShut
  {NoStop}%
\bibitem [{\citenamefont {{Downing}}\ \emph {et~al.}(2010)\citenamefont
  {{Downing}}, \citenamefont {{Benacquista}}, \citenamefont {{Giersz}},\ and\
  \citenamefont {{Spurzem}}}]{2010MNRAS.407.1946D}%
  \BibitemOpen
  \bibfield  {author} {\bibinfo {author} {\bibfnamefont {J.~M.~B.}\
  \bibnamefont {{Downing}}}, \bibinfo {author} {\bibfnamefont {M.~J.}\
  \bibnamefont {{Benacquista}}}, \bibinfo {author} {\bibfnamefont
  {M.}~\bibnamefont {{Giersz}}}, \ and\ \bibinfo {author} {\bibfnamefont
  {R.}~\bibnamefont {{Spurzem}}},\ }\bibfield  {title} {\enquote {\bibinfo
  {title} {{Compact binaries in star clusters - I. Black hole binaries inside
  globular clusters}},}\ }\href {\doibase 10.1111/j.1365-2966.2010.17040.x}
  {\bibfield  {journal} {\bibinfo  {journal} {\mnras}\ }\textbf {\bibinfo
  {volume} {407}},\ \bibinfo {pages} {1946--1962} (\bibinfo {year} {2010})},\
  \Eprint {http://arxiv.org/abs/0910.0546} {arXiv:0910.0546 [astro-ph.SR]}
  \BibitemShut {NoStop}%
\bibitem [{\citenamefont {Rodriguez}\ \emph {et~al.}(2015)\citenamefont
  {Rodriguez}, \citenamefont {Morscher}, \citenamefont {Pattabiraman},
  \citenamefont {Chatterjee}, \citenamefont {Haster},\ and\ \citenamefont
  {Rasio}}]{Rodriguez:2015oxa}%
  \BibitemOpen
  \bibfield  {author} {\bibinfo {author} {\bibfnamefont {Carl~L.}\ \bibnamefont
  {Rodriguez}}, \bibinfo {author} {\bibfnamefont {Meagan}\ \bibnamefont
  {Morscher}}, \bibinfo {author} {\bibfnamefont {Bharath}\ \bibnamefont
  {Pattabiraman}}, \bibinfo {author} {\bibfnamefont {Sourav}\ \bibnamefont
  {Chatterjee}}, \bibinfo {author} {\bibfnamefont {Carl-Johan}\ \bibnamefont
  {Haster}}, \ and\ \bibinfo {author} {\bibfnamefont {Frederic~A.}\
  \bibnamefont {Rasio}},\ }\bibfield  {title} {\enquote {\bibinfo {title}
  {{Binary Black Hole Mergers from Globular Clusters: Implications for Advanced
  LIGO}},}\ }\href {\doibase 10.1103/PhysRevLett.115.051101} {\bibfield
  {journal} {\bibinfo  {journal} {Phys. Rev. Lett.}\ }\textbf {\bibinfo
  {volume} {115}},\ \bibinfo {pages} {051101} (\bibinfo {year} {2015})},\
  \bibinfo {note} {[Erratum: Phys.Rev.Lett. 116, 029901 (2016)]},\ \Eprint
  {http://arxiv.org/abs/1505.00792} {arXiv:1505.00792 [astro-ph.HE]}
  \BibitemShut {NoStop}%
\bibitem [{\citenamefont {Rodriguez}\ \emph {et~al.}(2016)\citenamefont
  {Rodriguez}, \citenamefont {Chatterjee},\ and\ \citenamefont
  {Rasio}}]{Rodriguez:2016kxx}%
  \BibitemOpen
  \bibfield  {author} {\bibinfo {author} {\bibfnamefont {Carl~L.}\ \bibnamefont
  {Rodriguez}}, \bibinfo {author} {\bibfnamefont {Sourav}\ \bibnamefont
  {Chatterjee}}, \ and\ \bibinfo {author} {\bibfnamefont {Frederic~A.}\
  \bibnamefont {Rasio}},\ }\bibfield  {title} {\enquote {\bibinfo {title}
  {{Binary Black Hole Mergers from Globular Clusters: Masses, Merger Rates, and
  the Impact of Stellar Evolution}},}\ }\href {\doibase
  10.1103/PhysRevD.93.084029} {\bibfield  {journal} {\bibinfo  {journal} {Phys.
  Rev. D}\ }\textbf {\bibinfo {volume} {93}},\ \bibinfo {pages} {084029}
  (\bibinfo {year} {2016})},\ \Eprint {http://arxiv.org/abs/1602.02444}
  {arXiv:1602.02444 [astro-ph.HE]} \BibitemShut {NoStop}%
\bibitem [{\citenamefont {Fragione}\ and\ \citenamefont
  {Kocsis}(2018)}]{Fragione:2018vty}%
  \BibitemOpen
  \bibfield  {author} {\bibinfo {author} {\bibfnamefont {Giacomo}\ \bibnamefont
  {Fragione}}\ and\ \bibinfo {author} {\bibfnamefont {Bence}\ \bibnamefont
  {Kocsis}},\ }\bibfield  {title} {\enquote {\bibinfo {title} {{Black hole
  mergers from an evolving population of globular clusters}},}\ }\href
  {\doibase 10.1103/PhysRevLett.121.161103} {\bibfield  {journal} {\bibinfo
  {journal} {Phys. Rev. Lett.}\ }\textbf {\bibinfo {volume} {121}},\ \bibinfo
  {pages} {161103} (\bibinfo {year} {2018})},\ \Eprint
  {http://arxiv.org/abs/1806.02351} {arXiv:1806.02351 [astro-ph.GA]}
  \BibitemShut {NoStop}%
\bibitem [{\citenamefont {Hong}\ \emph {et~al.}(2018)\citenamefont {Hong},
  \citenamefont {Vesperini}, \citenamefont {Askar}, \citenamefont {Giersz},
  \citenamefont {Szkudlarek},\ and\ \citenamefont {Bulik}}]{hong2018}%
  \BibitemOpen
  \bibfield  {author} {\bibinfo {author} {\bibfnamefont {Jongsuk}\ \bibnamefont
  {Hong}}, \bibinfo {author} {\bibfnamefont {Enrico}\ \bibnamefont
  {Vesperini}}, \bibinfo {author} {\bibfnamefont {Abbas}\ \bibnamefont
  {Askar}}, \bibinfo {author} {\bibfnamefont {Mirek}\ \bibnamefont {Giersz}},
  \bibinfo {author} {\bibfnamefont {Magdalena}\ \bibnamefont {Szkudlarek}}, \
  and\ \bibinfo {author} {\bibfnamefont {Tomasz}\ \bibnamefont {Bulik}},\
  }\bibfield  {title} {\enquote {\bibinfo {title} {Binary black hole mergers
  from globular clusters: the impact of globular cluster properties},}\ }\href
  {\doibase 10.1093/mnras/sty2211} {\bibfield  {journal} {\bibinfo  {journal}
  {Monthly Notices of the Royal Astronomical Society}\ }\textbf {\bibinfo
  {volume} {480}},\ \bibinfo {pages} {5645–5656} (\bibinfo {year}
  {2018})}\BibitemShut {NoStop}%
\bibitem [{\citenamefont {Haster}\ \emph {et~al.}(2016)\citenamefont {Haster},
  \citenamefont {Antonini}, \citenamefont {Kalogera},\ and\ \citenamefont
  {Mandel}}]{haster2016}%
  \BibitemOpen
  \bibfield  {author} {\bibinfo {author} {\bibfnamefont {Carl-Johan}\
  \bibnamefont {Haster}}, \bibinfo {author} {\bibfnamefont {Fabio}\
  \bibnamefont {Antonini}}, \bibinfo {author} {\bibfnamefont {Vicky}\
  \bibnamefont {Kalogera}}, \ and\ \bibinfo {author} {\bibfnamefont {Ilya}\
  \bibnamefont {Mandel}},\ }\bibfield  {title} {\enquote {\bibinfo {title}
  {N-body dynamics of intermediate mass-ratio inspirals in star clusters},}\
  }\href {\doibase 10.3847/0004-637x/832/2/192} {\bibfield  {journal} {\bibinfo
   {journal} {The Astrophysical Journal}\ }\textbf {\bibinfo {volume} {832}},\
  \bibinfo {pages} {192} (\bibinfo {year} {2016})}\BibitemShut {NoStop}%
\bibitem [{\citenamefont {Park}\ \emph {et~al.}(2017)\citenamefont {Park},
  \citenamefont {Kim}, \citenamefont {Lee}, \citenamefont {Bae},\ and\
  \citenamefont {Belczynski}}]{park2017}%
  \BibitemOpen
  \bibfield  {author} {\bibinfo {author} {\bibfnamefont {Dawoo}\ \bibnamefont
  {Park}}, \bibinfo {author} {\bibfnamefont {Chunglee}\ \bibnamefont {Kim}},
  \bibinfo {author} {\bibfnamefont {Hyung~Mok}\ \bibnamefont {Lee}}, \bibinfo
  {author} {\bibfnamefont {Yeong-Bok}\ \bibnamefont {Bae}}, \ and\ \bibinfo
  {author} {\bibfnamefont {Krzysztof}\ \bibnamefont {Belczynski}},\ }\bibfield
  {title} {\enquote {\bibinfo {title} {Black hole binaries dynamically formed
  in globular clusters},}\ }\href {\doibase 10.1093/mnras/stx1015} {\bibfield
  {journal} {\bibinfo  {journal} {Monthly Notices of the Royal Astronomical
  Society}\ }\textbf {\bibinfo {volume} {469}},\ \bibinfo {pages} {4665–4674}
  (\bibinfo {year} {2017})}\BibitemShut {NoStop}%
\bibitem [{\citenamefont {Pavlík}\ \emph {et~al.}(2018)\citenamefont
  {Pavlík}, \citenamefont {Jeřábková}, \citenamefont {Kroupa},\ and\
  \citenamefont {Baumgardt}}]{pavlik2018}%
  \BibitemOpen
  \bibfield  {author} {\bibinfo {author} {\bibfnamefont {Václav}\ \bibnamefont
  {Pavlík}}, \bibinfo {author} {\bibfnamefont {Tereza}\ \bibnamefont
  {Jeřábková}}, \bibinfo {author} {\bibfnamefont {Pavel}\ \bibnamefont
  {Kroupa}}, \ and\ \bibinfo {author} {\bibfnamefont {Holger}\ \bibnamefont
  {Baumgardt}},\ }\bibfield  {title} {\enquote {\bibinfo {title} {The black
  hole retention fraction in star clusters},}\ }\href {\doibase
  10.1051/0004-6361/201832919} {\bibfield  {journal} {\bibinfo  {journal}
  {Astronomy \& Astrophysics}\ }\textbf {\bibinfo {volume} {617}},\ \bibinfo
  {pages} {A69} (\bibinfo {year} {2018})}\BibitemShut {NoStop}%
\bibitem [{\citenamefont {Fishbach}\ \emph {et~al.}(2017)\citenamefont
  {Fishbach}, \citenamefont {Holz},\ and\ \citenamefont
  {Farr}}]{Fishbach:2017dwv}%
  \BibitemOpen
  \bibfield  {author} {\bibinfo {author} {\bibfnamefont {Maya}\ \bibnamefont
  {Fishbach}}, \bibinfo {author} {\bibfnamefont {Daniel~E.}\ \bibnamefont
  {Holz}}, \ and\ \bibinfo {author} {\bibfnamefont {Ben}\ \bibnamefont
  {Farr}},\ }\bibfield  {title} {\enquote {\bibinfo {title} {{Are LIGO's Black
  Holes Made From Smaller Black Holes?}}}\ }\href {\doibase
  10.3847/2041-8213/aa7045} {\bibfield  {journal} {\bibinfo  {journal}
  {Astrophys. J. Lett.}\ }\textbf {\bibinfo {volume} {840}},\ \bibinfo {pages}
  {L24} (\bibinfo {year} {2017})},\ \Eprint {http://arxiv.org/abs/1703.06869}
  {arXiv:1703.06869 [astro-ph.HE]} \BibitemShut {NoStop}%
\bibitem [{\citenamefont {Kovetz}\ \emph {et~al.}(2018)\citenamefont {Kovetz},
  \citenamefont {Cholis}, \citenamefont {Kamionkowski},\ and\ \citenamefont
  {Silk}}]{Kovetz:2018vly}%
  \BibitemOpen
  \bibfield  {author} {\bibinfo {author} {\bibfnamefont {Ely~D.}\ \bibnamefont
  {Kovetz}}, \bibinfo {author} {\bibfnamefont {Ilias}\ \bibnamefont {Cholis}},
  \bibinfo {author} {\bibfnamefont {Marc}\ \bibnamefont {Kamionkowski}}, \ and\
  \bibinfo {author} {\bibfnamefont {Joseph}\ \bibnamefont {Silk}},\ }\bibfield
  {title} {\enquote {\bibinfo {title} {{Limits on Runaway Growth of
  Intermediate Mass Black Holes from Advanced LIGO}},}\ }\href {\doibase
  10.1103/PhysRevD.97.123003} {\bibfield  {journal} {\bibinfo  {journal} {Phys.
  Rev. D}\ }\textbf {\bibinfo {volume} {97}},\ \bibinfo {pages} {123003}
  (\bibinfo {year} {2018})},\ \Eprint {http://arxiv.org/abs/1803.00568}
  {arXiv:1803.00568 [astro-ph.HE]} \BibitemShut {NoStop}%
\bibitem [{\citenamefont {Antonini}\ \emph {et~al.}(2019)\citenamefont
  {Antonini}, \citenamefont {Gieles},\ and\ \citenamefont
  {Gualandris}}]{Antonini:2018auk}%
  \BibitemOpen
  \bibfield  {author} {\bibinfo {author} {\bibfnamefont {Fabio}\ \bibnamefont
  {Antonini}}, \bibinfo {author} {\bibfnamefont {Mark}\ \bibnamefont {Gieles}},
  \ and\ \bibinfo {author} {\bibfnamefont {Alessia}\ \bibnamefont
  {Gualandris}},\ }\bibfield  {title} {\enquote {\bibinfo {title} {{Black hole
  growth through hierarchical black hole mergers in dense star clusters:
  implications for gravitational wave detections}},}\ }\href {\doibase
  10.1093/mnras/stz1149} {\bibfield  {journal} {\bibinfo  {journal} {Mon. Not.
  Roy. Astron. Soc.}\ }\textbf {\bibinfo {volume} {486}},\ \bibinfo {pages}
  {5008--5021} (\bibinfo {year} {2019})},\ \Eprint
  {http://arxiv.org/abs/1811.03640} {arXiv:1811.03640 [astro-ph.HE]}
  \BibitemShut {NoStop}%
\bibitem [{\citenamefont {Fishbach}\ and\ \citenamefont
  {Holz}(2017)}]{Fishbach:2017zga}%
  \BibitemOpen
  \bibfield  {author} {\bibinfo {author} {\bibfnamefont {Maya}\ \bibnamefont
  {Fishbach}}\ and\ \bibinfo {author} {\bibfnamefont {Daniel~E.}\ \bibnamefont
  {Holz}},\ }\bibfield  {title} {\enquote {\bibinfo {title} {{Where Are LIGO's
  Big Black Holes?}}}\ }\href {\doibase 10.3847/2041-8213/aa9bf6} {\bibfield
  {journal} {\bibinfo  {journal} {Astrophys. J. Lett.}\ }\textbf {\bibinfo
  {volume} {851}},\ \bibinfo {pages} {L25} (\bibinfo {year} {2017})},\ \Eprint
  {http://arxiv.org/abs/1709.08584} {arXiv:1709.08584 [astro-ph.HE]}
  \BibitemShut {NoStop}%
\bibitem [{\citenamefont {Fishbach}\ and\ \citenamefont
  {Holz}(2020)}]{Fishbach:2019bbm}%
  \BibitemOpen
  \bibfield  {author} {\bibinfo {author} {\bibfnamefont {Maya}\ \bibnamefont
  {Fishbach}}\ and\ \bibinfo {author} {\bibfnamefont {Daniel~E.}\ \bibnamefont
  {Holz}},\ }\bibfield  {title} {\enquote {\bibinfo {title} {{Picky Partners:
  The Pairing of Component Masses in Binary Black Hole Mergers}},}\ }\href
  {\doibase 10.3847/2041-8213/ab7247} {\bibfield  {journal} {\bibinfo
  {journal} {Astrophys. J. Lett.}\ }\textbf {\bibinfo {volume} {891}},\
  \bibinfo {pages} {L27} (\bibinfo {year} {2020})},\ \Eprint
  {http://arxiv.org/abs/1905.12669} {arXiv:1905.12669 [astro-ph.HE]}
  \BibitemShut {NoStop}%
\bibitem [{\citenamefont {Gerosa}\ \emph {et~al.}(2020)\citenamefont {Gerosa},
  \citenamefont {Vitale},\ and\ \citenamefont {Berti}}]{Gerosa:2020bjb}%
  \BibitemOpen
  \bibfield  {author} {\bibinfo {author} {\bibfnamefont {Davide}\ \bibnamefont
  {Gerosa}}, \bibinfo {author} {\bibfnamefont {Salvatore}\ \bibnamefont
  {Vitale}}, \ and\ \bibinfo {author} {\bibfnamefont {Emanuele}\ \bibnamefont
  {Berti}},\ }\bibfield  {title} {\enquote {\bibinfo {title} {{Astrophysical
  implications of GW190412 as a remnant of a previous black-hole merger}},}\
  }\href@noop {} {\  (\bibinfo {year} {2020})},\ \Eprint
  {http://arxiv.org/abs/2005.04243} {arXiv:2005.04243 [astro-ph.HE]}
  \BibitemShut {NoStop}%
\bibitem [{\citenamefont {Kimball}\ \emph {et~al.}(2020)\citenamefont
  {Kimball}, \citenamefont {Talbot}, \citenamefont {Berry}, \citenamefont
  {Carney}, \citenamefont {Zevin}, \citenamefont {Thrane},\ and\ \citenamefont
  {Kalogera}}]{Kimball:2020opk}%
  \BibitemOpen
  \bibfield  {author} {\bibinfo {author} {\bibfnamefont {Chase}\ \bibnamefont
  {Kimball}}, \bibinfo {author} {\bibfnamefont {Colm}\ \bibnamefont {Talbot}},
  \bibinfo {author} {\bibfnamefont {Christopher~P.L.}\ \bibnamefont {Berry}},
  \bibinfo {author} {\bibfnamefont {Matthew}\ \bibnamefont {Carney}}, \bibinfo
  {author} {\bibfnamefont {Michael}\ \bibnamefont {Zevin}}, \bibinfo {author}
  {\bibfnamefont {Eric}\ \bibnamefont {Thrane}}, \ and\ \bibinfo {author}
  {\bibfnamefont {Vicky}\ \bibnamefont {Kalogera}},\ }\bibfield  {title}
  {\enquote {\bibinfo {title} {{Black hole genealogy: Identifying hierarchical
  mergers with gravitational waves}},}\ }\href@noop {} {\  (\bibinfo {year}
  {2020})},\ \Eprint {http://arxiv.org/abs/2005.00023} {arXiv:2005.00023
  [astro-ph.HE]} \BibitemShut {NoStop}%
\bibitem [{\citenamefont {Rodriguez}\ \emph {et~al.}(2020)\citenamefont
  {Rodriguez} \emph {et~al.}}]{Rodriguez:2020viw}%
  \BibitemOpen
  \bibfield  {author} {\bibinfo {author} {\bibfnamefont {Carl~L.}\ \bibnamefont
  {Rodriguez}} \emph {et~al.},\ }\bibfield  {title} {\enquote {\bibinfo {title}
  {{GW190412 as a Third-Generation Black Hole Merger from a Super Star
  Cluster}},}\ }\href@noop {} {\  (\bibinfo {year} {2020})},\ \Eprint
  {http://arxiv.org/abs/2005.04239} {arXiv:2005.04239 [astro-ph.HE]}
  \BibitemShut {NoStop}%
\bibitem [{\citenamefont {{Peters}}(1964)}]{peters1964}%
  \BibitemOpen
  \bibfield  {author} {\bibinfo {author} {\bibfnamefont {P.~C.}\ \bibnamefont
  {{Peters}}},\ }\bibfield  {title} {\enquote {\bibinfo {title} {{Gravitational
  Radiation and the Motion of Two Point Masses}},}\ }\href {\doibase
  10.1103/PhysRev.136.B1224} {\bibfield  {journal} {\bibinfo  {journal}
  {Physical Review}\ }\textbf {\bibinfo {volume} {136}},\ \bibinfo {pages}
  {1224--1232} (\bibinfo {year} {1964})}\BibitemShut {NoStop}%
\bibitem [{\citenamefont {Mandel}\ \emph {et~al.}(2008)\citenamefont {Mandel},
  \citenamefont {Brown}, \citenamefont {Gair},\ and\ \citenamefont
  {Miller}}]{mandel2008}%
  \BibitemOpen
  \bibfield  {author} {\bibinfo {author} {\bibfnamefont {Ilya}\ \bibnamefont
  {Mandel}}, \bibinfo {author} {\bibfnamefont {Duncan~A.}\ \bibnamefont
  {Brown}}, \bibinfo {author} {\bibfnamefont {Jonathan~R.}\ \bibnamefont
  {Gair}}, \ and\ \bibinfo {author} {\bibfnamefont {M.~Coleman}\ \bibnamefont
  {Miller}},\ }\bibfield  {title} {\enquote {\bibinfo {title} {Rates and
  characteristics of intermediate mass ratio inspirals detectable by advanced
  ligo},}\ }\href {\doibase 10.1086/588246} {\bibfield  {journal} {\bibinfo
  {journal} {The Astrophysical Journal}\ }\textbf {\bibinfo {volume} {681}},\
  \bibinfo {pages} {1431–1447} (\bibinfo {year} {2008})}\BibitemShut
  {NoStop}%
\bibitem [{\citenamefont {{Quinlan}}\ and\ \citenamefont
  {{Shapiro}}(1989)}]{quinlanShapiro1989}%
  \BibitemOpen
  \bibfield  {author} {\bibinfo {author} {\bibfnamefont {Gerald~D.}\
  \bibnamefont {{Quinlan}}}\ and\ \bibinfo {author} {\bibfnamefont {Stuart~L.}\
  \bibnamefont {{Shapiro}}},\ }\bibfield  {title} {\enquote {\bibinfo {title}
  {{Dynamical Evolution of Dense Clusters of Compact Stars}},}\ }\href
  {\doibase 10.1086/167745} {\bibfield  {journal} {\bibinfo  {journal} {\apj}\
  }\textbf {\bibinfo {volume} {343}},\ \bibinfo {pages} {725} (\bibinfo {year}
  {1989})}\BibitemShut {NoStop}%
\bibitem [{\citenamefont {Mouri}\ and\ \citenamefont
  {Taniguchi}(2002{\natexlab{a}})}]{Mouri:2002mc}%
  \BibitemOpen
  \bibfield  {author} {\bibinfo {author} {\bibfnamefont {Hideaki}\ \bibnamefont
  {Mouri}}\ and\ \bibinfo {author} {\bibfnamefont {Yoshiaki}\ \bibnamefont
  {Taniguchi}},\ }\bibfield  {title} {\enquote {\bibinfo {title} {{Runaway
  merging of black holes: analytical constraint on the time\ scale}},}\ }\href
  {\doibase 10.1086/339472} {\bibfield  {journal} {\bibinfo  {journal}
  {Astrophys. J. Lett.}\ }\textbf {\bibinfo {volume} {566}},\ \bibinfo {pages}
  {L17--L20} (\bibinfo {year} {2002}{\natexlab{a}})},\ \Eprint
  {http://arxiv.org/abs/astro-ph/0201102} {arXiv:astro-ph/0201102} \BibitemShut
  {NoStop}%
\bibitem [{\citenamefont {{Hills}}(1983)}]{1983AJ.....88.1269H}%
  \BibitemOpen
  \bibfield  {author} {\bibinfo {author} {\bibfnamefont {J.~G.}\ \bibnamefont
  {{Hills}}},\ }\bibfield  {title} {\enquote {\bibinfo {title} {{The effect of
  low-velocity, low-mass intruders (collisionless\ gas) on the dynamical
  evolution of a binary system}},}\ }\href {\doibase 10.1086/113418} {\bibfield
   {journal} {\bibinfo  {journal} {\aj}\ }\textbf {\bibinfo {volume} {88}},\
  \bibinfo {pages} {1269--1283} (\bibinfo {year} {1983})}\BibitemShut {NoStop}%
\bibitem [{\citenamefont {Sesana}\ \emph
  {et~al.}(2006{\natexlab{a}})\citenamefont {Sesana}, \citenamefont {Haardt},\
  and\ \citenamefont {Madau}}]{Sesana:2006xw}%
  \BibitemOpen
  \bibfield  {author} {\bibinfo {author} {\bibfnamefont {Alberto}\ \bibnamefont
  {Sesana}}, \bibinfo {author} {\bibfnamefont {Francesco}\ \bibnamefont
  {Haardt}}, \ and\ \bibinfo {author} {\bibfnamefont {Piero}\ \bibnamefont
  {Madau}},\ }\bibfield  {title} {\enquote {\bibinfo {title} {{Interaction of
  massive black hole binaries with their stellar env\ ironment. 1. Ejection of
  hypervelocity stars}},}\ }\href {\doibase 10.1086/507596} {\bibfield
  {journal} {\bibinfo  {journal} {Astrophys. J.}\ }\textbf {\bibinfo {volume}
  {651}},\ \bibinfo {pages} {392--400} (\bibinfo {year}
  {2006}{\natexlab{a}})},\ \Eprint {http://arxiv.org/abs/astro-ph/0604299}
  {arXiv:astro-ph/0604299} \BibitemShut {NoStop}%
\bibitem [{\citenamefont {Rodriguez}\ \emph {et~al.}(2018)\citenamefont
  {Rodriguez}, \citenamefont {Amaro-Seoane}, \citenamefont {Chatterjee},
  \citenamefont {Kremer}, \citenamefont {Rasio}, \citenamefont {Samsing},
  \citenamefont {Ye},\ and\ \citenamefont {Zevin}}]{rodriguez2018}%
  \BibitemOpen
  \bibfield  {author} {\bibinfo {author} {\bibfnamefont {Carl~L.}\ \bibnamefont
  {Rodriguez}}, \bibinfo {author} {\bibfnamefont {Pau}\ \bibnamefont
  {Amaro-Seoane}}, \bibinfo {author} {\bibfnamefont {Sourav}\ \bibnamefont
  {Chatterjee}}, \bibinfo {author} {\bibfnamefont {Kyle}\ \bibnamefont
  {Kremer}}, \bibinfo {author} {\bibfnamefont {Frederic~A.}\ \bibnamefont
  {Rasio}}, \bibinfo {author} {\bibfnamefont {Johan}\ \bibnamefont {Samsing}},
  \bibinfo {author} {\bibfnamefont {Claire~S.}\ \bibnamefont {Ye}}, \ and\
  \bibinfo {author} {\bibfnamefont {Michael}\ \bibnamefont {Zevin}},\
  }\bibfield  {title} {\enquote {\bibinfo {title} {Post-newtonian dynamics in
  dense star clusters: Formation, masses, and merger rates of highly-eccentric
  black hole binaries},}\ }\href {\doibase 10.1103/physrevd.98.123005}
  {\bibfield  {journal} {\bibinfo  {journal} {Physical Review D}\ }\textbf
  {\bibinfo {volume} {98}} (\bibinfo {year} {2018}),\
  10.1103/physrevd.98.123005}\BibitemShut {NoStop}%
\bibitem [{\citenamefont {Samsing}(2018)}]{samsing2018}%
  \BibitemOpen
  \bibfield  {author} {\bibinfo {author} {\bibfnamefont {Johan}\ \bibnamefont
  {Samsing}},\ }\bibfield  {title} {\enquote {\bibinfo {title} {Eccentric black
  hole mergers forming in globular clusters},}\ }\href {\doibase
  10.1103/physrevd.97.103014} {\bibfield  {journal} {\bibinfo  {journal}
  {Physical Review D}\ }\textbf {\bibinfo {volume} {97}} (\bibinfo {year}
  {2018}),\ 10.1103/physrevd.97.103014}\BibitemShut {NoStop}%
\bibitem [{\citenamefont {Miller}\ and\ \citenamefont
  {Hamilton}(2002)}]{miller2002}%
  \BibitemOpen
  \bibfield  {author} {\bibinfo {author} {\bibfnamefont {M.~Coleman}\
  \bibnamefont {Miller}}\ and\ \bibinfo {author} {\bibfnamefont {Douglas~P.}\
  \bibnamefont {Hamilton}},\ }\bibfield  {title} {\enquote {\bibinfo {title}
  {Four‐body effects in globular cluster black hole coalescence},}\ }\href
  {\doibase 10.1086/341788} {\bibfield  {journal} {\bibinfo  {journal} {The
  Astrophysical Journal}\ }\textbf {\bibinfo {volume} {576}},\ \bibinfo {pages}
  {894–898} (\bibinfo {year} {2002})}\BibitemShut {NoStop}%
\bibitem [{\citenamefont {{Arca-Sedda}}\ \emph {et~al.}(2018)\citenamefont
  {{Arca-Sedda}}, \citenamefont {{Li}},\ and\ \citenamefont
  {{Kocsis}}}]{arcaSedda2018}%
  \BibitemOpen
  \bibfield  {author} {\bibinfo {author} {\bibfnamefont {Manuel}\ \bibnamefont
  {{Arca-Sedda}}}, \bibinfo {author} {\bibfnamefont {Gongjie}\ \bibnamefont
  {{Li}}}, \ and\ \bibinfo {author} {\bibfnamefont {Bence}\ \bibnamefont
  {{Kocsis}}},\ }\bibfield  {title} {\enquote {\bibinfo {title} {{Ordering the
  chaos: stellar black hole mergers from non-hierarchical triples}},}\
  }\href@noop {} {\bibfield  {journal} {\bibinfo  {journal} {arXiv e-prints}\
  ,\ \bibinfo {eid} {arXiv:1805.06458}} (\bibinfo {year} {2018})},\ \Eprint
  {http://arxiv.org/abs/1805.06458} {arXiv:1805.06458 [astro-ph.HE]}
  \BibitemShut {NoStop}%
\bibitem [{\citenamefont {Antognini}\ and\ \citenamefont
  {Thompson}(2016)}]{antognini2016}%
  \BibitemOpen
  \bibfield  {author} {\bibinfo {author} {\bibfnamefont {Joseph M.~O.}\
  \bibnamefont {Antognini}}\ and\ \bibinfo {author} {\bibfnamefont {Todd~A.}\
  \bibnamefont {Thompson}},\ }\bibfield  {title} {\enquote {\bibinfo {title}
  {Dynamical formation and scattering of hierarchical triples: cross-sections,
  kozai–lidov oscillations, and collisions},}\ }\href {\doibase
  10.1093/mnras/stv2938} {\bibfield  {journal} {\bibinfo  {journal} {Monthly
  Notices of the Royal Astronomical Society}\ }\textbf {\bibinfo {volume}
  {456}},\ \bibinfo {pages} {4219–4246} (\bibinfo {year} {2016})}\BibitemShut
  {NoStop}%
\bibitem [{\citenamefont {Catalog}()}]{GLOBCLUST}%
  \BibitemOpen
  \bibfield  {author} {\bibinfo {author} {\bibfnamefont {GLOBCLUST Milky Way
  Globular~Clusters}\ \bibnamefont {Catalog}},\ }\bibfield  {title} {\enquote
  {\bibinfo {title} {https://heasarc.gsfc.nasa.gov/w3browse/all/globclust
  .html},}\ }\href@noop {} {\ }\BibitemShut {NoStop}%
\bibitem [{\citenamefont {{King}}(1962)}]{king1962}%
  \BibitemOpen
  \bibfield  {author} {\bibinfo {author} {\bibfnamefont {Ivan}\ \bibnamefont
  {{King}}},\ }\bibfield  {title} {\enquote {\bibinfo {title} {{The structure
  of star clusters. I. an empirical density law}},}\ }\href {\doibase
  10.1086/108756} {\bibfield  {journal} {\bibinfo  {journal} {\aj}\ }\textbf
  {\bibinfo {volume} {67}},\ \bibinfo {pages} {471} (\bibinfo {year}
  {1962})}\BibitemShut {NoStop}%
\bibitem [{\citenamefont {{Collins}}(1978)}]{collins1978}%
  \BibitemOpen
  \bibfield  {author} {\bibinfo {author} {\bibfnamefont {II}~\bibnamefont
  {{Collins}}, \bibfnamefont {G.~W.}},\ }\href@noop {} {\emph {\bibinfo {title}
  {{The virial theorem in stellar astrophysics}}}}\ (\bibinfo {year}
  {1978})\BibitemShut {NoStop}%
\bibitem [{\citenamefont {{Plummer}}(1911)}]{plummer1911}%
  \BibitemOpen
  \bibfield  {author} {\bibinfo {author} {\bibfnamefont {H.~C.}\ \bibnamefont
  {{Plummer}}},\ }\bibfield  {title} {\enquote {\bibinfo {title} {{On the
  problem of distribution in globular star clusters}},}\ }\href {\doibase
  10.1093/mnras/71.5.460} {\bibfield  {journal} {\bibinfo  {journal} {\mnras}\
  }\textbf {\bibinfo {volume} {71}},\ \bibinfo {pages} {460--470} (\bibinfo
  {year} {1911})}\BibitemShut {NoStop}%
\bibitem [{\citenamefont {{Dejonghe}}(1987)}]{dejonghe1987}%
  \BibitemOpen
  \bibfield  {author} {\bibinfo {author} {\bibfnamefont {Herwig}\ \bibnamefont
  {{Dejonghe}}},\ }\bibfield  {title} {\enquote {\bibinfo {title} {{A
  completely analytical family of anisotropic Plummer models}},}\ }\href
  {\doibase 10.1093/mnras/224.1.13} {\bibfield  {journal} {\bibinfo  {journal}
  {\mnras}\ }\textbf {\bibinfo {volume} {224}},\ \bibinfo {pages} {13--39}
  (\bibinfo {year} {1987})}\BibitemShut {NoStop}%
\bibitem [{\citenamefont {{Harris}}(1996)}]{harris1996}%
  \BibitemOpen
  \bibfield  {author} {\bibinfo {author} {\bibfnamefont {William~E.}\
  \bibnamefont {{Harris}}},\ }\bibfield  {title} {\enquote {\bibinfo {title}
  {{A Catalog of Parameters for Globular Clusters in the Milky Way}},}\ }\href
  {\doibase 10.1086/118116} {\bibfield  {journal} {\bibinfo  {journal} {\aj}\
  }\textbf {\bibinfo {volume} {112}},\ \bibinfo {pages} {1487} (\bibinfo {year}
  {1996})}\BibitemShut {NoStop}%
\bibitem [{\citenamefont {Kroupa}(2002)}]{kroupa2002}%
  \BibitemOpen
  \bibfield  {author} {\bibinfo {author} {\bibfnamefont {P.}~\bibnamefont
  {Kroupa}},\ }\bibfield  {title} {\enquote {\bibinfo {title} {The initial mass
  function of stars: Evidence for uniformity in variable systems},}\ }\href
  {\doibase 10.1126/science.1067524} {\bibfield  {journal} {\bibinfo  {journal}
  {Science}\ }\textbf {\bibinfo {volume} {295}},\ \bibinfo {pages} {82–91}
  (\bibinfo {year} {2002})}\BibitemShut {NoStop}%
\bibitem [{\citenamefont {Kruckow}\ \emph {et~al.}(2018)\citenamefont
  {Kruckow}, \citenamefont {Tauris}, \citenamefont {Langer}, \citenamefont
  {Kramer},\ and\ \citenamefont {Izzard}}]{Kruckow:2018slo}%
  \BibitemOpen
  \bibfield  {author} {\bibinfo {author} {\bibfnamefont {Matthias~U.}\
  \bibnamefont {Kruckow}}, \bibinfo {author} {\bibfnamefont {Thomas~M.}\
  \bibnamefont {Tauris}}, \bibinfo {author} {\bibfnamefont {Norbert}\
  \bibnamefont {Langer}}, \bibinfo {author} {\bibfnamefont {Michael}\
  \bibnamefont {Kramer}}, \ and\ \bibinfo {author} {\bibfnamefont {Robert~G.}\
  \bibnamefont {Izzard}},\ }\bibfield  {title} {\enquote {\bibinfo {title}
  {{Progenitors of gravitational wave mergers: Binary evolution with the
  stellar grid-based code ComBinE}},}\ }\href {\doibase 10.1093/mnras/sty2190}
  {\bibfield  {journal} {\bibinfo  {journal} {Mon. Not. Roy. Astron. Soc.}\
  }\textbf {\bibinfo {volume} {481}},\ \bibinfo {pages} {1908--1949} (\bibinfo
  {year} {2018})},\ \Eprint {http://arxiv.org/abs/1801.05433} {arXiv:1801.05433
  [astro-ph.SR]} \BibitemShut {NoStop}%
\bibitem [{\citenamefont {Giacobbo}\ and\ \citenamefont
  {Mapelli}(2018)}]{Giacobbo:2018etu}%
  \BibitemOpen
  \bibfield  {author} {\bibinfo {author} {\bibfnamefont {Nicola}\ \bibnamefont
  {Giacobbo}}\ and\ \bibinfo {author} {\bibfnamefont {Michela}\ \bibnamefont
  {Mapelli}},\ }\bibfield  {title} {\enquote {\bibinfo {title} {{The
  progenitors of compact-object binaries: impact of metallicity, common
  envelope and natal kicks}},}\ }\href {\doibase 10.1093/mnras/sty1999}
  {\bibfield  {journal} {\bibinfo  {journal} {Mon. Not. Roy. Astron. Soc.}\
  }\textbf {\bibinfo {volume} {480}},\ \bibinfo {pages} {2011--2030} (\bibinfo
  {year} {2018})},\ \Eprint {http://arxiv.org/abs/1806.00001} {arXiv:1806.00001
  [astro-ph.HE]} \BibitemShut {NoStop}%
\bibitem [{\citenamefont {{Spitzer}}(1969)}]{1969ApJ...158L.139S}%
  \BibitemOpen
  \bibfield  {author} {\bibinfo {author} {\bibfnamefont {Jr.}\ \bibnamefont
  {{Spitzer}}, \bibfnamefont {Lyman}},\ }\bibfield  {title} {\enquote {\bibinfo
  {title} {{Equipartition and the Formation of Compact Nuclei in Spherical
  Stellar Systems}},}\ }\href {\doibase 10.1086/180451} {\bibfield  {journal}
  {\bibinfo  {journal} {\apjl}\ }\textbf {\bibinfo {volume} {158}},\ \bibinfo
  {pages} {L139} (\bibinfo {year} {1969})}\BibitemShut {NoStop}%
\bibitem [{\citenamefont {{Heggie}}(1975)}]{heggie1975}%
  \BibitemOpen
  \bibfield  {author} {\bibinfo {author} {\bibfnamefont {D.~C.}\ \bibnamefont
  {{Heggie}}},\ }\bibfield  {title} {\enquote {\bibinfo {title} {{Binary
  evolution in stellar dynamics.}}}\ }\href {\doibase 10.1093/mnras/173.3.729}
  {\bibfield  {journal} {\bibinfo  {journal} {\mnras}\ }\textbf {\bibinfo
  {volume} {173}},\ \bibinfo {pages} {729--787} (\bibinfo {year}
  {1975})}\BibitemShut {NoStop}%
\bibitem [{\citenamefont {Ivanova}\ \emph {et~al.}(2010)\citenamefont
  {Ivanova}, \citenamefont {Kologera},\ and\ \citenamefont {van~der
  Sluys}}]{ivanova2010}%
  \BibitemOpen
  \bibfield  {author} {\bibinfo {author} {\bibfnamefont {Natalia}\ \bibnamefont
  {Ivanova}}, \bibinfo {author} {\bibfnamefont {Vicky}\ \bibnamefont
  {Kologera}}, \ and\ \bibinfo {author} {\bibfnamefont {Marc}\ \bibnamefont
  {van~der Sluys}},\ }\bibfield  {title} {\enquote {\bibinfo {title} {Evolution
  of binaries in dense stellar systems},}\ }\href {\doibase 10.1063/1.3536357}
  {\  (\bibinfo {year} {2010}),\ 10.1063/1.3536357}\BibitemShut {NoStop}%
\bibitem [{\citenamefont {Ivanova}\ \emph {et~al.}(2005)\citenamefont
  {Ivanova}, \citenamefont {Belczynski}, \citenamefont {Fregeau},\ and\
  \citenamefont {Rasio}}]{ivanova2005}%
  \BibitemOpen
  \bibfield  {author} {\bibinfo {author} {\bibfnamefont {N.}~\bibnamefont
  {Ivanova}}, \bibinfo {author} {\bibfnamefont {K.}~\bibnamefont {Belczynski}},
  \bibinfo {author} {\bibfnamefont {J.~M.}\ \bibnamefont {Fregeau}}, \ and\
  \bibinfo {author} {\bibfnamefont {F.~A.}\ \bibnamefont {Rasio}},\ }\bibfield
  {title} {\enquote {\bibinfo {title} {The evolution of binary fractions in
  globular clusters},}\ }\href {\doibase 10.1111/j.1365-2966.2005.08804.x}
  {\bibfield  {journal} {\bibinfo  {journal} {Monthly Notices of the Royal
  Astronomical Society}\ }\textbf {\bibinfo {volume} {358}},\ \bibinfo {pages}
  {572–584} (\bibinfo {year} {2005})}\BibitemShut {NoStop}%
\bibitem [{\citenamefont {{Hills}}\ and\ \citenamefont
  {{Fullerton}}(1980)}]{hills1980}%
  \BibitemOpen
  \bibfield  {author} {\bibinfo {author} {\bibfnamefont {J.~G.}\ \bibnamefont
  {{Hills}}}\ and\ \bibinfo {author} {\bibfnamefont {L.~W.}\ \bibnamefont
  {{Fullerton}}},\ }\bibfield  {title} {\enquote {\bibinfo {title} {{Computer
  simulations of close encounters between single stars and hard binaries}},}\
  }\href {\doibase 10.1086/112798} {\bibfield  {journal} {\bibinfo  {journal}
  {\aj}\ }\textbf {\bibinfo {volume} {85}},\ \bibinfo {pages} {1281--1291}
  (\bibinfo {year} {1980})}\BibitemShut {NoStop}%
\bibitem [{\citenamefont {Mouri}\ and\ \citenamefont
  {Taniguchi}(2002{\natexlab{b}})}]{mouri2002}%
  \BibitemOpen
  \bibfield  {author} {\bibinfo {author} {\bibfnamefont {Hideaki}\ \bibnamefont
  {Mouri}}\ and\ \bibinfo {author} {\bibfnamefont {Yoshiaki}\ \bibnamefont
  {Taniguchi}},\ }\bibfield  {title} {\enquote {\bibinfo {title} {Runaway
  merging of black holes: Analytical constraint on the timescale},}\ }\href
  {\doibase 10.1086/339472} {\bibfield  {journal} {\bibinfo  {journal} {The
  Astrophysical Journal}\ }\textbf {\bibinfo {volume} {566}},\ \bibinfo {pages}
  {L17–L20} (\bibinfo {year} {2002}{\natexlab{b}})}\BibitemShut {NoStop}%
\bibitem [{\citenamefont {{Turner}}(1977)}]{turner1977}%
  \BibitemOpen
  \bibfield  {author} {\bibinfo {author} {\bibfnamefont {M.}~\bibnamefont
  {{Turner}}},\ }\bibfield  {title} {\enquote {\bibinfo {title} {{Gravitational
  radiation from point-masses in unbound orbits: Newtonian results.}}}\ }\href
  {\doibase 10.1086/155501} {\bibfield  {journal} {\bibinfo  {journal} {\apj}\
  }\textbf {\bibinfo {volume} {216}},\ \bibinfo {pages} {610--619} (\bibinfo
  {year} {1977})}\BibitemShut {NoStop}%
\bibitem [{\citenamefont {Cholis}\ \emph {et~al.}(2016)\citenamefont {Cholis},
  \citenamefont {Kovetz}, \citenamefont {Ali-Haïmoud}, \citenamefont {Bird},
  \citenamefont {Kamionkowski}, \citenamefont {Muñoz},\ and\ \citenamefont
  {Raccanelli}}]{cholis2016}%
  \BibitemOpen
  \bibfield  {author} {\bibinfo {author} {\bibfnamefont {Ilias}\ \bibnamefont
  {Cholis}}, \bibinfo {author} {\bibfnamefont {Ely~D.}\ \bibnamefont {Kovetz}},
  \bibinfo {author} {\bibfnamefont {Yacine}\ \bibnamefont {Ali-Haïmoud}},
  \bibinfo {author} {\bibfnamefont {Simeon}\ \bibnamefont {Bird}}, \bibinfo
  {author} {\bibfnamefont {Marc}\ \bibnamefont {Kamionkowski}}, \bibinfo
  {author} {\bibfnamefont {Julian~B.}\ \bibnamefont {Muñoz}}, \ and\ \bibinfo
  {author} {\bibfnamefont {Alvise}\ \bibnamefont {Raccanelli}},\ }\bibfield
  {title} {\enquote {\bibinfo {title} {Orbital eccentricities in primordial
  black hole binaries},}\ }\href {\doibase 10.1103/physrevd.94.084013}
  {\bibfield  {journal} {\bibinfo  {journal} {Physical Review D}\ }\textbf
  {\bibinfo {volume} {94}} (\bibinfo {year} {2016}),\
  10.1103/physrevd.94.084013}\BibitemShut {NoStop}%
\bibitem [{\citenamefont {O’Leary}\ \emph {et~al.}(2009)\citenamefont
  {O’Leary}, \citenamefont {Kocsis},\ and\ \citenamefont
  {Loeb}}]{oleary2009}%
  \BibitemOpen
  \bibfield  {author} {\bibinfo {author} {\bibfnamefont {Ryan~M.}\ \bibnamefont
  {O’Leary}}, \bibinfo {author} {\bibfnamefont {Bence}\ \bibnamefont
  {Kocsis}}, \ and\ \bibinfo {author} {\bibfnamefont {Abraham}\ \bibnamefont
  {Loeb}},\ }\bibfield  {title} {\enquote {\bibinfo {title} {Gravitational
  waves from scattering of stellar-mass black holes in galactic nuclei},}\
  }\href {\doibase 10.1111/j.1365-2966.2009.14653.x} {\bibfield  {journal}
  {\bibinfo  {journal} {Monthly Notices of the Royal Astronomical Society}\
  }\textbf {\bibinfo {volume} {395}},\ \bibinfo {pages} {2127–2146} (\bibinfo
  {year} {2009})}\BibitemShut {NoStop}%
\bibitem [{\citenamefont {Quinlan}(1996)}]{quinlan1996}%
  \BibitemOpen
  \bibfield  {author} {\bibinfo {author} {\bibfnamefont {Gerald~D.}\
  \bibnamefont {Quinlan}},\ }\bibfield  {title} {\enquote {\bibinfo {title}
  {The dynamical evolution of massive black hole binaries i. hardening in a
  fixed stellar background},}\ }\href {\doibase 10.1016/s1384-1076(96)00003-6}
  {\bibfield  {journal} {\bibinfo  {journal} {New Astronomy}\ }\textbf
  {\bibinfo {volume} {1}},\ \bibinfo {pages} {35–56} (\bibinfo {year}
  {1996})}\BibitemShut {NoStop}%
\bibitem [{\citenamefont {Sesana}\ \emph
  {et~al.}(2006{\natexlab{b}})\citenamefont {Sesana}, \citenamefont {Haardt},\
  and\ \citenamefont {Madau}}]{sesana2006}%
  \BibitemOpen
  \bibfield  {author} {\bibinfo {author} {\bibfnamefont {Alberto}\ \bibnamefont
  {Sesana}}, \bibinfo {author} {\bibfnamefont {Francesco}\ \bibnamefont
  {Haardt}}, \ and\ \bibinfo {author} {\bibfnamefont {Piero}\ \bibnamefont
  {Madau}},\ }\bibfield  {title} {\enquote {\bibinfo {title} {Interaction of
  massive black hole binaries with their stellar environment. i. ejection of
  hypervelocity stars},}\ }\href {\doibase 10.1086/507596} {\bibfield
  {journal} {\bibinfo  {journal} {The Astrophysical Journal}\ }\textbf
  {\bibinfo {volume} {651}},\ \bibinfo {pages} {392–400} (\bibinfo {year}
  {2006}{\natexlab{b}})}\BibitemShut {NoStop}%
\bibitem [{\citenamefont {{Sigurdsson}}\ and\ \citenamefont
  {{Phinney}}(1993)}]{sigurdsson1993}%
  \BibitemOpen
  \bibfield  {author} {\bibinfo {author} {\bibfnamefont {Steinn}\ \bibnamefont
  {{Sigurdsson}}}\ and\ \bibinfo {author} {\bibfnamefont {E.~S.}\ \bibnamefont
  {{Phinney}}},\ }\bibfield  {title} {\enquote {\bibinfo {title}
  {{Binary--Single Star Interactions in Globular Clusters}},}\ }\href {\doibase
  10.1086/173190} {\bibfield  {journal} {\bibinfo  {journal} {\apj}\ }\textbf
  {\bibinfo {volume} {415}},\ \bibinfo {pages} {631} (\bibinfo {year}
  {1993})}\BibitemShut {NoStop}%
\bibitem [{\citenamefont {Biava}\ \emph {et~al.}(2019)\citenamefont {Biava},
  \citenamefont {Colpi}, \citenamefont {Capelo}, \citenamefont {Bonetti},
  \citenamefont {Volonteri}, \citenamefont {Tamfal}, \citenamefont {Mayer},\
  and\ \citenamefont {Sesana}}]{biava2019}%
  \BibitemOpen
  \bibfield  {author} {\bibinfo {author} {\bibfnamefont {Nadia}\ \bibnamefont
  {Biava}}, \bibinfo {author} {\bibfnamefont {Monica}\ \bibnamefont {Colpi}},
  \bibinfo {author} {\bibfnamefont {Pedro~R}\ \bibnamefont {Capelo}}, \bibinfo
  {author} {\bibfnamefont {Matteo}\ \bibnamefont {Bonetti}}, \bibinfo {author}
  {\bibfnamefont {Marta}\ \bibnamefont {Volonteri}}, \bibinfo {author}
  {\bibfnamefont {Tomas}\ \bibnamefont {Tamfal}}, \bibinfo {author}
  {\bibfnamefont {Lucio}\ \bibnamefont {Mayer}}, \ and\ \bibinfo {author}
  {\bibfnamefont {Alberto}\ \bibnamefont {Sesana}},\ }\bibfield  {title}
  {\enquote {\bibinfo {title} {The lifetime of binary black holes in sérsic
  galaxy models},}\ }\href {\doibase 10.1093/mnras/stz1614} {\bibfield
  {journal} {\bibinfo  {journal} {Monthly Notices of the Royal Astronomical
  Society}\ }\textbf {\bibinfo {volume} {487}},\ \bibinfo {pages} {4985–4994}
  (\bibinfo {year} {2019})}\BibitemShut {NoStop}%
\bibitem [{\citenamefont {Samsing}\ \emph {et~al.}(2019)\citenamefont
  {Samsing}, \citenamefont {D'Orazio}, \citenamefont {Kremer}, \citenamefont
  {Rodriguez},\ and\ \citenamefont {Askar}}]{samsing2019}%
  \BibitemOpen
  \bibfield  {author} {\bibinfo {author} {\bibfnamefont {Johan}\ \bibnamefont
  {Samsing}}, \bibinfo {author} {\bibfnamefont {Daniel~J.}\ \bibnamefont
  {D'Orazio}}, \bibinfo {author} {\bibfnamefont {Kyle}\ \bibnamefont {Kremer}},
  \bibinfo {author} {\bibfnamefont {Carl~L.}\ \bibnamefont {Rodriguez}}, \ and\
  \bibinfo {author} {\bibfnamefont {Abbas}\ \bibnamefont {Askar}},\ }\href@noop
  {} {\enquote {\bibinfo {title} {Gravitational-wave captures of single black
  holes in globular clusters},}\ } (\bibinfo {year} {2019}),\ \Eprint
  {http://arxiv.org/abs/1907.11231} {arXiv:1907.11231 [astro-ph.HE]}
  \BibitemShut {NoStop}%
\bibitem [{\citenamefont {Samsing}\ and\ \citenamefont
  {Hotokezaka}(2020)}]{Samsing:2020qqd}%
  \BibitemOpen
  \bibfield  {author} {\bibinfo {author} {\bibfnamefont {Johan}\ \bibnamefont
  {Samsing}}\ and\ \bibinfo {author} {\bibfnamefont {Kenta}\ \bibnamefont
  {Hotokezaka}},\ }\bibfield  {title} {\enquote {\bibinfo {title} {{Populating
  the Black Hole Mass Gaps In Stellar Clusters: General Relations and Upper
  Limits}},}\ }\href@noop {} {\  (\bibinfo {year} {2020})},\ \Eprint
  {http://arxiv.org/abs/2006.09744} {arXiv:2006.09744 [astro-ph.HE]}
  \BibitemShut {NoStop}%
\bibitem [{\citenamefont {Ye}\ \emph {et~al.}(2020)\citenamefont {Ye},
  \citenamefont {Fong}, \citenamefont {Kremer}, \citenamefont {Rodriguez},
  \citenamefont {Chatterjee}, \citenamefont {Fragione},\ and\ \citenamefont
  {Rasio}}]{Ye:2019xvf}%
  \BibitemOpen
  \bibfield  {author} {\bibinfo {author} {\bibfnamefont {Claire~S.}\
  \bibnamefont {Ye}}, \bibinfo {author} {\bibfnamefont {Wen-fai}\ \bibnamefont
  {Fong}}, \bibinfo {author} {\bibfnamefont {Kyle}\ \bibnamefont {Kremer}},
  \bibinfo {author} {\bibfnamefont {Carl~L.}\ \bibnamefont {Rodriguez}},
  \bibinfo {author} {\bibfnamefont {Sourav}\ \bibnamefont {Chatterjee}},
  \bibinfo {author} {\bibfnamefont {Giacomo}\ \bibnamefont {Fragione}}, \ and\
  \bibinfo {author} {\bibfnamefont {Frederic~A.}\ \bibnamefont {Rasio}},\
  }\bibfield  {title} {\enquote {\bibinfo {title} {{On the Rate of Neutron Star
  Binary Mergers from Globular Clusters}},}\ }\href {\doibase
  10.3847/2041-8213/ab5dc5} {\bibfield  {journal} {\bibinfo  {journal}
  {Astrophys. J. Lett.}\ }\textbf {\bibinfo {volume} {888}},\ \bibinfo {pages}
  {L10} (\bibinfo {year} {2020})},\ \Eprint {http://arxiv.org/abs/1910.10740}
  {arXiv:1910.10740 [astro-ph.HE]} \BibitemShut {NoStop}%
\bibitem [{\citenamefont {Ade}\ \emph {et~al.}(2016)\citenamefont {Ade} \emph
  {et~al.}}]{Ade:2015xua}%
  \BibitemOpen
  \bibfield  {author} {\bibinfo {author} {\bibfnamefont {P.A.R.}\ \bibnamefont
  {Ade}} \emph {et~al.} (\bibinfo {collaboration} {Planck}),\ }\bibfield
  {title} {\enquote {\bibinfo {title} {{Planck 2015 results. XIII. Cosmological
  parameters}},}\ }\href {\doibase 10.1051/0004-6361/201525830} {\bibfield
  {journal} {\bibinfo  {journal} {Astron. Astrophys.}\ }\textbf {\bibinfo
  {volume} {594}},\ \bibinfo {pages} {A13} (\bibinfo {year} {2016})},\ \Eprint
  {http://arxiv.org/abs/1502.01589} {arXiv:1502.01589 [astro-ph.CO]}
  \BibitemShut {NoStop}%
\bibitem [{\citenamefont {Hogg}(1999)}]{Hogg:1999ad}%
  \BibitemOpen
  \bibfield  {author} {\bibinfo {author} {\bibfnamefont {David~W.}\
  \bibnamefont {Hogg}},\ }\bibfield  {title} {\enquote {\bibinfo {title}
  {{Distance measures in cosmology}},}\ }\href@noop {} {\  (\bibinfo {year}
  {1999})},\ \Eprint {http://arxiv.org/abs/astro-ph/9905116}
  {arXiv:astro-ph/9905116} \BibitemShut {NoStop}%
\bibitem [{\citenamefont {Xu}\ and\ \citenamefont {Zhang}(2016)}]{Xu:2016grp}%
  \BibitemOpen
  \bibfield  {author} {\bibinfo {author} {\bibfnamefont {Yue-Yao}\ \bibnamefont
  {Xu}}\ and\ \bibinfo {author} {\bibfnamefont {Xin}\ \bibnamefont {Zhang}},\
  }\bibfield  {title} {\enquote {\bibinfo {title} {{Comparison of dark energy
  models after Planck 2015}},}\ }\href {\doibase
  10.1140/epjc/s10052-016-4446-5} {\bibfield  {journal} {\bibinfo  {journal}
  {Eur. Phys. J. C}\ }\textbf {\bibinfo {volume} {76}},\ \bibinfo {pages} {588}
  (\bibinfo {year} {2016})},\ \Eprint {http://arxiv.org/abs/1607.06262}
  {arXiv:1607.06262 [astro-ph.CO]} \BibitemShut {NoStop}%
\bibitem [{\citenamefont {{Kozai}}(1962)}]{kozai1962}%
  \BibitemOpen
  \bibfield  {author} {\bibinfo {author} {\bibfnamefont {Yoshihide}\
  \bibnamefont {{Kozai}}},\ }\bibfield  {title} {\enquote {\bibinfo {title}
  {{Secular perturbations of asteroids with high inclination and
  eccentricity}},}\ }\href {\doibase 10.1086/108790} {\bibfield  {journal}
  {\bibinfo  {journal} {\aj}\ }\textbf {\bibinfo {volume} {67}},\ \bibinfo
  {pages} {591--598} (\bibinfo {year} {1962})}\BibitemShut {NoStop}%
\bibitem [{\citenamefont {{Lidov}}(1962)}]{1962P&SS....9..719L}%
  \BibitemOpen
  \bibfield  {author} {\bibinfo {author} {\bibfnamefont {M.~L.}\ \bibnamefont
  {{Lidov}}},\ }\bibfield  {title} {\enquote {\bibinfo {title} {{The evolution
  of orbits of artificial satellites of planets under the action of
  gravitational perturbations of external bodies}},}\ }\href {\doibase
  10.1016/0032-0633(62)90129-0} {\bibfield  {journal} {\bibinfo  {journal}
  {\planss}\ }\textbf {\bibinfo {volume} {9}},\ \bibinfo {pages} {719--759}
  (\bibinfo {year} {1962})}\BibitemShut {NoStop}%
\bibitem [{\citenamefont {Naoz}(2016)}]{Naoz_2016}%
  \BibitemOpen
  \bibfield  {author} {\bibinfo {author} {\bibfnamefont {Smadar}\ \bibnamefont
  {Naoz}},\ }\bibfield  {title} {\enquote {\bibinfo {title} {The eccentric
  kozai-lidov effect and its applications},}\ }\href {\doibase
  10.1146/annurev-astro-081915-023315} {\bibfield  {journal} {\bibinfo
  {journal} {Annual Review of Astronomy and Astrophysics}\ }\textbf {\bibinfo
  {volume} {54}},\ \bibinfo {pages} {441–489} (\bibinfo {year}
  {2016})}\BibitemShut {NoStop}%
\end{thebibliography}%

\appendix

\section{Numerical setup for the BBH evolution in the 3rd-body channel}
\label{Appendix_numerical}

In this appendix we present the numerical  setup of our calculations for the case of
the evolution of a BBH due to GW emission and interactions with 3rd-bodies.

\begin{figure}[h]
    \centering
    \includegraphics[width=8.5cm,height=5.2cm]{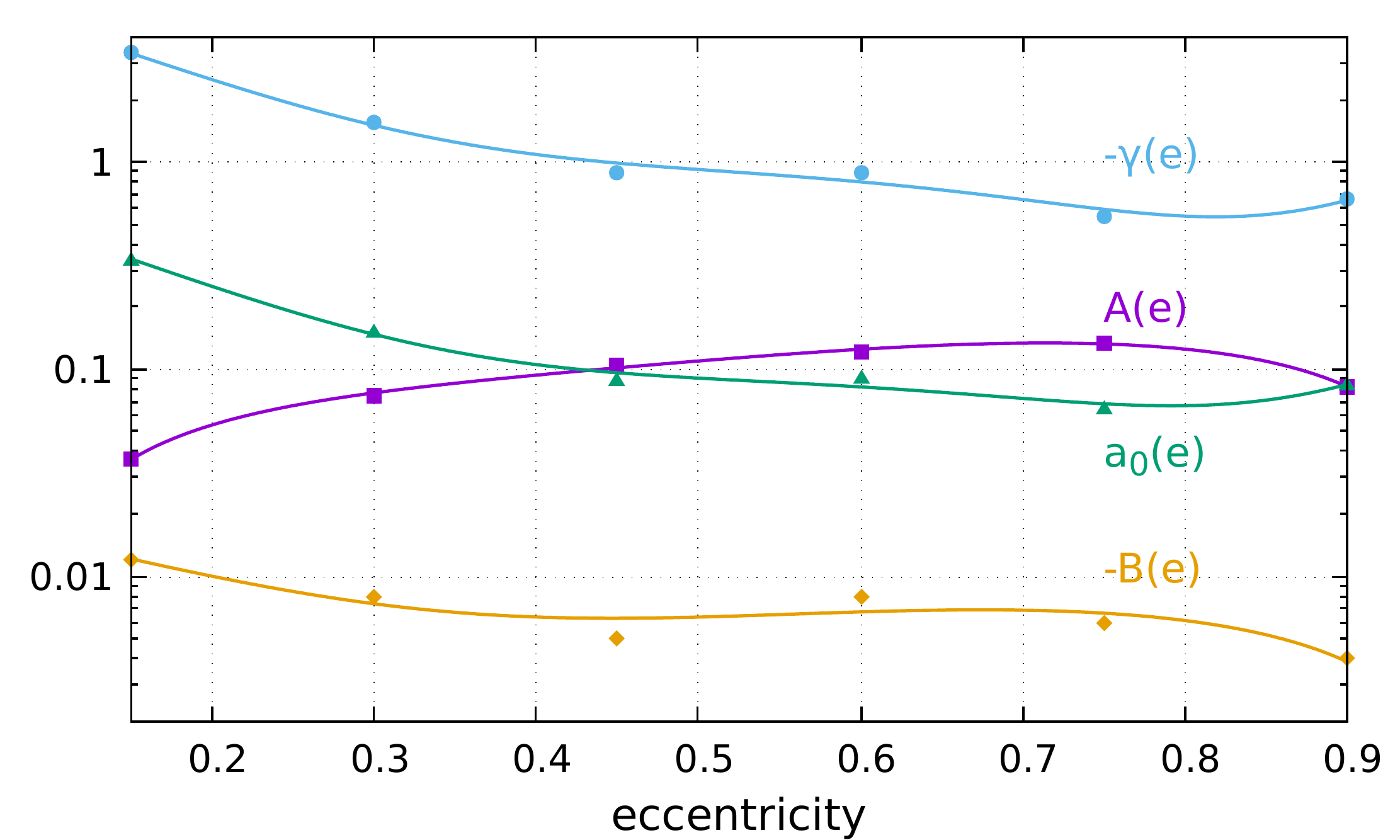}
    \includegraphics[width=8.5cm,height=5.2cm]{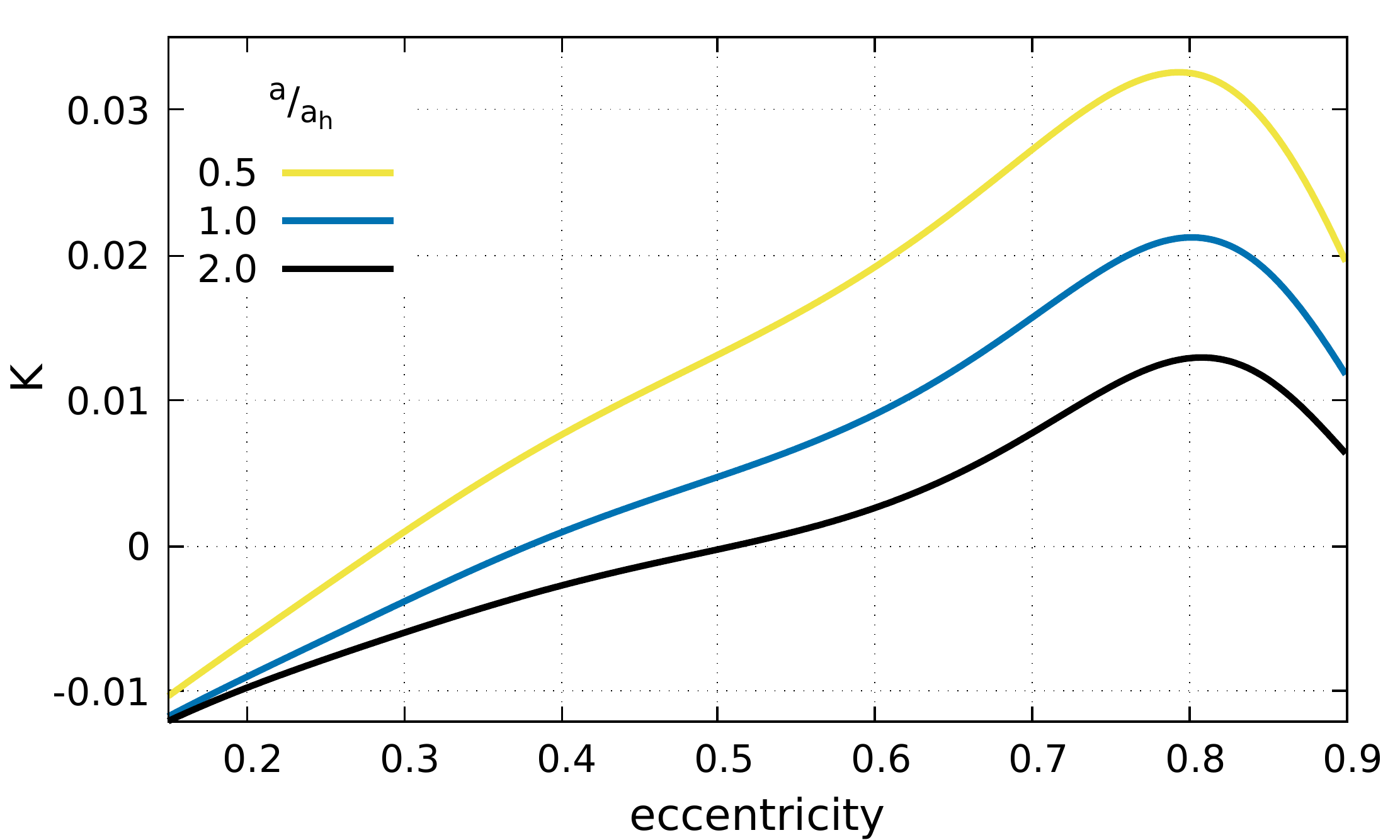}
    \caption{\textit{Top panel:} Cubic  fits on the data (solid circles, squares, triangles and diamonds) in table 3 of \cite{Sesana:2006xw} for $q=1$. \textit{Bottom panel:} The eccentricity growth rate $K(a,e)=A(e)(1+a/a_0(e))^{\gamma(e)}+B(e)$ with the appropriate fitting parameters shown for three values of semi-major axis. Also, $a_h$ is the hardness semi-major axis.}
    \label{KRATE}
\end{figure}

In Eqs.~\ref{SMAevol} and~\ref{ECCevol} 
the term describing the 3rd-body interactions with stars
has a proportionality coefficient $K$.
That coefficient is not constant but depends on the value of the binary's 
eccentricity and semi-major axis. We rely on the numerical work 
by \cite{Sesana:2006xw} to include that dependence.  
In Fig.~\ref{KRATE} we show 
the fitting functions we use for the evolution of a 
BBH as were provided in \cite{Sesana:2006xw}. 
We implement a cubic fit on the data in their table 3. 
These fits are accurate only for values of eccentricity on the 
interval $e\in[0.15,0.9]$. 

An important parameter we should set in our simulations 
is the time-step, $dt$. 
The first term in the right hand side of Eq.~\ref{SMAevol} sets an upper 
limit on the time-step we should use, since it relies on 
Eq.~\ref{avgEnergyVar} 
which describes a single interaction. 
Our time-step has to be bigger than 
the period of the binary so that the Peters term remains secular. 
This sets a lower bound. 
We can choose our the numerical time-step to be that of the 
local interaction timescale,
$dt=T_{\textrm{int}}\sim30 \, \textrm{Myr}$ for 47 Tuc 
as in Eq.~\ref{interTime}, when the binary's semi-major axis 
is $a_0$. 
We remind the reader that at all times, the period of the BBH 
is no more than a few years 
and as the binary tightens its $T_{\textrm{int}}$ increases.

Regarding the end of the evolution of the BBHs, we choose 
$a_{\textrm{end}}=0.01$ AU. 
This corresponds to 1.58 Myr until merger due to GW emission; 
a time much smaller than the Gyrs it takes for the binary to evolve.

\section{The ejection of binaries from globular clusters}
\label{Appendix_ejection}

In this appendix we make a short note on the ejection of a binary 
from a GC considering soft interactions.
During each encounter of a 
binary $A-B$ with a third object $C$, on a statistical average 
the perturber acquires a high kick and by 
momentum conservation the binary recoils. 
This recoil velocity is random and the distribution 
is spherically uniform. 
To a good approximation we can treat 
the binary as having its typical velocity dispersion in each encounter. 
By energy considerations one can show that the final 
relative velocity during the interaction of a binary $A-B$ with 
a third body $C$,
is, \cite{sigurdsson1993},
\begin{equation}
    \sigma_{A-B,C}{'}=\sqrt{\sigma_{A-B,C}^2+2{2\over\mu_{A-B,C}}\cdot\Delta E_b},
\end{equation}
where $\Delta E_b$ is given by Eq.~\ref{avgEnergyVar}.
Assuming that $m_A+m_B\gg m_C$ we can
estimate the recoil velocity, 
from momentum conservation, 
as, $v_{\textrm{recoil}}\approx{m_C\over m_A+m_B}\ v_{A-B,C}{'}$.
The binary will be kept in the cluster as long as the kick 
it acquires is not enough to eject it from the GC. 
The safest condition that the binary remains in the 
GC is that even if the recoil it acquires happens 
to be exactly opposite to its velocity 
with respect to the center of the GC it 
still has not surpassed the escape 
velocity threshold, i.e., 
$v_{\textrm{esc}}>v_{\textrm{kick}}^{\textrm{min}}=v_{\textrm{recoil}}-\sigma_{A-B,C}$,
where we will use $v_{\textrm{esc}}=2\sigma_{\textrm{star}}$ is the escape 
velocity. Finally, we obtain our 
condition for the critical semi-major axis to be, for $m_A=m_B=10 \, M_{\odot}$ and $m_C=1 \, M_{\odot}$,
\begin{equation}
    a>a_{ej}\approx0.0128\times\left({H\over15}\right)\left({20 \, \textrm{km/s}\over\sigma_{\textrm{star}}}\right)^2 \textrm{AU},
\end{equation}
i.e. it should not be smaller than a critical 
value of roughly $a_{ej}\sim10^{-2}$ AU 
at core radius in 47 Tuc. 
This value is only slightly bigger than $a_{\textrm{end}}$. 
Therefore, the binary’s orbit is quite tight when it 
is ejected. 
We note again that these statements 
are valid in our context where BBHs mostly encounter $1\, M_{\odot}$ 
stars and interact softly yet efficiently.
If the BBH encounters another BH, 
the ejection point occurs at a larger semi-major axis, 
about $O(10)$ larger than our result, 
as in \cite{samsing2018},
\begin{equation}
    a_{ej}\simeq 0.266 \times\left({m_{BH}\over \, 10M_{\odot}}\right) \left({40 \, \textrm{km/s}\over v_{\textrm{esc}}}\right)^2 \, \textrm{AU}.
\end{equation}

\end{document}